\documentclass[11pt]{rsauthor}
\usepackage{geometry}                
\usepackage[parfill]{parskip}    
\usepackage{graphicx,lineno}
\usepackage{amsmath,amssymb}
\usepackage{epstopdf}
\usepackage{natbib}
\usepackage{appendix}

\def\nr0{n_{r}^{\hspace{-0.9ex}^{\circ}}}

     \def\aap{A\&A}
     \def\apjl{ApJL}
     \def\apjs{ApJS}

\title{Atmospheric electrification in dusty, reactive gases in the solar system and beyond}

\author{Christiane Helling$^1$, R. Giles Harrison$^2$, Farideh Honary$^3$, Declan A. Diver$^4$,   Karen Aplin$^5$, Ian Dobbs-Dixon$^6$, Ute Ebert$^7$, Shu-ichiro Inutsuka$^8$,  Francisco J. Gordillo-Vazquez$^9$, Stuart Littlefair$^{10}$
 \\ $^1$ SUPA, School of Physics \& Astronomy, University of St Andrews,\\ North Haugh, KY16 9SS, UK\\
 $^2$ Department of Meteorology, The University of Reading, UK\\
 $^3$ Department of Physics, Lancaster University, Lancaster, UK\\
 $^4$ SUPA, School of Physics \& Astronomy, University of Glasgow, Glasgow G12 8QQ, UK\\
 $^5$ Department of Physics, University of Oxford, Denys Wilkinson Building, Keble Road, Oxford OX1 3RH, UK\\
 $^6$ NYU Abu Dhabi P.O. Box 129188 Abu Dhabi, UAE\\
 $^7$ Centrum Wiskunde \& Informatica, Amsterdam, The Netherlands\\
$^8$ Department of Physics, Nagoya University, Nagoya, Aichi 464-8602, Japan\\
 $^9$ Instituto de Astrof{\'i}sica de Andaluc{\'i}a
P.O. Box 3004, 18080, Granada, Spain\\
 $^{10}$ Department of Physics and Astronomy, University of Sheffield, Sheffield S3 7RH, UK\\
 }
\date{\today}                                           

\begin{document}
\maketitle
\tableofcontents

{\bf Abstract:} Detailed observations of the solar system planets
reveal a wide variety of local atmospheric conditions. Astronomical
observations have revealed a variety of extrasolar planets none of
which resembles any of the solar system planets in full.  Instead, the
most massive amongst the extrasolar planets, the gas giants, appear
very similar to the class of (young) Brown Dwarfs which are amongst
the oldest objects in the universe.  Despite of this diversity, solar
system planets, extrasolar planets and Brown Dwarfs have broadly
similar global temperatures between 300K and 2500K. In consequence,
clouds of different chemical species form in their atmospheres.  While
the details of these clouds differ, the fundamental physical processes
are the same. Further to this, all these objects were observed to
produce radio and X-ray emission. While both kinds of radiation are
well studied on Earth and to a lesser extent on the solar system
planets, the occurrence of emission that potentially originate from
accelerated electrons on Brown Dwarfs, extrasolar planets and
protoplanetary disks is not well understood yet.  This paper offers an
interdisciplinary view on electrification processes and their feedback
on their hosting environment in meteorology, volcanology, planetology
and research on extrasolar planets and planet formation.

\begin{figure}[ht]
\centerline{
\includegraphics[width=\textwidth]{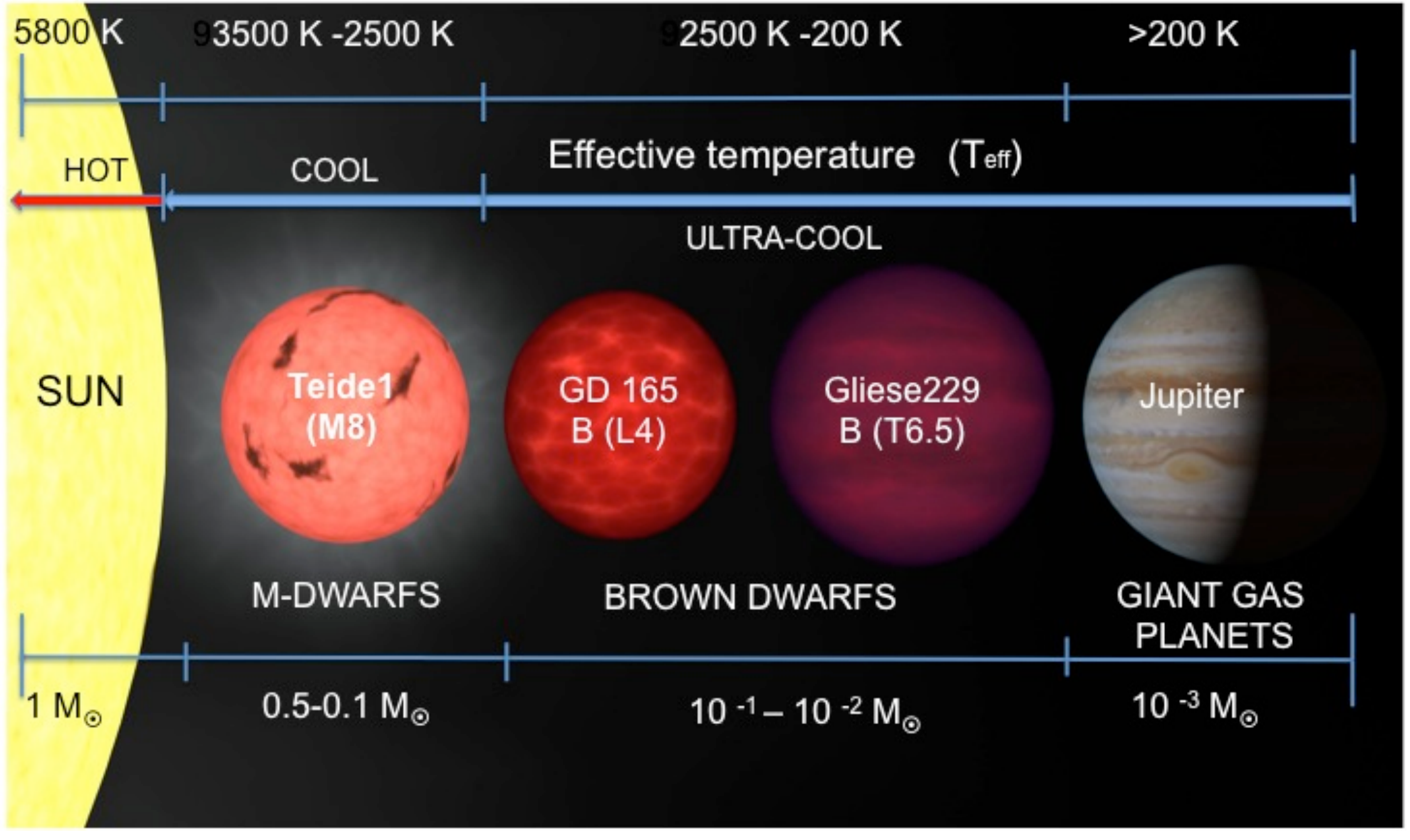}}
\caption{The large context: Planets are the coldest and smallest objects in the universe
  known to possess a cloud-forming and potential life protecting
  atmosphere. Brown Dwarfs are as cool as planets but
  they form like stars (like the Sun) through the collapse of a
  gravitational unstable interstellar cloud. Planets (like Jupiter and
  Earth) form as by-product of star formation in protoplanetary
  disks. Note that the lower temperature boundary is not yet well determined.}
\label{fig:SMDP}
\end{figure}

\newpage

\section{Introduction}
The Earth and the solar system planets were the only planetary
objects known until the discovery of the first brown dwarf GD165B
(\citealt{1988Natur.336..656B}) and the first extrasolar planet in
1992 (orbiting the pulsar PSR1257+12,
\cite{1992Natur.355..145W}). Earth, Jupiter, and Saturn are cloudy
solar system planets for which atmospheric discharges in form of
lightning is confirmed observationally in radio and in optical
wavelengths. Space exploration and ground based observations have
shown that lightning is a process universal in the solar system, but
also that charge and discharge processes occur in a large diversity on
solar system planets. Charging and discharging  processes are
essential for our understanding of the origin of our planet and maybe
even for the origin of life: It is believed that charged dust is
required to form planets and that lightning opens chemical paths to
the formation of biomolecules. The purpose of this paper is 
to point out overlapping interests in electrifying media that
contain liquid and solid particles in meteorology, volcanology, solar
system objects, extrasolar planets, brown dwarfs and protoplanetary
disks. We therefore provide a selective overview of atmospheric
electrification processes and related electrical phenomena based on
knowledge from solar system and Earth observations, and on lab-based
research in combination with relevant findings and development in
research on extrasolar planets, brown dwarfs and protoplanetary
disks. We hope to stimulate a closer interaction between these
communities.

The last few decades have taken us from a Universe with only a single
planetary system known, to one with thousands, and maybe millions, of
such systems.  We are now entering the time when we explore theories
and results derived for the solar system and for Earth in application
to unknown worlds.  Figure~\ref{fig:SMDP} places Jupiter, one of the
solar system giant gas planets, into the astrophysical context:
Jupiter (right) is compared to the coolest stellar objects (M-dwarfs
and Brown Dwarfs).  Brown Dwarfs bridge the stellar (represented by
the Sun in Fig.~\ref{fig:SMDP}) and the planetary regime as their
atmospheres can be as cold as those of planets but they form like
stars. The Sun (left) is surrounded by hot plasma (corona) while
planets are enveloped in a cold cloud forming atmosphere some of which
exhibit electrical phenomena as part of a global electric circuit. The
Sun is intensively studied by satellites like
SOHO\footnote{http://sci.esa.int/soho/} and
HINODE\footnote{http://www.nasa.gov/mission$\_$pages/hinode} leading
to efforts like SWIFF for space weather forecasting
(\citealt{lapenta2013}). Comparable high-resolution monitoring is
neither feasible for solar system planets, moons or comets nor for
extrasolar objects. Instead, experimental work on Earth, Earth
observation, modelling and comparative studies for the solar system
and extrasolar objects need to be combined; examples for Earth studied
as extrasolar planet are e.g. in \cite{kitzmann2010,kaltenegger2013}
and \cite{hod2016}.

\begin{figure}
{\ }\\*[-4cm]
\begin{tabular}{p{4.0cm} p{4.5cm} p{4.0cm}}
\centering  Earth (NASA) & \centering Saturn (ESA) & \begin{minipage}{3.7cm}\centering Brown Dwarfs \end{minipage}\\
\begin{minipage}{3.7cm}
\centering
\includegraphics[width=3.2cm]{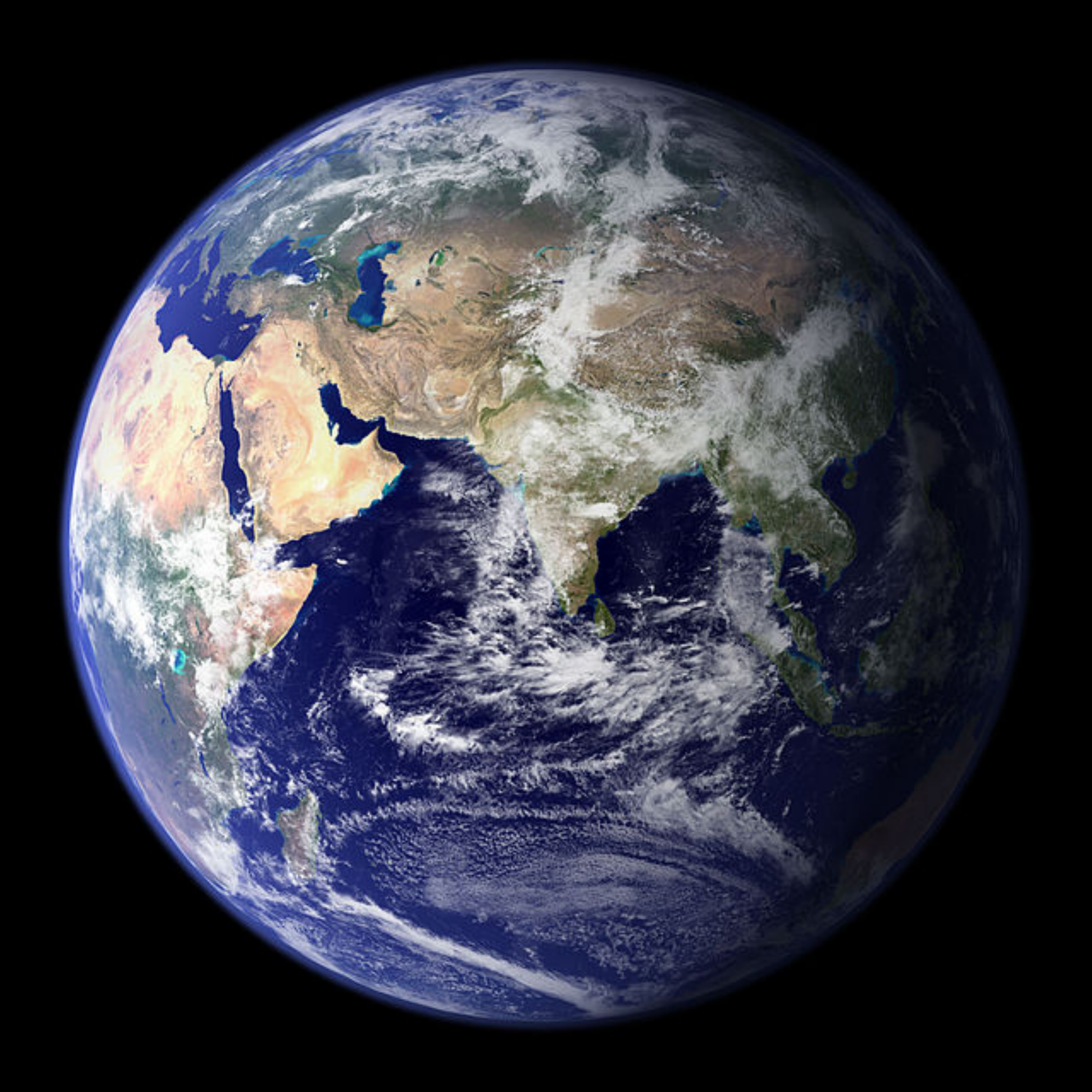} 
\end{minipage}
&
\begin{minipage}{4.0cm}
\centering
 \includegraphics[width=3.2cm]{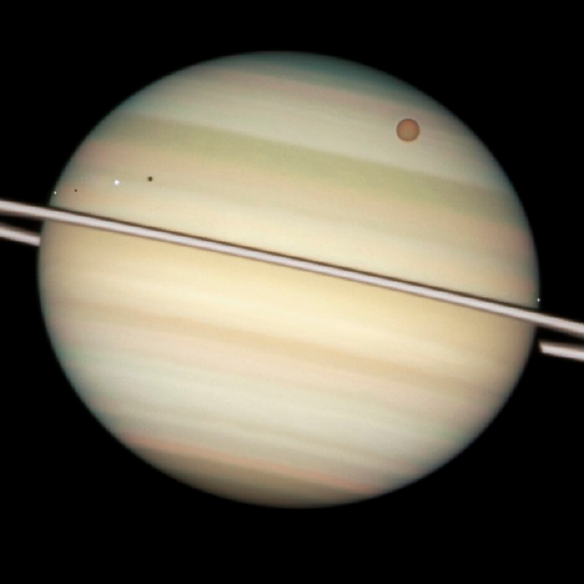}
\end{minipage}
&  
\begin{minipage}{3.7cm}
\centering
\includegraphics[width=4.2cm]{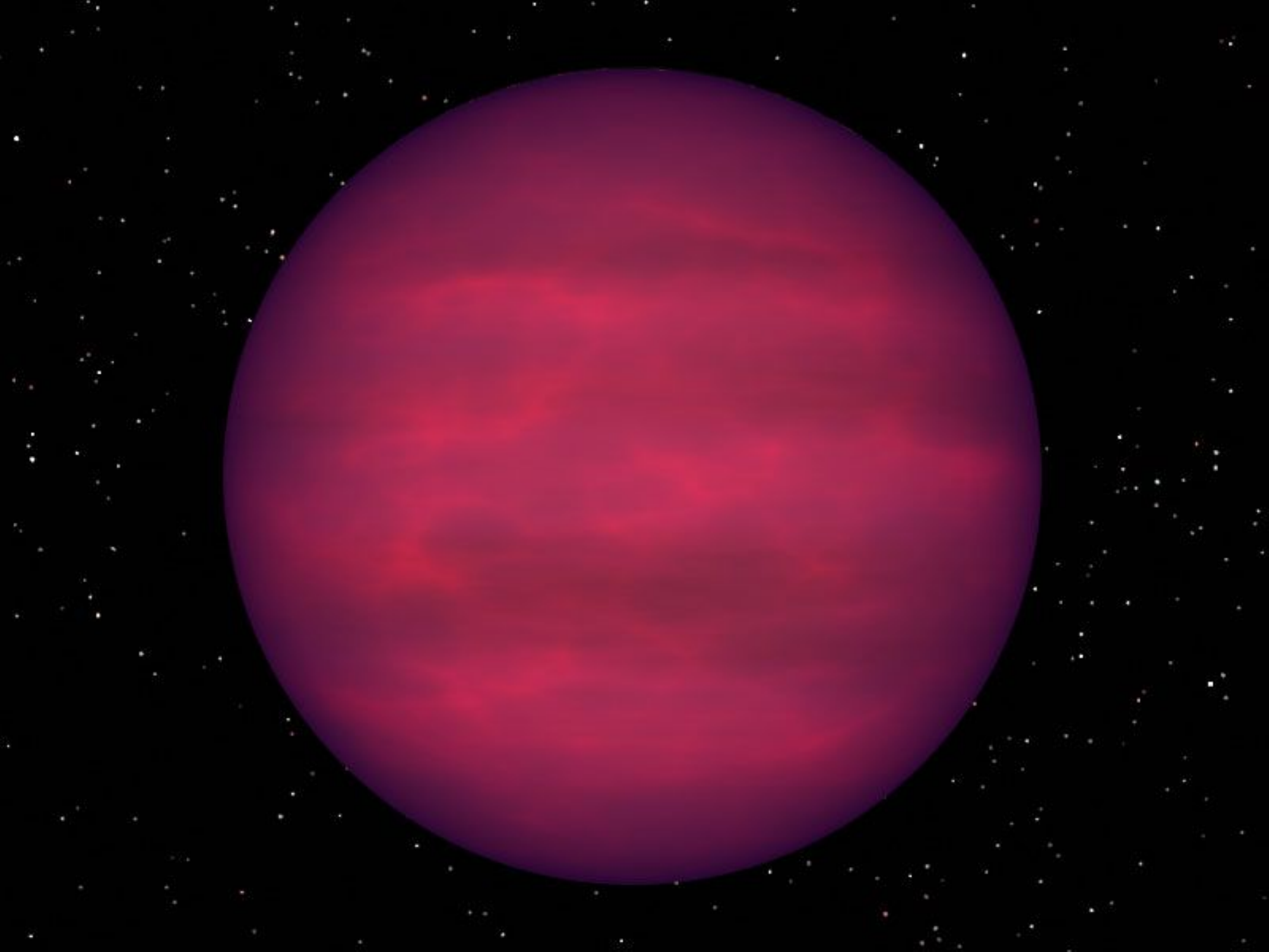}
\end{minipage}\\
\multicolumn{3}{c}{\ }\\*[-0.2cm]
\multicolumn{3}{c}{radiation fluxes $F(\lambda)$:}\\*[-0.2cm]
\begin{minipage}{3.7cm}
\centering
\hspace*{-1cm}\includegraphics[width=4.6cm]{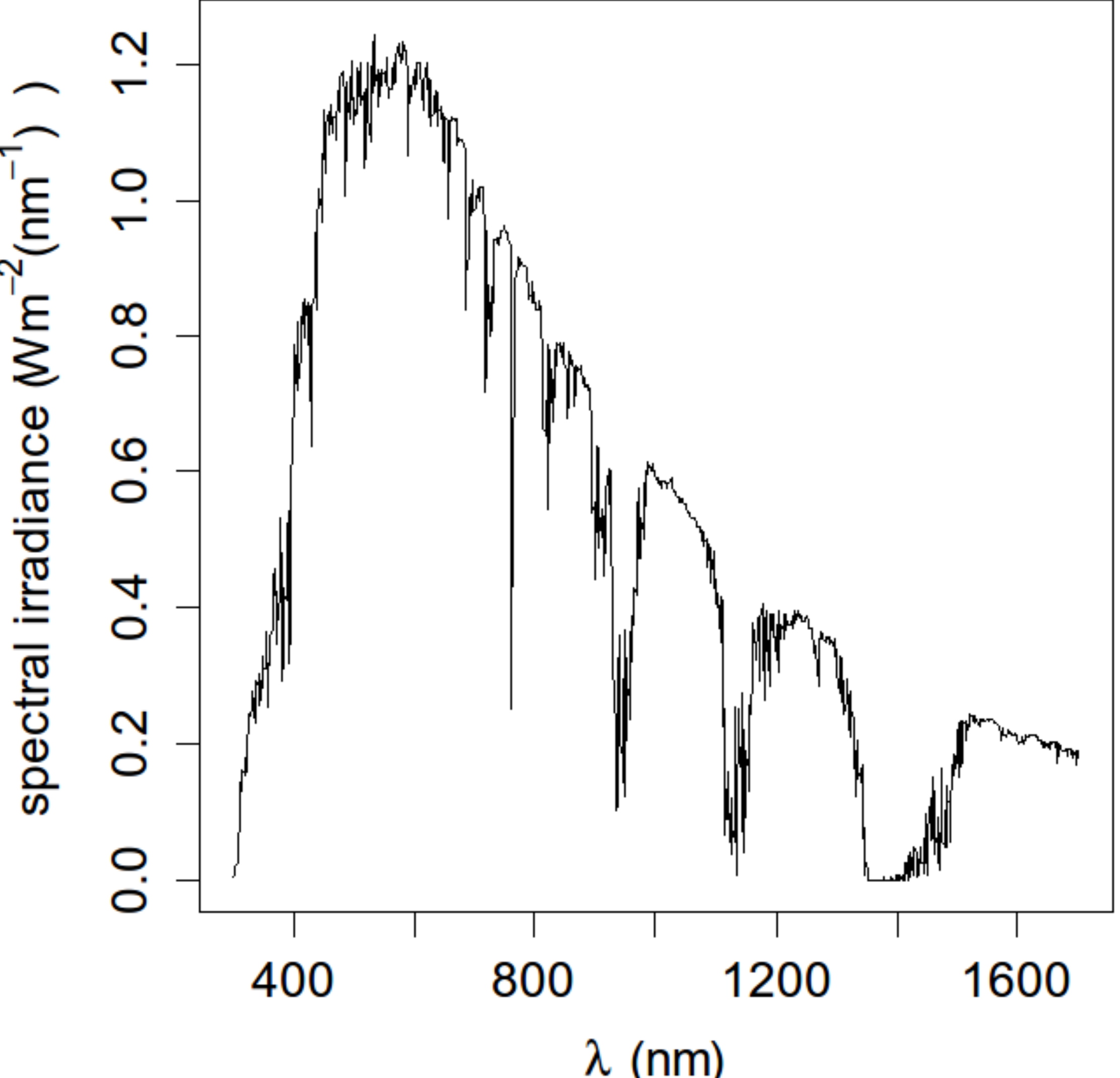}
\end{minipage}
& 
\begin{minipage}{4.0cm}
\centering
\vspace*{-0.2cm}
\hspace*{-0.5cm}\includegraphics[width=4.65cm]{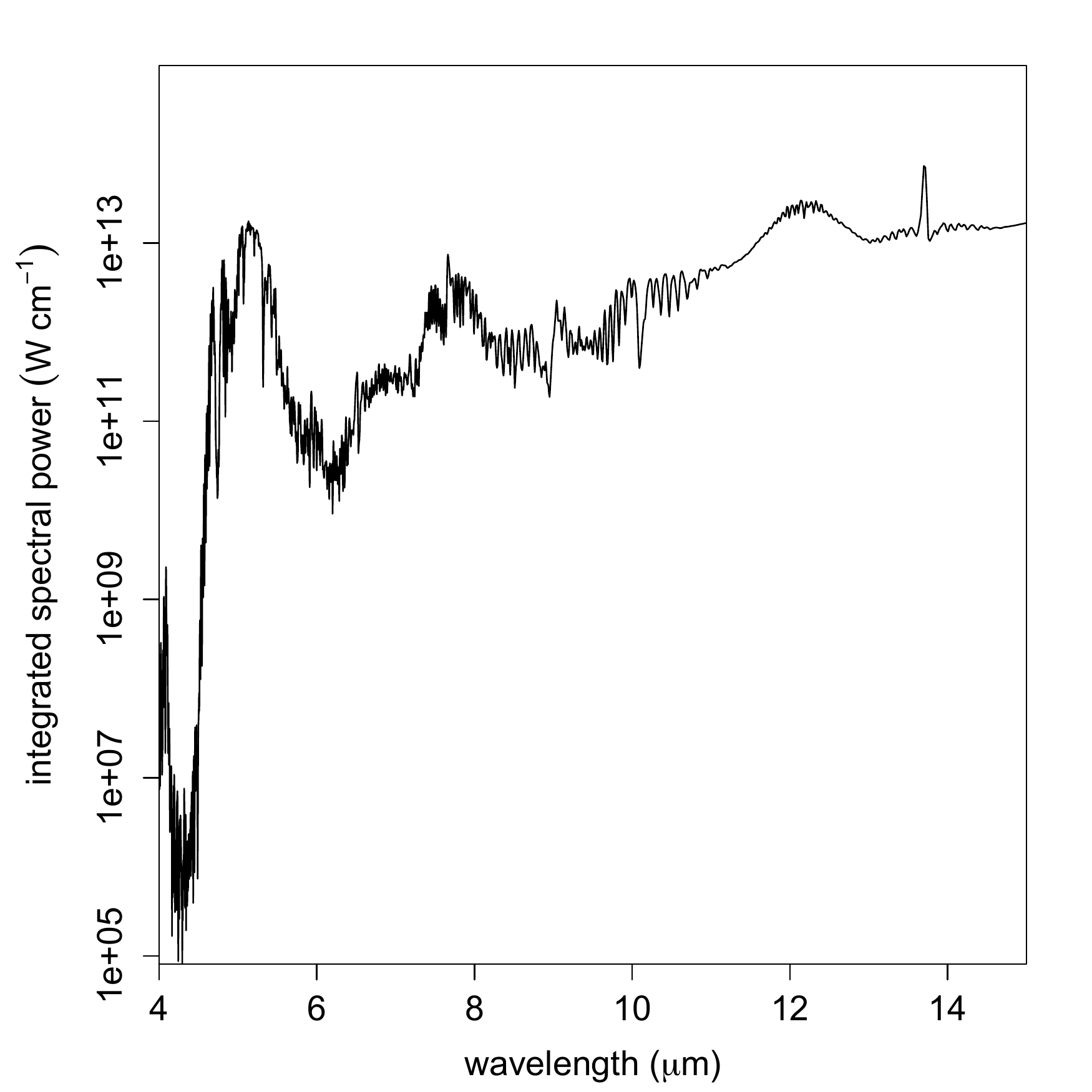} 
\end{minipage}
&
\begin{minipage}{4.0cm}
\centering
\vspace*{-0.2cm}
\hspace*{-2.0cm}\includegraphics[width=7.5cm]{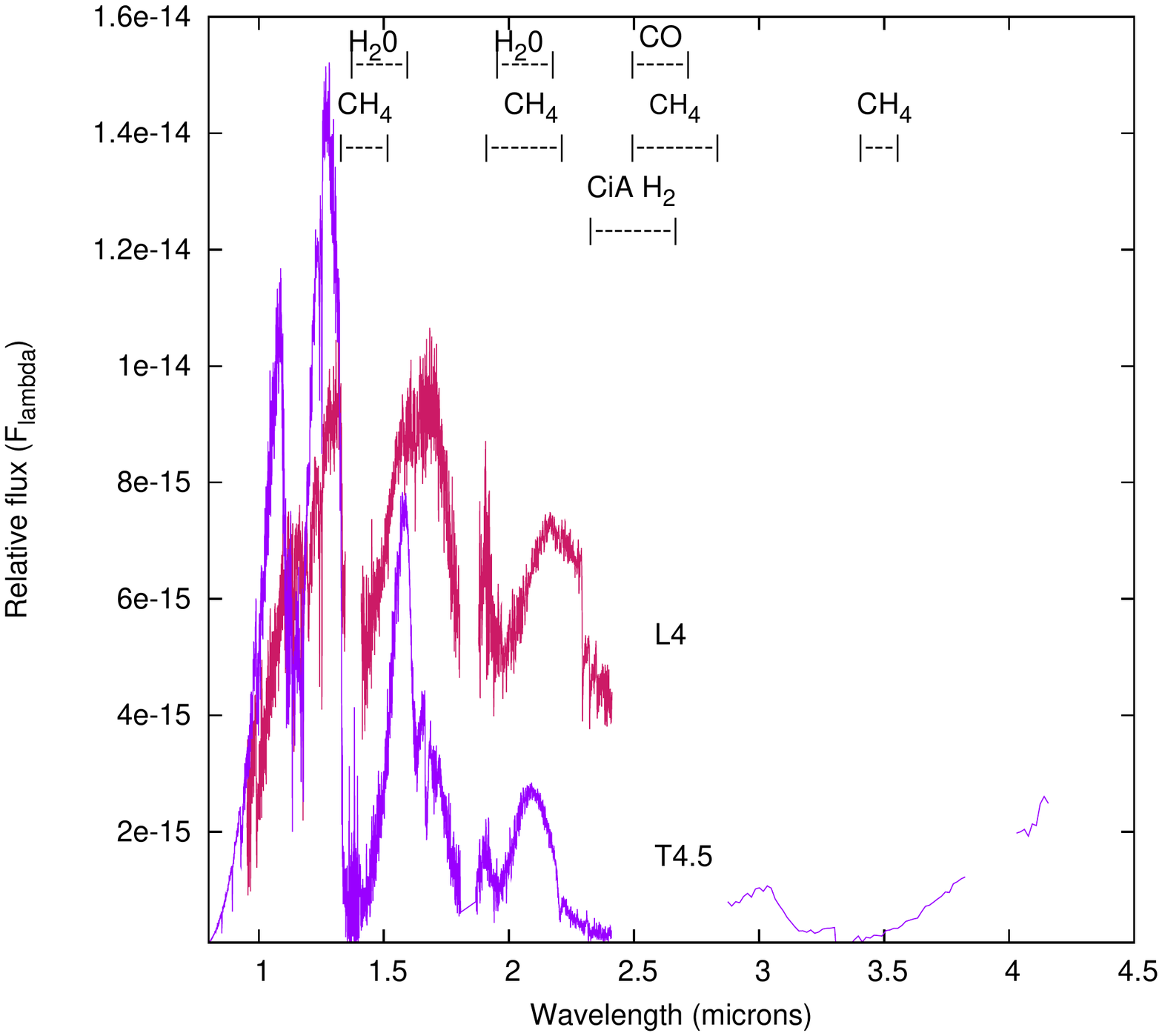}
\end{minipage}
\\
\multicolumn{3}{c}{\ }\\*[-0.7cm]
\multicolumn{3}{c}{atmospheric ($T_{\rm gas}$, p$_{\rm gas}$)-structures:}\\*[0.2cm]
\begin{minipage}{3.7cm}
\centering
\hspace*{-1.3cm}\includegraphics[width=4.6cm]{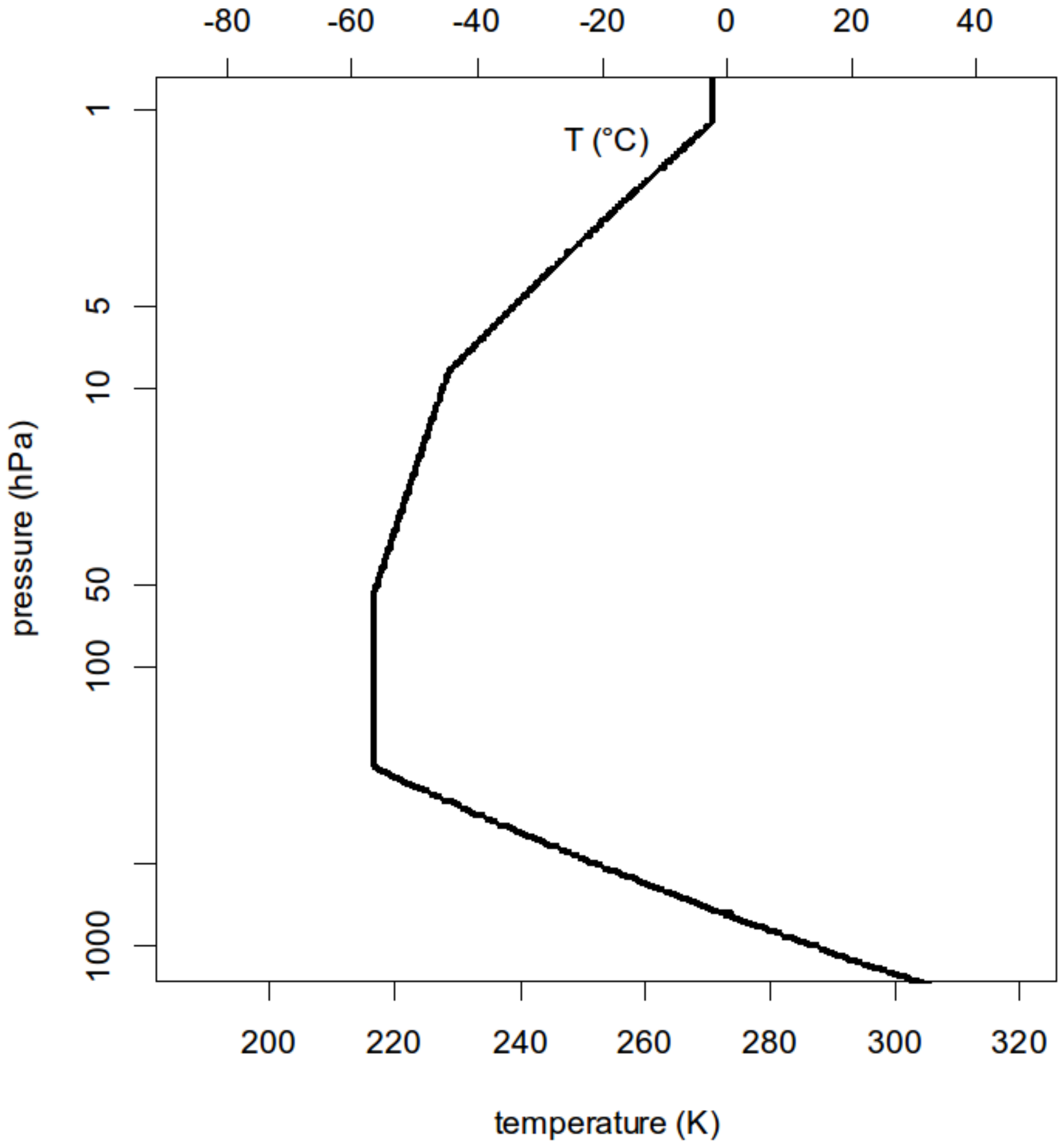}
\end{minipage}
&
\begin{minipage}{4.0cm}
\centering
\vspace*{-0.2cm}
\hspace*{-0.8cm}\includegraphics[width=5.3cm]{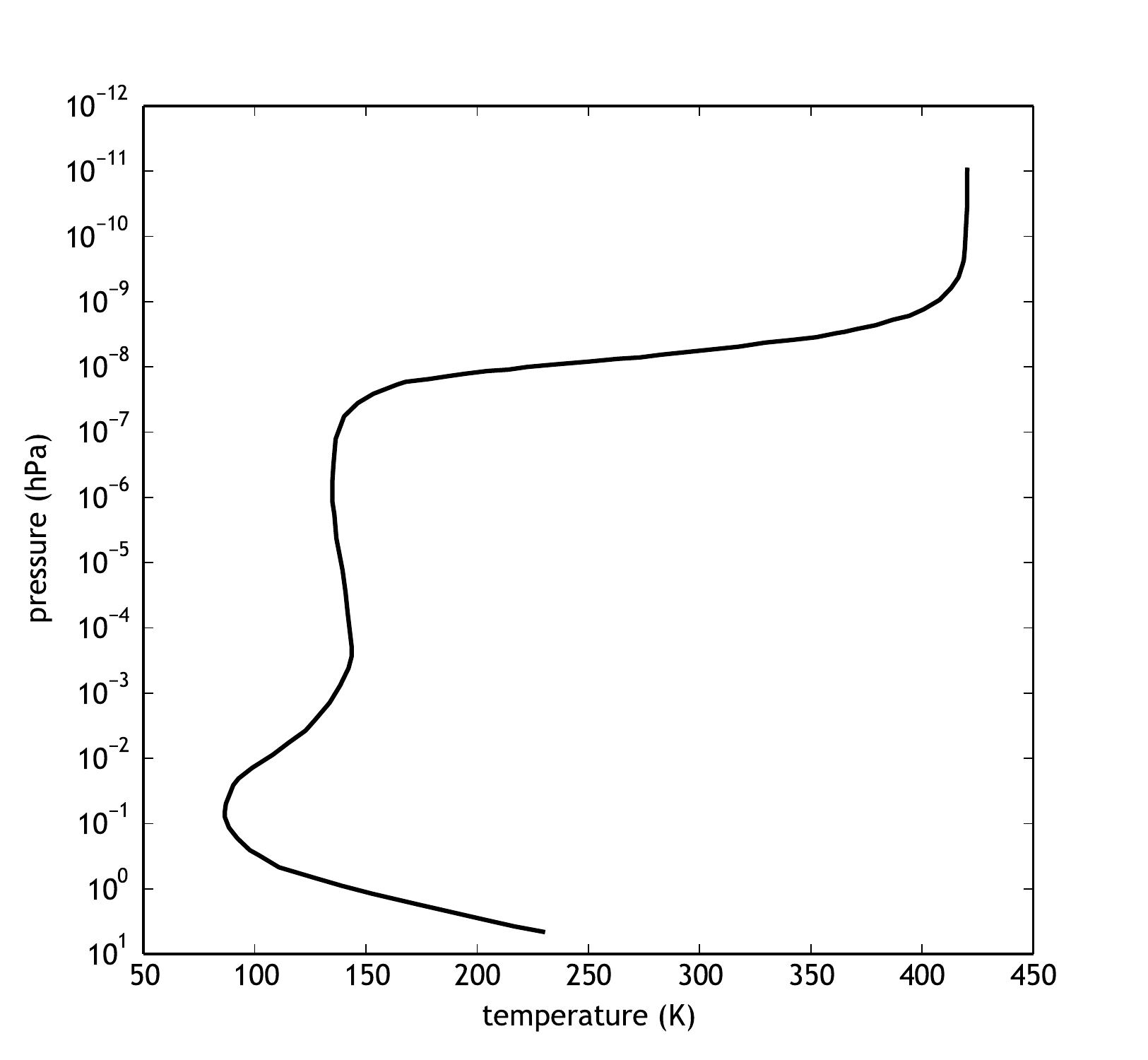} 
\end{minipage}
&
\begin{minipage}{4.0cm}
\centering
\vspace*{0.25cm}
\hspace*{-0.6cm}\includegraphics[width=5.0cm]{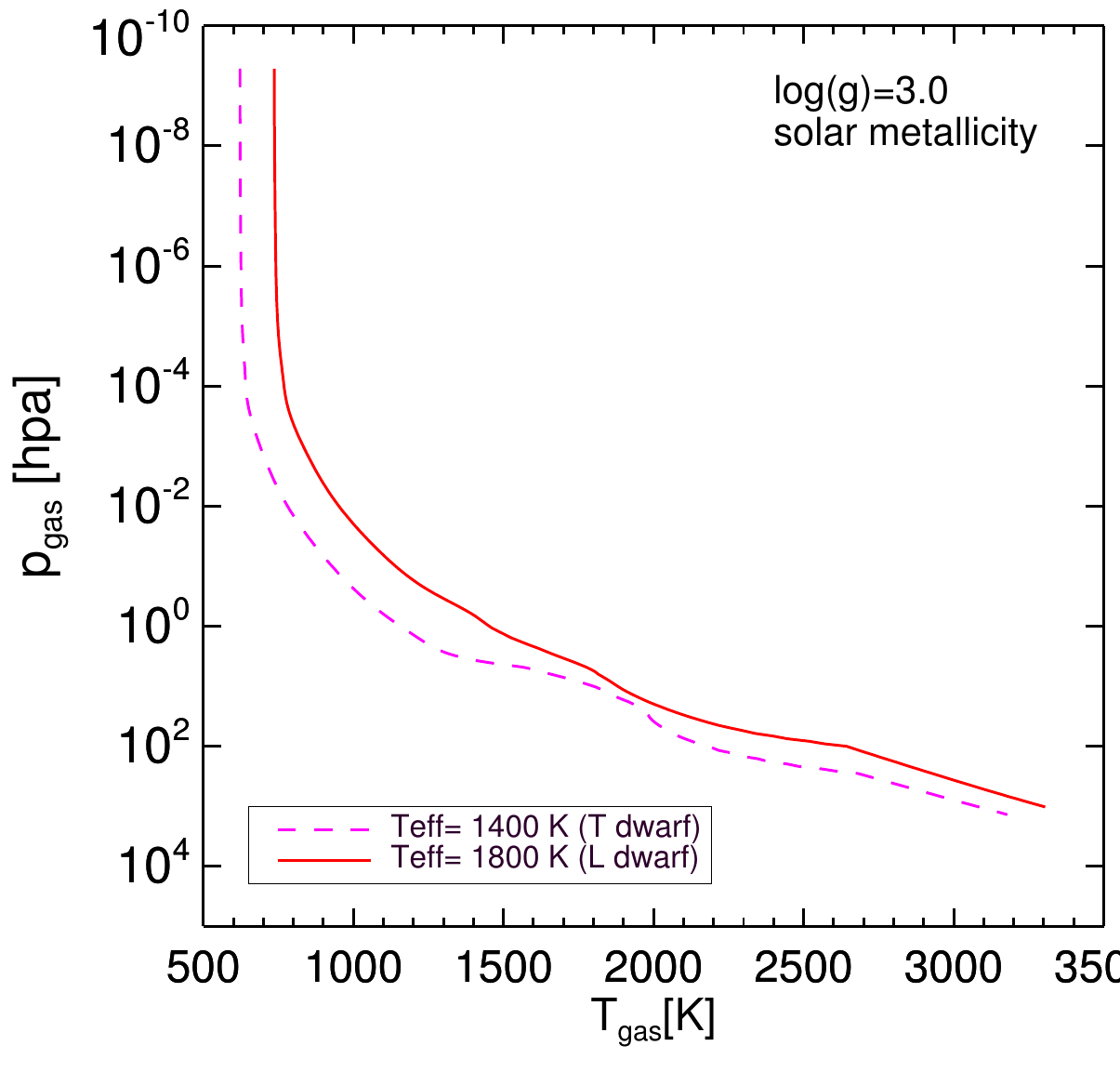}
\end{minipage}
\\
\multicolumn{3}{c}{\ }\\*[-0.3cm]
\multicolumn{3}{c}{local degree of ionization $f_{\rm e}=\frac{p_{\rm e}}{p_{\rm gas}}$}\\*[0.1cm]
\begin{minipage}{3.7cm}
\vspace*{0.3cm}
\centering \hspace*{-1.3cm}\includegraphics[width=4.6cm]{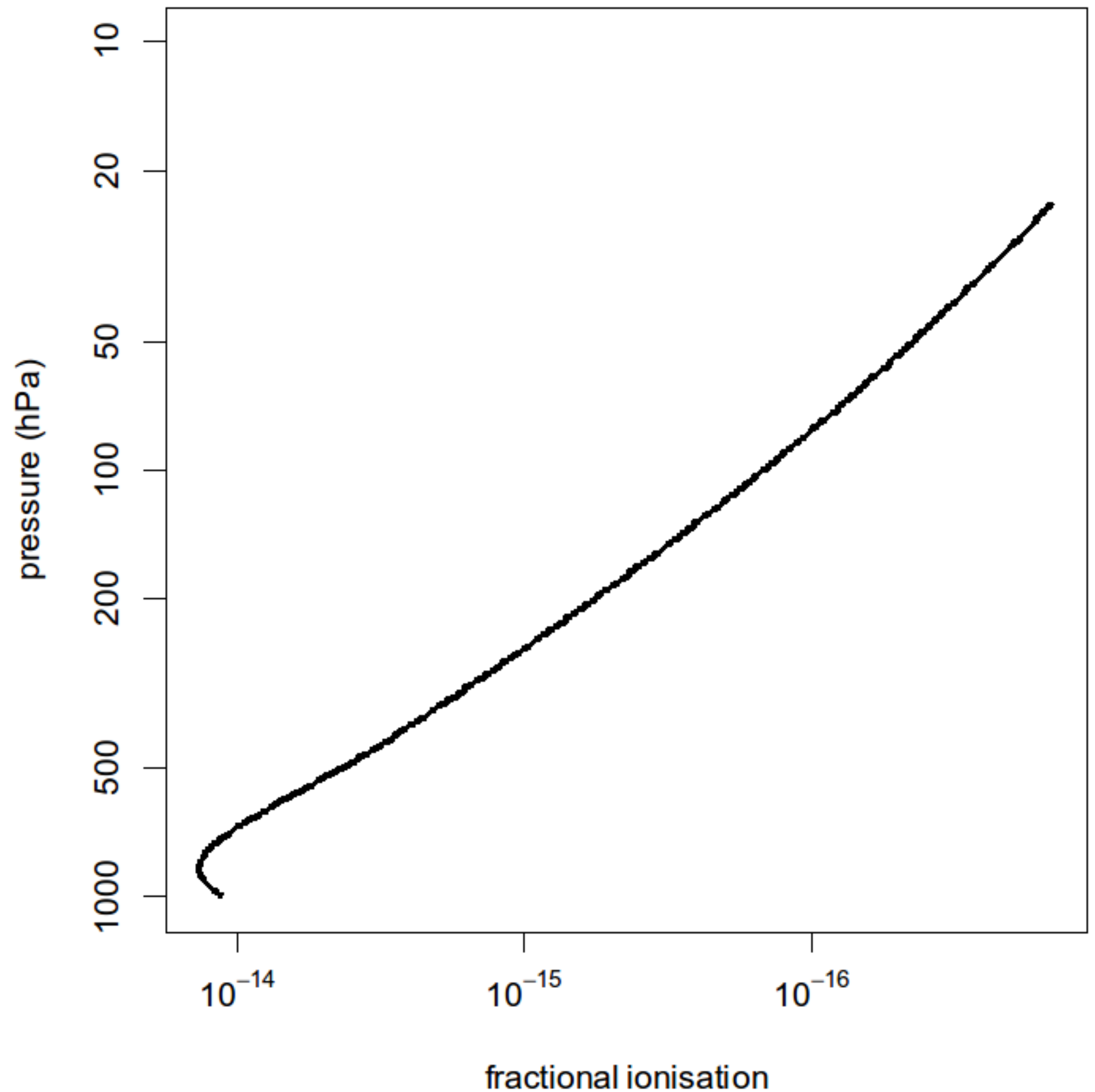}
\end{minipage}
&
\begin{minipage}{4.0cm}
\centering
\vspace*{-0.2cm}
\hspace*{-0.8cm}\includegraphics[width=5.3cm]{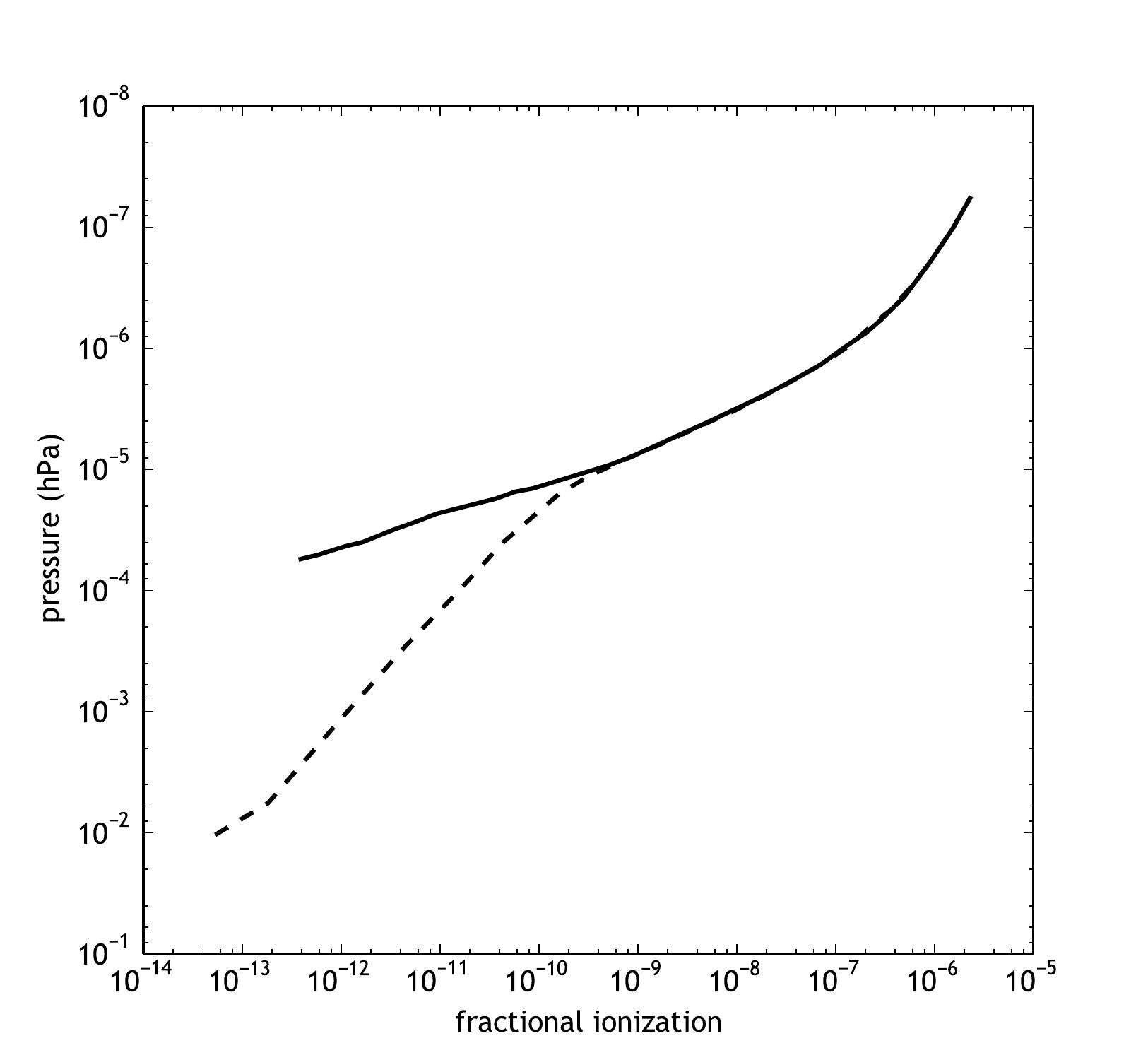} 
\end{minipage}
& 
\begin{minipage}{4.0cm}
\centering
\vspace*{0.25cm}
\hspace*{-0.6cm}\includegraphics[width=5cm]{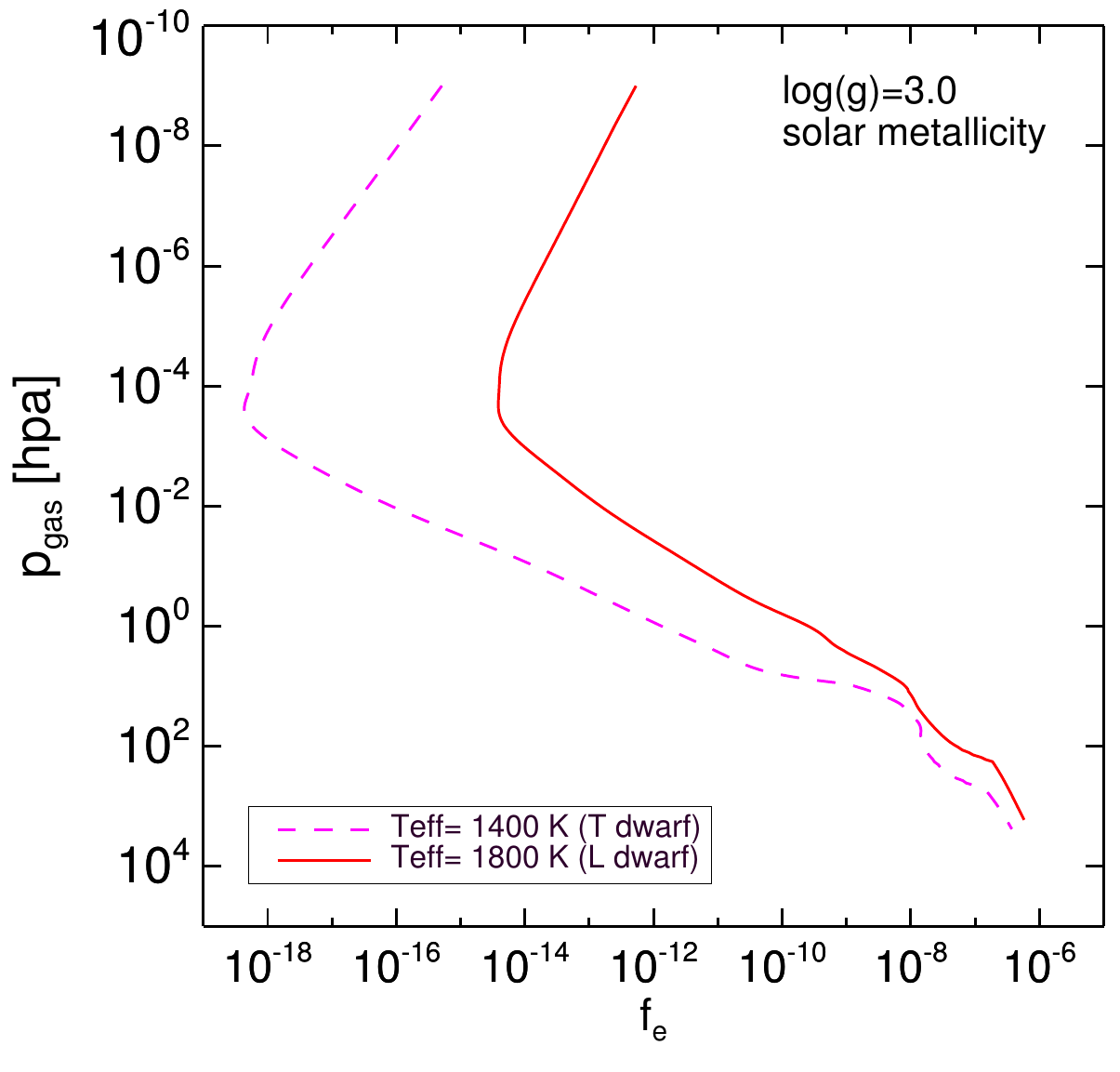}
\end{minipage}
\end{tabular}
\caption{This figure shows the spectrum of emitted radiation, $F(\lambda)$, 
the temperature profile as a function of pressure going up into the
  atmosphere, ($T_{\rm gas}$, p$_{\rm gas}$), and the degree of
  ionisation, $f_{\rm e}$, as a function of pressure for planet Earth, for Saturn
  and for two brown dwarfs. The Saturn thermodynamical data are
  from \cite{moses2000}, \cite{moore2004}
  (solid line) and \cite{galand2009} (dashed line) were used to derive
  the degree of ionization [courtesy: Alejandro Luque]. Saturn's
  disc-integrated spectrum is based on the latest profiles of atmospheric
  temperature and gaseous composition derived from analysis of Cassini
  Composite Infrared Spectrometer spectra (\citealt{irwin2008,
  fletcher2012}; courtesy: Leigh Fletcher).  The brown dwarf spectra
   are from \cite{cushing2005} [courtesy: Sarah Casewell], the
  atmosphere models from \cite{witte2011} [courtesy: Isabel
  Rodrigues-Barrera].}\
\label{tab:comp}
\end{figure}%
Figure~\ref{tab:comp}
compares images, spectra (disk-integrated radiation flux), atmospheric
(T$_{\rm gas}$, p$_{\rm gas}$)-structures, and the local degrees of
gas ionization for Earth, Saturn and two types of Brown Dwarfs (L-type
(pink) -- hotter, and T-type (purple) -- cooler).  All data for Earth
are from observations, the Saturn data are derived from
Cassini\footnote{http://sci.esa.int/cassini-huygens/} spacecraft
observation, the brown dwarf spectra are observed with SpeX on
IRTF\footnote{http://irtfweb.ifa.hawaii.edu/~spex/}
(\citealt{cushing2005}), and the (T$_{\rm gas}$, p$_{\rm gas}$)- and
the $f_e$-structure are results from atmosphere simulations. 
$f_e$ refers to the local degree of ionisation and is defined as
$f_{\rm e}=p_{\rm e}/p_{\rm gas}$ with $p_{\rm e}$ and $p_{\rm
gas}$ the local electron and the local gas pressure, respectively.
The Earth image is a photograph taken from the International Space
Station. The Saturn image is a visible light image taken by the
Cassini space craft, and the brown dwarf image is an artist's
impression based on atmosphere simulations. No direct image exists for
any brown dwarf because the nearest brown dwarfs (the binary system
Luhman 16) is 6.59 light years away from Earth.  All three classes of
objects have chemically and dynamically active atmospheres that form
clouds and that  may be undergoing local charge and discharge
events. Their local atmospheric conditions differ, including the
chemical composition, as result of their formation history and the
irradiation received from a host star.  Interdisciplinary research
combining plasma physics, meteorology, volcanology, solar
system exploration and astrophysics as suggested in
(\citealt{fuelle2013}) is required to study weather phenomena on
Earth, solar system planets and on extrasolar planetary objects also
in view of upcoming space missions like
CHEOPS\footnote{http://sci.esa.int/cheops/}, 
PLATO\footnote{http://sci.esa.int/plato/} and JWST\footnote{http://jwst.nasa.gov/}.

Plasma and discharge experiments are essential in providing a controlled environment
in contrast to observation of atmospheric phenomena. Such experiments can involve the
three different mass components constituting an atmospheric gas:
electrons, ions, and dust particles with their masses $m_{\rm
e-}<m_{\rm ion}<m_{\rm d}$. The mass differences result in different
spatial effects like ion acoustic waves and plasma crystals. An
atmospheric environment that is only partially ionized may show plasma
character on only local scales compared to the global scale of the
comet, moon, planet, brown dwarfs or protoplanetary disk. One
potentially far reaching example for the origin of life on Earth are
volcanoes (\citealt{johnsons2008}) which can produce significant
electrostatic charging and subsequent lightning during eruption
(Sect.~\ref{s:eldis_solsys}\ref{ss:volc}) on Earth and maybe also on
Jupiter's moon Io for example.  In volcanoes but also in terrestrial
clouds, particles of similar mass govern the charge and discharge
processes and plasmas form during violent discharge
only. Understanding dust charging processes is important for space
exploration because the local ionization changes as result of the
variability of the solar wind hitting the moon's or an asteroid's
surface. A spacecraft landing, like Philae, the Rosetta lander, has a
very similar effect (Sect.~\ref{s:solsys}). In situ measurements from
the chemically active Earth atmosphere offer insight in charge and
discharge processes, their local properties and their global changes
(Sect~\ref{s:eldis_solsys}\ref{ss:ionization}). While plasma
experiments are conducted in a controlled laboratory environment,
measurements inside the uncontrollable Earth's natural atmospheric
environment lead to an understanding of the vertical and horizontal
ionization where the relative importance of electrons, ions and dust,
hence their total mass relation, changes with atmospheric height. For
example, the fair weather current is carried by ions only due to the
lack of free electrons between 0--60\,km. Understanding the {\it
Wilson Global circuit} (Sect.~\ref{s:eldis_solsys}\ref{ss:globcirc})
helps the understanding of the Earth weather and climate.  Such
observations allow an understanding of atmospheric processes on Earth
that can only be gained for solar system and extrasolar bodies by
intensive modelling efforts guided by observations and experiments.

Section~\ref{s:FCI} provides a short background summary on charge
processes of discrete solid or liquid surfaces in atmospheric gases,
the link to laboratory works and an example of related  plasma technology
development. Section~\ref{s:FCI} further sets the stage for this
interdisciplinary paper by defining terms used in later
sections.

 Section~\ref{s:eldis_solsys} summarizes charging and discharging
 processes in the terrestrial atmosphere, including processes in the
 atmospheres of other solar system planets.  Section~\ref{s:solsys}
 reviews charging processes on moon and asteroids in the presence of
 solar wind and space plasmas, but without substantial neutral
 atmospheres. Section~\ref{s:exopl} provides insight into astronomical
 observations that suggest that mineral-cloud forming atmospheres of
 brown dwarfs and extrasolar planets are also electrically active,
 that different ionization processes will electrically activate
 different parts of such atmospheres, and that similar processes are
 expected to act in protoplanetary disks. Section~\ref{s:concl}
 concludes this paper. Each section ends with a list of future works/
 open questions when suitable.

\section{Setting the stage for interdisciplinary exchange}\label{s:FCI}

This section outlines the key concept of this interdisciplinary paper
and it provides definitions of terms used in
Sects.~\ref{s:eldis_solsys}\,-~\ref{s:exopl}. This section links to
laboratory experiments which have driven the understanding of ionised
atmosphere gases that contain or form dust particles or liquid
droplets. One example of plasma technology development is included to
demonstrate the impact of this paper's theme also beyond academic
research. This section deals with the smallest scales where charge
processes act, later sections will address topics related to
successively larger-scale charge processes in the terrestrial
atmosphere, on the Moon and asteroids, and also outside the solar
system in extrasolar planets, brown dwarfs and protoplanetary disks.

\subsection{Fundamental charging  processes}\label{ss:FCP}


 The key concepts in this paper depend on the accumulation and
  dissipation of electrical charge on discrete solid or liquid
  surfaces suspended in atmospheric gases. The free charge on the
  surfaces can arise from two primary mechanisms (in the planetary
  atmosphere context): processes involving (i) friction (triboelectric
  charging); and (ii) the transport of free charge (plasma
  processes). More details on processes specific to various
  environments like Earth atmosphere, volcanoes or extrasolar planets
  are provided in the respective subsections
  (e.g. Sects.~\ref{s:eldis_solsys}\ref{ss:ionization},~\ref{ss:Ute} and ~\ref{ss:volc}).

\subsubsection{Classical frictional charging}\label{ss:FCPa}
Transiently contacting surfaces can lead to charge accumulation, by
producing either a surplus or a deficit of electrons compared to
the neutral case. Indeed, there is evidence that fragments
of polymer chains can be exchanged by colliding particles
(\citealt{sau2008}), leaving net charges on the
surfaces. This process is termed triboelectric charging, and has a
very long history of practical application (\citealt{C4RA09604E}),
even if the underlying processes are still not entirely resolved. 
Originally, {\it contact electrification} was used to refer to electrostatic charge transfer resulting from contact, including contact modes such as detachment, sliding, rolling, impact, etc. The specific charge processes related to rubbing was only later termed as tribo-electrification.
Such charging is an inevitable consequence of the frictional interaction
between hard surfaces: electrons transfer (by some process) from one
surface to the other, leading to charged surfaces. For example, dust
entrained in strong, collisional flows (such as volcanic eruptions or
mineral clouds in extrasolar planets,
Sects.~\ref{s:eldis_solsys}\ref{ss:volc} and~\ref{s:exopl}) will
acquire charges of different polarity (negative and positive) directly
from the inter-grain collisions themselves. Such macroscopic particles
can include ice crystals in atmospheric clouds, where the diversity of
growth rates (and consequent dynamics) of crystals influences the
polarity of charge transfer, and leads to such clouds becoming charge
separated by the relative drift of the charged particles
(\citealt{sau2008}). Charge accumulation and
separation can lead to energetic relaxation, in the form of lightning.

\subsubsection{Plasma charging}\label{ss:FCPb}
There is an additional mechanism for forcing charge onto a surface, in
possibly much larger quantities than can be acquired by triboelectric or contact
processes: plasma charging. A plasma is a gas in which a fraction of
the molecules are ionised, leading to an abundance of free charge
existing as an additional `gas' component. Though neutral overall,
there is a natural scale-length over which the plasma can create large
potential differences caused by charge population fluctuations: this
is because free electrons are light and mobile compared to the heavier
positive ions, and therefore the electrons can temporarily escape
their charged counterparts, leading to charge densities appearing for
short intervals, and over restricted distances (this is explained in
detail in subsequent sections below). Should an isolated solid (dust
or crystal) or liquid (aerosol) surface be introduced into this
plasma, these natural fluctuations in the charge distribution will
cause such surfaces to acquire surplus free charge, forced onto it by
the action of the plasma itself. Isolated surfaces exposed to plasma
will quickly (typically on a microsecond timescale or less) charge up
to reach the plasma or floating potential (\citealt{Khrapak_etal_2012,
  Khrapak_Morfill_2008, Hutchinson_Patacchini_2007}), by the action of
a continuous electron current to the surface from the ambient plasma,
which rapidly establishes a negative charge before the compensating
positive ion current can respond. Ultimately there is a balance
reached, but one that reflects the relative electron mobility over the
ions.  Since there is so much more free charge available in a plasma
compared to triboelectric processes, then there is an enhanced
capacity for dust exposed to plasma discharges to store considerable
surface charge in comparison to purely collisional interactions
between grains: since the plasma surface charge reflects the plasma
conditions, and not just the grain chemistry and collisionality, then
the plasma is an independent and effective agent for creating charged
particles.

\subsubsection{Defining general terms}

 After a summary of the principal mechanisms for charging surfaces in
 gases in Sects.~\ref{ss:FCPa} and~\ref{ss:FCPb}, the most important vocabulary used
 throughout the paper is defined below to allow a better understanding
 of the links between the interdisciplinary topics in
 Sects.~\ref{s:eldis_solsys}\,-~\ref{s:exopl}.  Appendix~\ref{s:AGl}
 provides a glossary.

{\it Dust particles, aerosols, droplets:} An important feature in many
charging processes is the presence of macroscopic particles such
  as dust, aerosols or droplet.  These are macroscopic particles
large enough to move under the influence of gravity. The particle
sizes can vary by orders of magnitude. They can be liquid or
solid. They can be composed of a mix of different materials that
changes with temperature.  {\it Aerosols} are suspended particles
  of either phase.  Dust is predominant on moon and asteroids, in
volcanic lightning and mineral clouds of extrasolar planets and brown dwarfs,
and as building blocks for planets in protoplanetary disks. Also
hydrometeors (droplets, graupel and ice particles, snowflakes
$\ldots$) could fall into this category, but are considered aerosols
in geoscience. Macroscopic particles as dust and aerosols can be
electrically charged which de-mobilizes the charge that previously
resided in the gas in form of electrons or ions. Dust, for example,
will acquire a negative total charge in the absence of external
influence like stellar UV radiation.

{\it Ionization} is the process of dissociating neutrals into charged
species, due to a variety of mechanisms: electron impact ionization,
Penning ionization (ionization through chemical reactions), direct
dissociation by strong electric fields, UV-photo-ionization.  The
total electric charge is conserved during ionization, but once the
charges are free they can move independently.  In air (the atmospheric
gas on Earth with its electronegative oxygen component) free electrons
are very short lived in the absence of strong electric fields. Ionized
air in the Earth troposphere and stratosphere consists of positive and
negative ions. The fair weather currents on Earth are ion currents
(see Sect.~\ref{s:eldis_solsys}\ref{ss:globcirc}).

{\it Plasma} is a gas consisting of charged particles. It is often
restricted to charged particle gases where collective phenomena, like
plasma oscillations, are more important than collisional phenomena. A
{\it plasma} is created if there is sufficient ionization of neutrals
that charged particle density becomes significant.
A plasma is characterized by the capacity to produce a collective
self-field that is significant when compared to any imposed field
(such as that produced by external electrodes, or induced by
collapsing magnetic fields, or by impinging electromagnetic
radiation). An electrically neutral medium is created that can respond
to an external electromagnetic field, but there is no spontaneous
charge separation in equilibrium on scale-lengths greater than the
Debye length\footnote{The Debye length is the length beyond which the
  Coulomb force of a charge can not affect other charges. Strictly,
  the Debye length is the e-folding distance within which charge
  neutrality is not guaranteed, because thermal fluctuations can
  displace electrons relative to positive ions, leaving a small net
  charge.}.  There is a significant distinction between plasmas which
are collisionless, and those which are collisional\footnote{These
  terms refer to approximations made in the plasma kinetic gas theory
  where the Boltzmann equation describes the evolution of the particle
  distribution function $f(\vec{x}, \vec{v}, t)$. Neglecting the
  collisional source term of the Boltzmann equation leads to the {\it
    collisionless Boltzmann equation} ({\it Vlasov equation}) from
  which then the MHD equations are derived, and the electric and
  magnetic field strength are derived as macroscopic quantities. In a
  collisional plasma, the full Boltzman equation is to be solved.}:
1) Collisionless plasmas consist mainly of positively charged ions and
of electrons or negatively charged ions, depending on the
electronegativity of the ionized gas. They interact through
electromagnetic fields rather than through mechanical collisions.
Examples are the magnetosphere and the interplanetary plasma
(Sect.~\ref{s:solsys}) where the assumption of ideal MHD holds.  2) In
a collision dominated plasma, the motion of charged particles is
dominated by collisions with neutral atoms and molecules, rather than
by the direct electromagnetic interaction with other charged
particles.  The transiently existing plasmas in the terrestrial
tropo-, strato- and mesosphere up to the E layer of the ionosphere
are mostly collision dominated plasmas, except for the highly ionized
and hot lightning return stroke channel.


{\it Charging or Charge separation} will be used for the process where
macroscopic particles like dust or aerosols are charged. This can
occur in particle collisions (in thundercloud electrification,
dust devils in deserts, volcanic lighting) in non-ionized atmospheres
or in vacuum, or by attaining charge from a plasma (e.g. in dusty
plasmas) spontaneously due the different mobility of the charged
species, in ambipolar diffusion, for example.

If mechanical forces (gravity, convection) that act on the charged
dust particles are stronger than the electric forces, charges can be
separated over a certain distance. An electric potential builds up
that can discharge by lightning and the related transient luminous
events.

{\it Electrification} is understood as the processes leading to
charging of dust or other macroscopic particles obeying both polarity
and charge conservation. As a result, a macroscopic electric field
can build up. Sometimes used synonymously with {\it Charging or Charge separation}.


{\it Discharging} is the process where the electric potential is
released by electric currents. This can happen continuously, or
through a rapid transition like the rapid growth of discharge channels
in lightning discharges. Emission of high energy radiation can be
associated with the rapid channel growth. 

\subsection{Charged dust in experimental work}\label{s:exp}

Dust in plasmas has a long history - one which is even more relevant
in contemporary planetary exploration. This section explores the
phenomena associated with dust interacting with ionization in the
ambient atmosphere to ensure non-equilibrium processes (both physics
and chemistry) have a significant and enduring influence on the
evolution of the atmosphere in general, including the dust itself. The
discussion here ranges over the impact of charged dust imposing a
long-range order in confined plasmas, through to micro-discharges
arising from binary encounters between freely-floating charged
aerosols, releasing low-energy free electrons into the ambient
atmosphere, with all the possibilities that this entails for molecular
activation by dissociative attachment and radical formation. The
common theme throughout is the capacity - literally - for dust to
retain the electrostatic memory of ambient discharges via free-charge
acquisition, and for that discharge legacy to be reshaped and realised
in potent form by harnessing hydrodynamical forces on fluid
timescales, rather plasma ones. In this way, transient plasma effects
can be stored, reconfigured and released on meaningful scales in such
a way as to have a tangible influence on large-scale evolution of
planetary atmospheres. The following sections discuss dust-plasma
interactions in  (i) laboratory plasma dust, where floating
particulates can be a help or a hazard in plasma applications,
including plasma crystals, and in  (ii) the dynamic evolution of charged aerosols,
where fluid deformation and evaporation can moderate the evolution of
encapsulated targets.

\subsubsection{The plasma laboratory: Dusty plasmas and plasma crystals}\label{ss:plph}

 Dusty plasmas have been studied in laboratory experiments for several
 decades.  
 \cite{1924Sci....60..392L}
 reported the observation of minute solid particles and aggregates in
 a laboratory streamer discharge and suggested the dust could play a
 role in ball lightning (see also \citealt{RakovUman2003} for a review). `Dusty plasmas'
 are sometimes referred to as `complex plasmas' although the latter
 description is more wide-ranging and can include other types of
 constituents and features such as sheaths
 (\citealt{1976RSPSA.348..221P}), quantum effects and dust. {\it Dusty
 plasma} is referred to in cases when collective behaviour of dust
 becomes important resulting in new types of waves and
 instabilities. This occurs when the Debye length and inter-particle
 distance are of the same order and the effects of neighbouring
 particles cannot be neglected, as opposed to the case when the Debye
 length is much less than the typical inter-particle distance
 (isolated charged dust).

The experimental research on dusty plasmas in laboratories has
(i)
been aimed at increasing fundamental understanding and
 (ii) also been
strongly motivated by the need to control the behaviour of dust in
plasmas that are used in industrial applications. Dust deposited from
within the plasmas that are involved in the semiconductor component
fabrication and materials processing industries can damage the
components and significantly affect the productivity of these
industries.  In contrast to the need to mitigate the potentially
harmful effects of dust in industrial plasma etching and deposition,
the capability to form and control dust in plasmas is being exploited
in the production of nanoparticles for the expanding nanoscience
industry.

Fundamental research programmes have explored phenomena such as dust
crystallisation and wave propagation within dusty laboratory
plasmas where a stationary and fully ionised gas is considered.  In laboratory experiments the earth's gravitational field 
influences the dusty plasma behaviour and while the vast majority of
experiments have been carried out in laboratories on the
surface of the earth, there have been some experiments on dusty
plasmas carried out in the near-weightless conditions within the
International Space Station. Whereas at sea level 2D dust crystals can
be produced, the low-gravity conditions are usually needed to produce
3D dust crystals.

Several types of waves, including longitudinal electron plasma waves
and ion acoustic waves (\citealt{1977RPPh...40.1305A}), can propagate
in dust-free plasmas formed from ionized gas and containing electrons
and ions as well as some neutral atoms and molecules. Additional
wave propagation modes appear if a magnetic field is applied to the
plasma.  While all of these waves are damped usually as they propagate
it is also possible for them to become growing waves, or instabilities
(\citealt{1977RPPh...40.1305A, 1981PhFl...24.1586K}), when
appropriately excited.  For example ion acoustic waves
(\citealt{1977RPPh...40.1305A}) can be driven unstable by passing a
current through the plasma, i.e they are triggered by a drift motion
of the electrons relative to the ions. In a dusty plasma the charged,
massive dust particles can produce new types of wave motion: The dust
ion-acoustic wave (DIAW) is a modified ion acoustic wave, where the
ions continue to provide the inertia and the presence of the
quasi-stationary charged dust particles modifies the normal ion
acoustic wave dispersion. In contrast to the DIAW, in the dust
acoustic wave (DAW) the dust particles move and provide the inertia
rather than the ions.  Both the DIAW and the DAW can be observed
because their frequencies are low enough for camera systems to resolve
the images of the wave propagation.

Measurement of dusty plasmas in the laboratory and comparison with
simulations using particle in cell (PIC) codes allows these codes to
be benchmarked against the laboratory experimental observations. PIC
code simulation of laboratory plasma experiments and comparison with
space measurements has proven successful in the case of auroral
kilometric radiation (\citealt{2008PPCF...50g4011S,
2008PPCF...50g4010M}) because of their capability to simulate the
onset and dynamics of microinstabilities in dusty plasmas.  The use of
PiC codes to simulate the behaviour of dusty plasmas in space
should prove equally fruitful in
obtaining detailed explanations of the formation, properties and
consequences in astrophysics 
(\citealt{2002idpp.book.....S, 2010cdpl.book.....F}).

\subsubsection{Delivering charges to microscopic particles}\label{ss:bac}
The evolutionary processes governing the dynamics and stability of
charged macroscopic water droplets in a discharge plasma are part of
an innovative collaborative project on bacteria detection
(\citealt{2014APS..GECDT3003R, mag2015}).
The technique of using droplet evaporation as a moderator for
charge deposition provides a method to precisely deliver a known
amount of charge to microscopic particles such as bacteria cells or
(cloud) condensation seeds. For that, aerosolised bacteria samples
will be passed through a discharge plasma to acquire significant
electrical charge which can be measured in the lab.  If the
charge-carrying aerosol evaporates, it's surface area decreases but
the aerosol retains the charge.  Ultimately, if the Coulomb force
overcomes the surface tension, then the droplet expels charge to bring
the retained charge back into the stability limit (the Rayleigh
limit $Q_{\rm r}(t)$), which is a function of its radius. Hence the droplet continues
to track the Rayleigh limit\footnote{The
Rayleigh limit, $Q_{\rm r}(t)$,  gives the limiting size of the surface electric field
that balances the surface tension: the latter provides the restoring
force to return the droplet to it's equilibrium spherical shape, and
so causes the perturbed droplet to oscillate. If the
distorted outer surface of the droplet carries sufficient
electric charge, then the local surface field may oppose the effect of
surface tension and thus prolong the restoration to equilibrium
profile, i.e. reduce the oscillation frequency. If there
is sufficient surface charge, then the deformation persists, and the
oscillation frequency is formally zero which defines the Rayleigh
limit. Exceeding the Rayleigh limit means that the droplet is unstable
to perturbation, and is forced to eject charge and mass. 
} as it evaporates. Once all the fluid has
gone and the interior seed particle (bacterium or grain) is revealed,
the charge placed on it is known. This is the charge consistent with
the Rayleigh limit at the radius of the grain.


\begin{figure}[h]
\centering
\includegraphics[width=.8\textwidth]{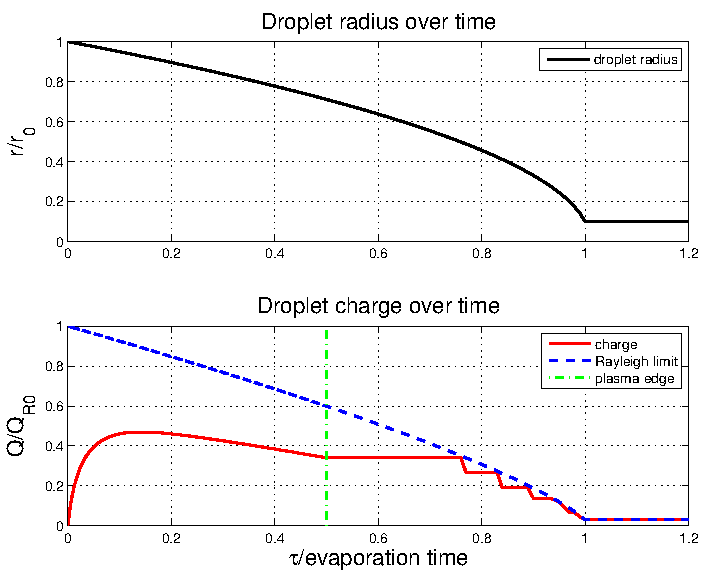}
\caption{The figures show the evolution of a liquid droplet that acquires 
a surface charge as a result of travelling through a plasma discharge.
The horizontal axis is time, normalised to the characteristic time
required to reduce (by evaporation) the droplet radius to one tenth of
its initial value.  The droplet spends 50\% of its evolution inside
the plasma; the green dotted line shows the time at which the droplet
leaves the discharg environment. {\bf Top:} The radius evolution as the
droplet evaporates. {\bf Bottom:} The charge (red line) and Rayleigh
limit (blue line) of an evaporating water droplet containing a
bacteria cell that is one-tenth of the initial droplet radius. Time is
normalised.  Outside the plasma, the charge on the droplet remains
relatively constant until the stability limit is reached, at which
point the droplet emits enough charge to remain stable and enters a
feedback cycle of emission and evaporation. The final charge deposited
on the bacterium is closely linked to the Rayleigh limit of the
minimally-encapsulating droplet (\citealt{bennet}).}
\label{mid_time_in_plasma}
\end{figure}

The charging mechanism can be described as follows
(\citealt{bennet}). Water droplets entering a plasma will form a
sheath between the droplet surface and the plasma, as a simple
consequence of the disparity in mobility between electrons and
ions. Electrons will collide more frequently with the drop surface and
remain there, causing it to acquire a negative surface charge. The
charged droplet will then attract positive ions from the plasma until
the electron and ion currents to the surface of the droplet reach
equilibrium; at this point, the droplet is at the plasma potential.

Suppose an initially stable water droplet has acquired charge by
passing through a plasma (or indeed by an alternative charging
mechanism; green vertical line in Fig.~\ref{mid_time_in_plasma}) and is now floating freely in air, having left the plasma
behind.  If the the initial droplet charge is less than the initial
Rayleigh limit, $Q_{\rm r_0}$, of the droplet, then the droplet is stable.  As
evaporation  proceeds outside the plasma, the droplet charge stays
roughly constant, while the Rayleigh limit, $Q_r(t)$,  evolves according to
\begin{equation}
Q_r(t) = \beta(t) Q_{r_0} \label{Q_R(t)},
\end{equation}
with  $Q
 (t=0) = \alpha\, Q_{r_0}, \,\alpha<1$ being the initial charge on the
 droplet, and $\beta(t) < 1$ for all $t>0$.  The initial values for the results in Fig.~\ref{mid_time_in_plasma} are:
 $\alpha(t=0)=0.0025$, $r_0=r(t=0)=10\mu$m, $Q(t=0)=10^4$ because the Rayleigh limit is
 $4\times 10^6$ e. $\beta=1$ at $t=0$;
 $\beta$ is not shown in Fig~\ref{mid_time_in_plasma}.
If $Q_r(t)$ decreases
far enough that $Q_r(t) \approx Q(t)$, then the droplet will become
unstable and emit sufficient charge to restore the stability condition
of $Q_r(t) > Q(t)$. Evaporation continues until once again the
stability condition is broken and more charge is emitted back into the
ambient gas. This feedback loop continues until the entire droplet has
evaporated.

As the droplet evaporates, both the droplet radius $r(t)$ and the
Rayleigh limit for the charges on the droplet, $Q_r(t)$, decrease. If
the droplet encapsulates a bacteria or dust grain, the evaporation
cannot proceed beyond a minimum radius $r_{m}$. The final charge on
the droplet of size $r_{m}$ at a final time, $t_f$, is then
\begin{eqnarray}
Q(t_{f}) &\approx& \beta(t_{f}) Q_{r0} = Q_r(t_{f})\nonumber \\
&\approx& 8\pi\sqrt{\gamma \varepsilon_0 r_{m}^3}. \label{QRfinal}
\end{eqnarray}

The upper limit of final droplet charge depends only on the minimum
radius of the particle, $r_{m}$,  left behind once the droplet has evaporated,
irrespective of the starting charge. This is assuming that the Rayleigh limit
is encountered at some intermediate point in the evaporative evolution
of the water mantle that forms the drop encapsulating a bacteria or dust grain. 

This is a valuable
process, since grains processed in this way carry the electrostatic
legacy of the plasma environment encountered earlier in their
history.  Such charged particles can either act as a source of
low-energy free charge injected into the atmosphere to produce
non-equilibrium electron-moderated chemical evolution of the latter
(for example, dissociative attachment producing radicals) or indeed a
constraining electrostatic environment stable over fluid length and
time scales.

\section{Electrification and discharging in terrestrial and planetary atmospheres}\label{s:eldis_solsys}

 When we aim to understand electrification and electric phenomena in
 weakly ionized atmospheres of extrasolar planets, a characterization of the
 phenomena on Earth and in the atmospheres of solar system planets can
 provide guideline and inspiration. This section therefore starts with
 an overview of the main electrical processes in the terrestrial atmosphere
 up to the ionosphere, the fair weather currents and the thunderstorms
 with transient luminous events and terrestrial gamma-ray flashes.
 Then we continue with  lightning phenomena in volcanic
 ash plumes and review a few processes in the atmospheres of other
 solar system planets. For more details see  \cite{RakovUman2003, Planets2008} and \cite{DwyerUman2014, betz2009, full2006, es2008}.

 Ionization and electric currents in the terrestrial atmosphere are
 driven by two main mechanisms: a) The atmosphere is very weakly
 ionized by external sources like Cosmic Rays and radioactivity
 (Sect.~\ref{s:eldis_solsys}\ref{ss:ionization}).  The resulting
 conductivity supports the fair weather currents that relax electric
 potentials in atmospheric regions far from thunderstorms.  b)
 Thunderclouds play a particular role in separating electric charges
 and in building up large electric potentials
 (Sect.~\ref{s:eldis_solsys}\ref{ss:Ute}). Cloud particles first
 exchange charge during collisions, and are then separated due to
 mechanical forces (such as gravity and convection) larger
 than the attractive electric forces between particles of opposite
 polarity.  For this reason, meteorologists use lightning flashes as
 indicators for strong turbulent convection in the atmosphere.  When
 these electric potentials suddenly discharge, a variety of ionized
 and conducting channels is formed through localized ionization
 processes (collisional, thermally driven or photon impact). In the
 first stage of a discharge, these ionization reactions are driven by
 strong electric fields and local field enhancement and are dominated
 by the impact of fast electrons on neutral atoms or molecules, while at later
 stages Ohmic heating and thermal equilibration create
 temperature driven ionization reactions.

\subsection{Ionization of the terrestrial atmosphere outside thunderstorm regions}\label{ss:ionization}

\begin{figure}
\begin{center}
\scalebox{0.4}{\includegraphics[angle=0]{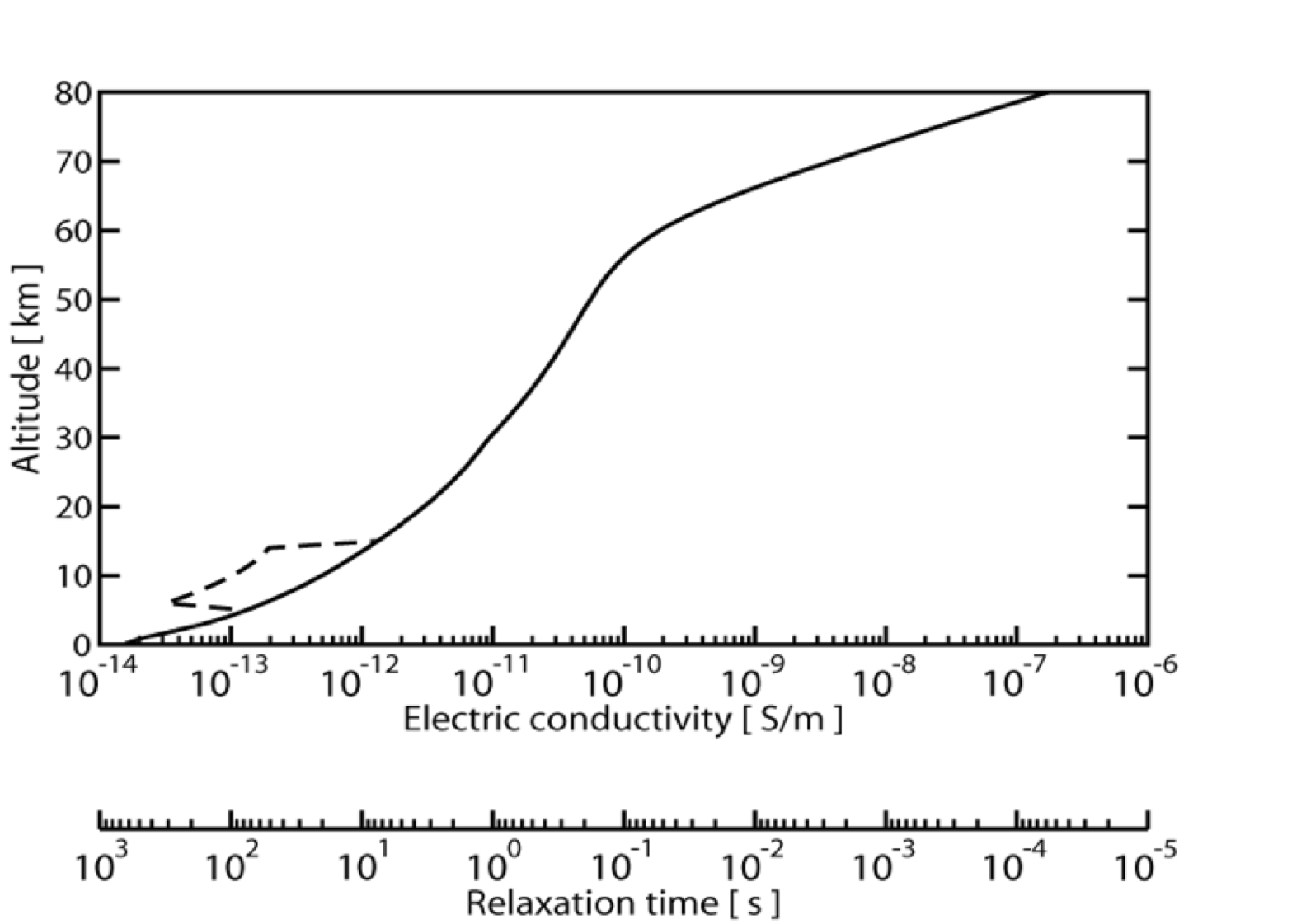}}
\caption{Vertical variation in electrical conductivity, $\sigma_{\rm t}$, of the terrestrial atmosphere, as represented in the model of \cite{rycroft2007}. The dashed line indicates the change of conductivity due to clouds. The equivalent electrical relaxation time is found from $\epsilon_0/\sigma_{\rm t}$,  where  $\epsilon_0$ is the permittivity of free space.}
\label{fig:GH3}
\end{center}
\end{figure}

In common with other solar system atmospheres
(\citealt{harrison2008}), the earth's lower atmosphere outside
thunderstorm regions is made electrically conductive by the ionising
action of high energy particles generated within the heliosphere
(e.g. solar energetic particles, SEPs) and beyond (e.g. galactic
cosmic rays, GCRs). A consequence of the terrestrial atmosphere's
small but finite conductivity ($\approx 10^{-14}$ S\,m$^{-1}$ in
surface air, see also Fig.~\ref{fig:GH3}) is that current flow can
occur through the atmosphere, between disturbed weather and fair
weather regions. Similar circumstances occur in other atmospheres,
depending on the existence of charge separation processes and the
atmospheric conductivity.

Ion production in the earth's lower atmosphere (i.e. the troposphere
and stratosphere) results from a combination of terrestrial and
extra-terrestrial sources. Near the planet's continental surfaces, the
effects of natural radioactivity contained within the soil and rocks,
or released in the form of radioactive gases such as radon, provide
the dominant source of ion production. At heights from 3 to 5 km above
the continents (i.e. above the boundary layer where eddy diffusion of
radon isotopes occurs which depend on orography), or over the oceans,
extra-terrestrial sources, principally GCR, dominate the ion
production, while SEPs and UV irradiation dominate the ionisation in
the ionosphere, but typically do not have sufficient energy to reach
the troposphere.

\begin{figure}
{\ }\\*[-4cm]
\begin{center}
\scalebox{0.6}{\includegraphics[angle=0]{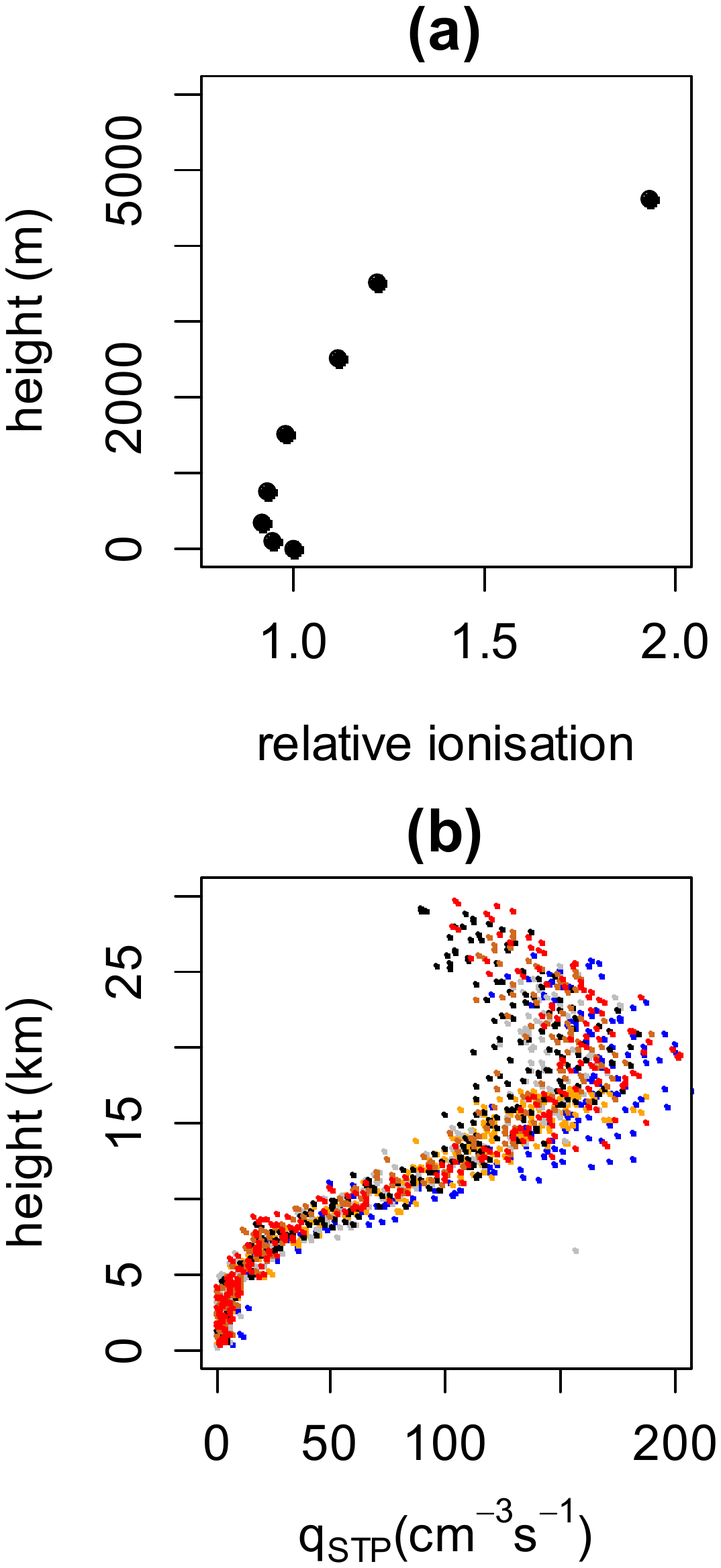}}\\*[-2cm]
\caption{Vertical profile of the ionisation rate, $q_{\rm STP}$ [cm$^{-3}$s$^{-1}$], 
per volume of terrestrial atmosphere normalised to standard
temperature and pressure, as (a) originally obtained by Hess (7th
August 1912), with the ionisation rate at each height shown relative
to the measured surface ionisation rate at sea level and, (b) from a
series of balloon flights (colours used to identify individual
flights) made from Reading, UK during 2013.
}
\label{fig:GH1}
\end{center}
\end{figure}

\paragraph{Balloon-borne measurements:} 
Vertical soundings of the ion production rate in the troposphere and
stratosphere (i.e. to about 35km) can be made using balloon-carried
instruments\footnote{The atmosphere above this altitude is sometimes
called ignorosphere, because above balloon and below satellite
altitudes it is very difficult to explore. In particular, the density
of free electrons in the lower ionosphere can now be measured only
indirectly through the pattern of electromagnetic radiation that is
emitted by lightning strokes and reflected by the ionosphere
(\citealt{Lay2010,Lay2013}).}.  Historically this was the original
airborne platform through which the existence of the cosmic source of
ionisation was confirmed, in a manned balloon flight made by Victor
Hess on 7th August 1912 (\citealt{hess1912}). This flight carried
ionisation chambers and fibre electrometers, in which the rate of
decay of the charged fibre was recorded visually and the ion
production rate inferred (\citealt{pfotzer1972}). Hess found that the
ion production rate initially diminished with height, but then began
to increase (Fig.~\ref{fig:GH1} (a)). This subsequent increase
indicated that ionisation was originating from
above. Figure~\ref{fig:GH1} (b) shows a profile of the ion production
rate per unit volume at standard temperature and pressure, $q_{\rm
STP}$, made using a modern balloon-carried Geiger counter (or
Geigersonde) launched from a mid-latitude site (details are given
in \citealt{harr2014}). This shows the same increase in ionisation
observed by Hess at the lower altitudes, but the modern balloons
extend the measurements to greater altitudes. A characteristic feature
is the maximum in ionisation at about 20km, first
observed \cite{regener1935}. The presence of the Regner-Pfotzer
maximum results from a balance between the energy of the incoming
particles, and the density of the atmosphere.

\begin{figure}
\begin{center}
\scalebox{0.4}{\includegraphics[angle=0]{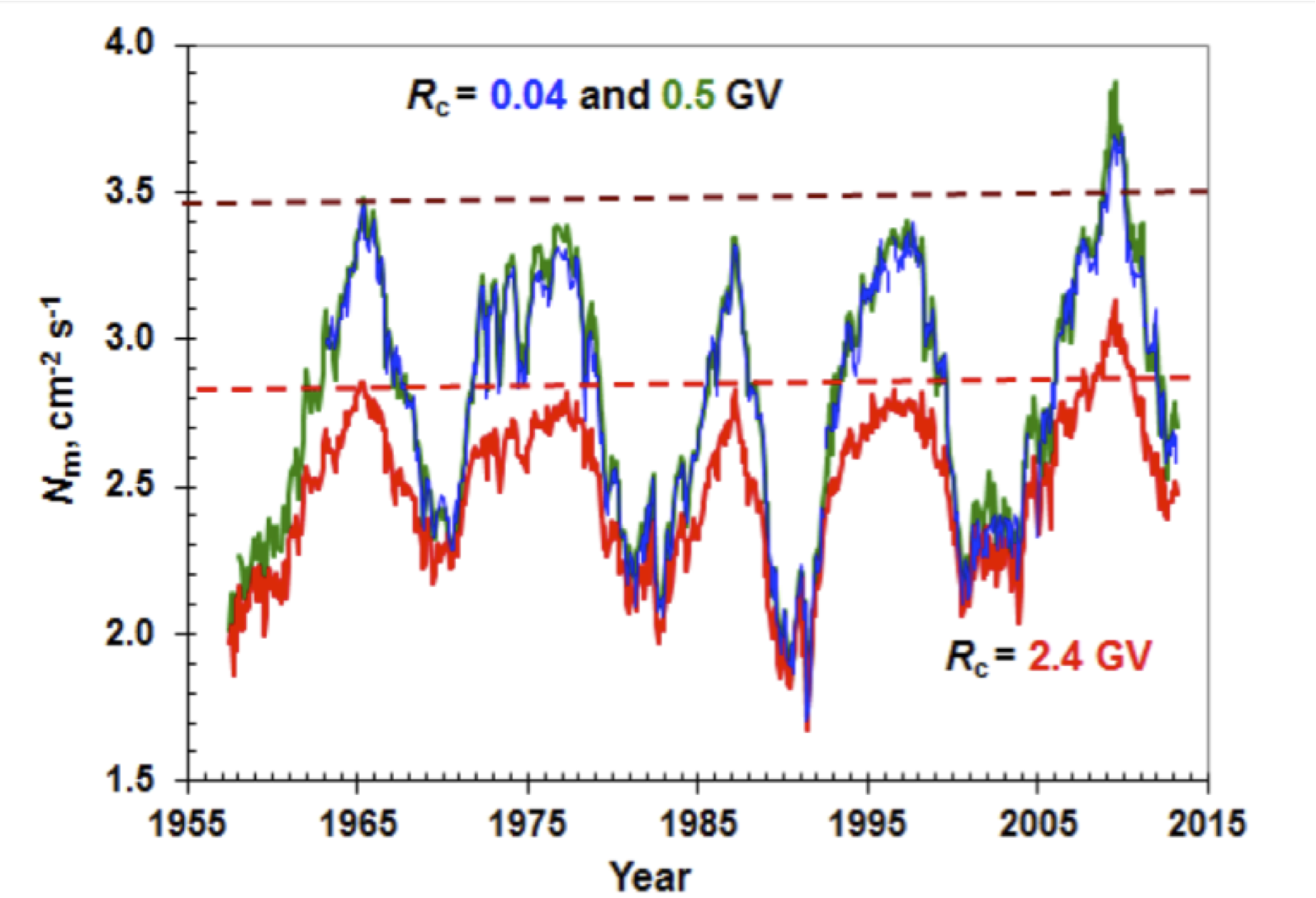}}
\caption{Time series of monthly averages of cosmic ray fluxes, $N_{\rm m}$ [cm$^{-2}$s$^{-1}$], measured at the height of the Regner-Pfotzer maximum. Curves show measurements made at northern polar latitude (geomagnetic rigidity $R_{\rm c} =$ 0.6 GV, green curve), southern polar latitude in Antarctica ($R_{\rm c} =$ 0.04 GV, blue curve) and at the mid-latitude location of Moscow ($R_{\rm c} =$ 2.4 GV, red curve). The  CR flux increase since 2010 can be seen from the comparison provided by the dashed lines, which mark the cosmic ray levels in1965
(from \citealt{stozhkov2013}).}
\label{fig:GH2}
\end{center}
\end{figure}

A long series of regular Geigersonde measurements has been made by the
Lebedev Institute in Moscow, using a variety of sites including
Moscow, Murmansk and Mirny (Antarctica). The value of this stable
long-term measurement series is considerable, as, by taking advantage
of the different geomagnetic latitudes of the sites concerned, it
allows features of the cosmic ray ionisation to be established. Cosmic
rays\footnote{'Cosmic rays' here also referes to energetic particles
originating from the solar wind and from outside the solar system
(galalctic cosmic rays). For more information on the astrophysical
context, on the origin of cosmic rays, the energy-dependent cosmic ray
flux, and their effect on atmospheric chemistry see,
e.g. \cite{rimmer2013, rimmer2014}, Rimmer, Helling \& Bilger (2014)  and for the solar
system see, e.g. \cite{baz2015} and referenes therein. See,
e.g. \cite{Ihongo2016} for the effect of the solar wind onto the galactic
cosmic ray flux at Earth.} follow the geomagnetic field lines, and the
lower energy particles are able to enter at higher latitudes (which is
expressed as a lower geomagnetic rigidity). The high energy CR
particles survive for longer (i.e. penetrate deeper in the atmosphere)
in the Earth atmosphere, while the low energy CR particles are
completely absorbed soon after they enter the atmosphere.
Figure~\ref{fig:GH2} shows a long times series of Geigersonde
measurements made at the Regner-Pfotzer maximum, from sites with
different rigidity (\citealt{stozhkov2013}). The 11 year (Schwabe)
cycle in solar activity is clearly present through the inverse
response in GCRs, and, at the high latitude sites, the exceptional
nature of the cosmic ray maximum in 2010/11 associated with the deep
solar minimum, is particularly apparent.

\paragraph{Atmospheric conductivity:}
Cosmic ray ionisation in the terrestrial atmosphere sustains a steady source of cluster ions, which provide the finite conductivity of air. The total conductivity, $\sigma_{\rm t}$, is given by
\begin{equation} 
\sigma_{\rm t} = e \, (\mu_+n_+ +\mu_- n_-)
\end{equation} 
where $\mu_{\pm}$ represents the mean mobility of positive or negative
ions present, $n_{\pm}$ the associated bipolar ion number
concentrations and $e$ is the elementary charge. Ions are removed by
attachment to aerosol particles and water droplets, reducing the
conductivity in these regions.  Both the mobility and concentration
vary with atmospheric properties and composition. The mobility of ions
depends on the environmental temperature and pressure and the ion
concentration is strongly affected by attachment to aerosol particles
and water droplets, reducing the conductivity accordingly where the
aerosol is abundant. This means that, in the earth's environment,
where aerosols are generated both naturally and through human
activities, the local air conductivity can show an anthropogenic
influence (\citealt{harrison2006,hugo2014}), allowing early indirect
conductivity measurements to provide an insight into historical air
pollution (\citealt{harrison2006, aplin2012A}). Together with
variations in the source rate, $q_{\rm STP}$, these lead to a
variation in the conductivity with height
(e.g. \citealt{harrison2003}). In the heights of the lower ionosphere,
where photo-ionisation also contributes appreciably, the conductivity
becomes substantially larger than the lower
atmosphere. Figure~\ref{fig:GH3} shows a vertical profile of the air's
conductivity, and a calculation of the relaxation timescale,
defined by $\epsilon_0 / \sigma_{\rm t}$. This is the e-folding timescale for
the discharge of an isolated particle in a conductive medium. This
provides an indication of how active (in terms of the rate of charge
separation) a charging process needs to be at different heights in the
atmosphere. In comparison with lower troposphere air with a
typical conductivity of $\approx 10^{-14}$ S\,m$^{-1}$ as reviewed
by \cite{rycroft2008}, the planetary surface has a greater electrical
conductivity, of at least $10^{-8}$ Sm$^{-1}$. This means the air
represents a low conductivity region sandwiched between upper and
lower boundaries having much greater conductivity.

\subsection{Thundercloud electrification, lightning and transient luminous events}\label{ss:Ute}

\paragraph{Ionic conductivity and ionic plasmas in the terrestrial atmosphere:}
Most electric phenomena in the terrestrial atmosphere are carried by
ions and aerosols; only in the strong transient electric fields of an
evolving discharge or in the ionosphere are more electrons free and
not attached to electronegative atoms, molecules or larger compounds
 consisting, e.g., of water molecules clustering around ions,
other aerosols, up to droplets from micro- to millimeter size.
Cosmic rays and radioactivity are external sources of ionization
(Sect.~\ref{s:eldis_solsys}\ref{ss:ionization}); they first
create electron ion pairs, then the electrons rapidly attach to
electronegative molecules (mostly to oxygen) leaving the positive and negative ions
in the atmosphere behind which carry the fair weather currents
(Sect.~\ref{s:eldis_solsys}\ref{ss:globcirc}).

\paragraph{The electric field in thunderclouds} 
builds up in two stages.
In the first stage macroscopic particles are
electrically charged, and in the second stage particles of different
polarity are separated by gravitation or other (mechanical) forces; in
order to separate particles with different polarities, these forces
need to be stronger than the electric attraction between charges of
different polarity, since otherwise the electric forces would
counteract the growth of the electric field. The possible charging
mechanisms at work within normal terrestrial thunderclouds are
reviewed, e.g., by \cite{jay1983} and \cite{sau2008}. An important
conclusion of these reviews is that charge is efficiently separated
between particles only in direct collisions.

Liquid droplets cannot experience collisions or fracture as charging
process as they would typically merge on contact, hence they do not
charge easily. However, frozen particles can collide and exchange
charge. Therefore, terrestrial water clouds get electrified mostly in
regions below the freezing temperature (\citealt{mas1953}), more
precisely at temperatures between 0 and $-40^o$C.  The dominant
charging mechanism is thought to occur when graupel and ice particles
collide. \cite{sau2008} reviews the evidence from Krehbiel's cloud
measurements in 1986 ''that ice crystals rebounding from riming
graupel\footnote{Riming graupel is a graupel particle coated with
water droplets that froze immediately when they collided with the ice
surface of the graupel. The surface structure of graupel deviates from
a perfect crystalline structure (e.g. \citealt{bl2009}).} in the
presence of super-cooled water is a requirement of the charge transfer
process''. This observation is consistent with lab measurements of
Saunders et al. that during collision essentially ''fast growing ice
surfaces charge positively, and conversely, sublimating (graupel)
surfaces charge negatively''. However, further dependencies on growth
velocities etc. need to be taken into account. The particle collisions
are mediated by gravity acting on large particles and by turbulent
convection within the cloud. Gravity will also move the heavy
positively charged graupel particles downward while the light positive
ice crystals move upward with the convective flow of the cloud air,
creating charge centers and electric fields within the cloud.  This
particular charging mechanism is based on the intrinsic polarization
of water molecules.  Macroscopic particles of different material can
charge quite efficiently, too, and create electric fields and
discharges. Both volcanic ash plumes, so-called dust devils in
terrestrial deserts and various granular media in the lab support
discharges, as is discussed further in
Sect.~\ref{s:eldis_solsys}\ref{ss:volc}.  The understanding of
charging processes in volcanic ash plumes might inspire further
progress on the long standing question of charging normal
thunderclouds (\citealt{yair2008}). Such normal water clouds mixed
with dust have recently been observed to exhibit particularly strong
and exceptional discharges (\citealt{fuellekrugebeam2013}).

Due to the attachment of ions to water droplets, electric charges in clouds are particularly immobile. The conductivity in the remaining gas phase
is therefore low before lightning activity starts.  This low
conductivity (hence low degree of ionization, see also
Fig.~\ref{tab:comp}) supports a high electric field up to the moment
of discharging.

\paragraph{The stages of lightning:}
Lightning is the sudden release of the electric potential energy
through the fast growth of a disperse network of ionized channels.  On
average, $44\pm 5$ lightning flashes (intracloud and cloud-to-ground
combined) occur around the globe every second
(\citealt{chr2003}). Moreover, according to OTD (Optical Transient
Detector) measurements, lightning occurs mainly over land areas with
an average land/ocean ratio of aproximately 10:1
(\citealt{chr2003}). The visible growing channels are called lightning
leaders; their path is prepared by streamer coronae. While streamers
are space charge driven ionization fronts, leaders maintain their
internal conductivity by increased temperature, molecular excitations
and ionization reactions in the discharge channel. If a
conducting channel connects cloud and ground, the return stroke
carries the largest current and is visible and audible as the
lightning stroke; but intra- and intercloud lightning are much
more likely. The stages of lightning have been described in many articles,
with varying emphasis on phenomena or physical mechanisms. A few
recent ones are by \cite{bara2000, coo2003, RakovUman2003, betz2009,
DwyerUman2014, coo2015}.

A long standing question is how lightning can be initiated because the
observed electric fields are below the classical break-down field
(where electron impact ionization overcomes electron attachment to
oxygen in the Earth atmosphere;
e.g. \citealt{treumann2008,hellingjsd2013}), and free electrons are
not available anywhere in the atmosphere.  \cite{Gurevich1992}
suggested that cosmic particle showers could supply free electrons,
and that relativistic run--away electron avalanches could develop in
an electric field below the classical breakdown
value. \cite{Gurevich2013} recently suggested that the interplay of a
cloud particle with Cosmic Rays could start the discharge.  A
quantitative analysis confirming this scenario is presented
by \cite{dubinova2015}.

Lightning occurs not only between cloud and ground, but also within
and between clouds. Also the 'bolt from the blue' is a phenomenon
where a lightning strike seems to appear out of a blue sky next
to a thundercloud. These strikes are an indication that lightning
leaders can leave the cloud also at its upper edge or in the sideward
direction and then turn downwards.

\paragraph{Transient Luminous Events:}
The full scale discharge activity associated with terrestrial water
clouds became known in the scientific literature only after 1989 when
the first Transient Luminous Events were described (for article
collections, see \cite{full2006,es2008}). Basically, electric
potential stored in a cloud can also discharge in the upward direction
as a jet up into the stratosphere or as a gigantic jet that
extends into the mesosphere. The primary  lightning can drive
secondary discharges, namely elves, halos and sprites in the E layer
of the ionosphere, and in the night time mesosphere (where the D layer
of the ionosphere is located during day time)\footnote{The electron
density at these altitudes is an important parameter for discharge
modeling. Only recently a method was developed to determine it
partially and indirectly (\citealt{Lay2010,Lay2013}).}. Elves and
halos are responses of the lower edge of the ionospheric E layer to
the electromagnetic pulse and the quasi-static potential of the parent
lightning stroke, while sprites propagate downward from the ionosphere
into the mesosphere (so-called column sprites) and sometimes back up
again (carrot
sprites; \citealt{Nielsen2008}, \citealt{luque2009}). Due to
similarity relations between discharges at different atmospheric
densities (\citealt{pasko2007,EbertJGR2010}), tens of kilometers long
sprite discharge channels in the thin upper atmosphere are physically
similar to cm size streamer discharges at normal temperature and
pressure - up to corrections due to different electron attachment and
detachment reactions that can explain long-delayed sprites
(\citealt{luque2012}).  Sprites are  pure streamer discharges
(\citealt{liu2004a,liu2004b}), and therefore are less complex than lightning
strokes with their streamer, leader and return stroke stages, 
evolving on very different scales of space, time and energy. Due to
the efforts of many authors in the past 20 years, the models for
streamer discharges are now becoming more quantitative, so that we now
approach the quantitative understanding  of sprite discharges
through detailed modelling and experimental efforts (\citealt{nij2014a})

\paragraph{Gamma-Ray Flashes and other high energy emissions from thunderstorms:}
In 1994, the BATSE\footnote{http://gammaray.nsstc.nasa.gov/batse/}
satellite detected gamma radiation from earth, and it was recognized
that this radiation came from thunderstorms
(\citealt{fish1994,fish1995}).  Later also beams of electrons (\citealt{dwyer2008}) and even
positrons (\citealt{Briggs2011}) were discovered from satellites . The
Fermi Gamma-Ray Space Telescope detected a clear positron annihilation
signal over Egypt from a thunderstorm over Zambia where the two events
were connected in space and time through a geomagnetic fields line
(that electrons and positrons follow sufficiently high in the
ionosphere where collisions with air molecules is
negligible; \citealt{Briggs2011}). High energy X-rays were also
detected from lightning leaders approaching ground and from long
sparks in the laboratory, see, e.g. \cite{Kochkin2012}. We refer to
the review by \cite{DwyerUman2014}.  It is clear that electrons are
accelerated into the run-away regime within the electric fields inside
and above the thunderstorm, where they continuously gain more energy
from the field than they can lose in collisions with neutral air
molecules. These collisions with molecules result in X- or gamma ray
emission (Bremsstrahlung). The gamma-rays are ionizing radiation and
generate electron positron pairs, or liberate neutrons or protons in
photonuclear reactions (\citealt{Babich2014}).

There are two basic mechanisms discussed in the literature for the
primary electron acceleration: either Galactic Cosmic Rays with
sufficient energy to penetrate deep into the atmosphere and to
generate relativistic run-away electrons avalanches (RREAs) in the
electric fields inside the thundercloud, or the acceleration of low
energy free electrons into the high-energy run-away regime at the tip
of a lightning leader where electric fields are very high. The review
by \cite{DwyerUman2014} favors the RREA mechanism, in agreement with
the previous model development by the first author. The alternative is
the runaway of thermal electrons at the leader tip suggested
by \cite{Xu2012}. Such detailed models depend on the model parameters
for the background cloud field and its geometry, on the altitude of the
lightning leader, but also on the collision cross-sections at the
required energies that are  not reliably available.

\cite{fuellekrugebeam2013} reported on the observation of two consecutive positive lightning discharges
where the first positive lightning discharge initiates sprite
streamers which discharge the lightning electromagnetic field above
the thundercloud. This was seen as a pulsed discharge event followed
by a high-energy electron beam. A small number of stratospheric,
charged aerosols was likely present as result of a Sahara dust storm
and forest fires in Spain providing a collimating electric field
geometry that accelerated the electrons.  This is the first
simultaneous detection of radio signatures from electrons accelerated
to thermal and relativistic energies above thunderclouds. 


\subsection{The Wilson Global Circuit}\label{ss:globcirc}

The vertical structure of conductivity in the atmosphere, with the
upper and lower conducting regions each able to sustain a local
potential, allows a vertical potential difference to exist between the
two regions. Investigations using balloon measurements from the late
1800s showed a variation in potential with atmospheric  height (\citealt{nicoll2012}), with
the upper conducting region being about 250kV positive with respect to
the lower conducting region. The finite conductivity of the
intermediate atmosphere between these charged regions allows a
vertical current to flow. This current was observed directly by CTR
Wilson (\citealt{wilson1906}) in fair weather conditions with no local charge
separation. CTR Wilson concluded that the current flow was likely to
be sustained by charge separation in distant disturbed weather
regions. Evidence supporting this is that the diurnal variation in
Universal Time (UT) near-surface electric field, measured under fair
weather conditions, is independent of where it is measured globally,
and shows strong similarities with the diurnal variation in active
global thunderstorm area (\citealt{whipple1936}). This diurnal
variation in surface atmospheric electric field is known as the
Carnegie curve, after the sailing vessel on which the original
defining measurements were made (\citealt{harr2013}).

The conceptual model that described the electrical transport across the
planet between disturbed weather and fair weather zones - the global
atmospheric electric circuit (\citealt{wilson1921, wilson1929}) - has provided a
fruitful description for investigation of terrestrial atmospheric
electrification, which may offer useful insights for other atmospheres
(\citealt{aplinh2008}). Although the original reasoning used to identify
the global circuit was based on current flow considerations, the wide
range of timescales of the contributing processes leads to a
distinction being made conventionally between the AC and DC global
circuit (\citealt{rycroft2012}).

\begin{figure}
\begin{center}
\scalebox{0.7}{\includegraphics[angle=0]{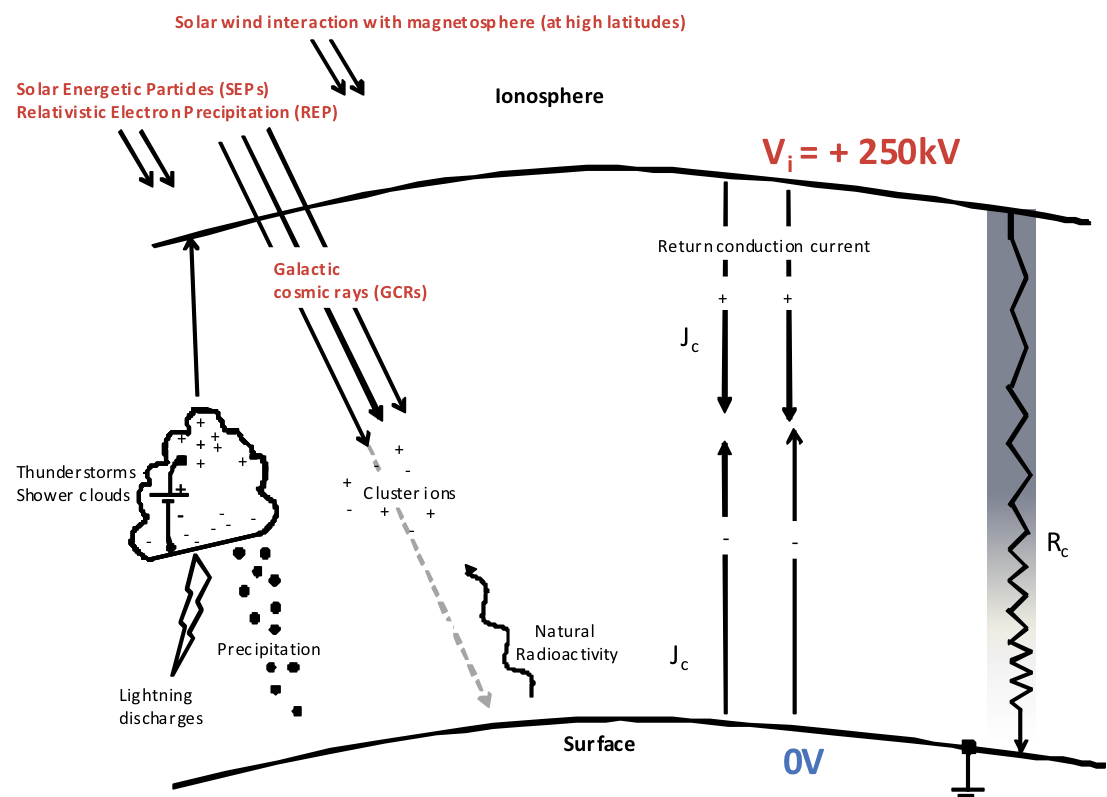}}
\caption{Schematic depiction of the role of ionization from Solar Energetic Particles (SEP), Relativistic Electron Precipitation (REP) and Galactic Cosmic Rays (GCR), in facilitating the current flow within the global atmospheric electric circuit. Natural sources of radioactivity include isotopes within the soil and the release of radon (from \cite{nicoll2014}).}
\label{fig:GH4}
\end{center}
\end{figure}
\paragraph{The AC global circuit:}
The upper and lower conducting regions of the terrestrial atmosphere
form a simple waveguide, in which electromagnetic waves can propagate,
as originally predicted by Schumann
(\citealt{schumann1952}). Lightning provides a source of such
electromagnetic radiation to excite waves in this cavity oscillator,
and natural resonances with a fundamental mode at about 8Hz as
predicted were first observed at the earth's surface in the 1960s
(\citealt{balser1960, rycroft1965}). These natural resonances in the
earth-ionosphere cavity (Q resonator) constitute the AC global
electric circuit. Somewhat surprisingly, resonances at 8, 14, 20 Hz
are also observed on satellites at altitudes of several hundred km,
above the ionosphere (\citealt{simoes2011,dudkin2014}). Although
the electric field measured is much smaller at a satellite platform
compared with ground based measurements (three orders of magnitude
smaller for the first Schumann peak), the fact that it is detectable
at all offers the possibility for fly-by measurements at other
planetary bodies.

\begin{figure}
\hspace*{-2cm}\scalebox{0.8}{\includegraphics[angle=0]{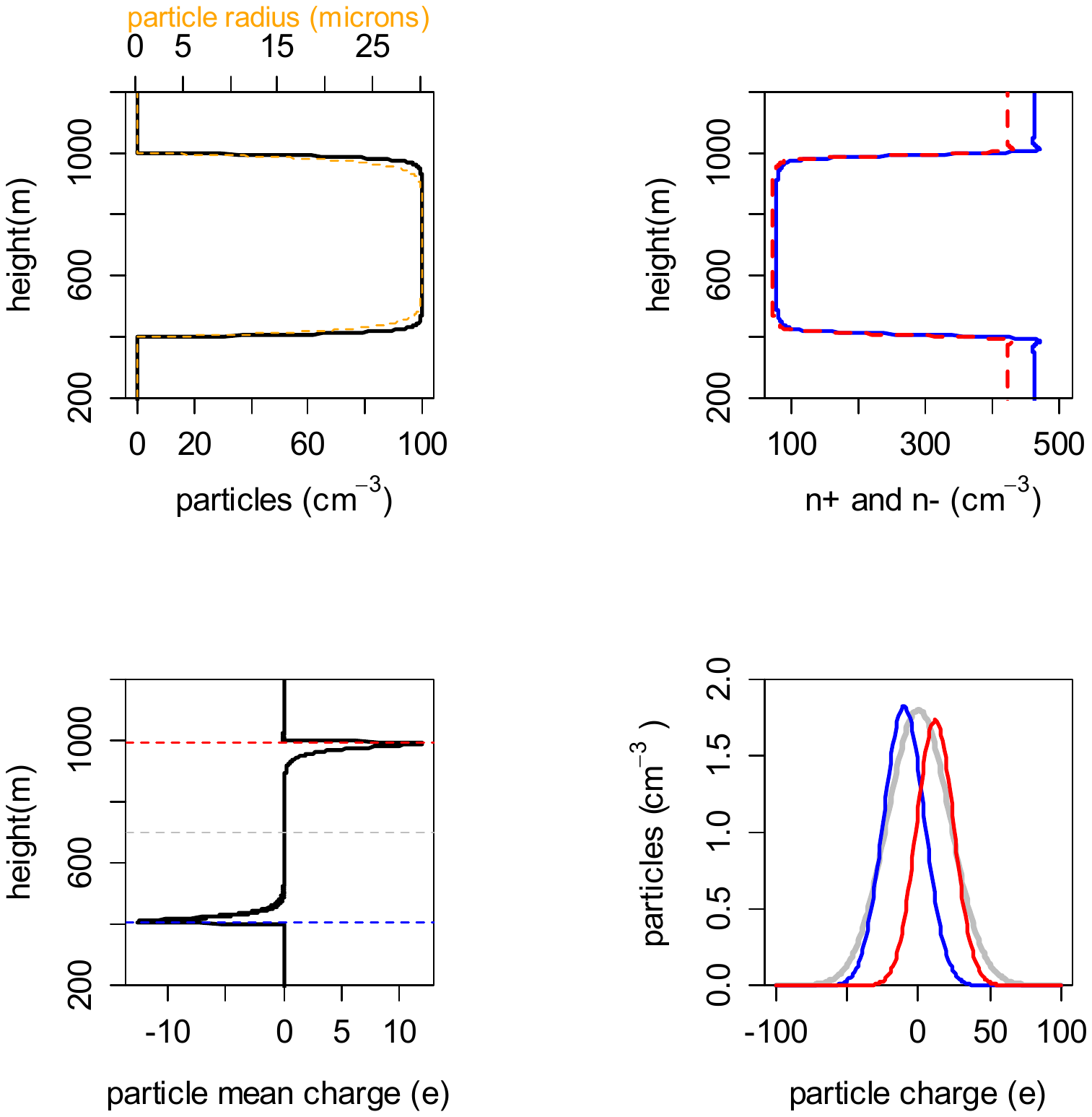}}\\*[-9cm]
\caption{Simulated effect of a horizontal layer of particles through which a current flows. The panel show profile of: (upper left panel) prescribed particle size and concentrations, (upper right panel) number concentrations of positive ($n_+$, dashed red line) and negative ($n_-$, solid blue line) small ions, (lower left panel) mean charge on particles and (lower right panel) particle charge distribution evaluated at the three positions marked on the lower left panel with dashed lines. (Assumptions: ion production rate 10 ions cm$^{-3}$s$^{-1}$, vertical conduction current density 2 pA m$^{-2}$.)}
\label{fig:GH5}
\end{figure}

\paragraph{The DC global circuit:}
Figure~\ref{fig:GH4} summarizes the DC current flow in the Wilson
global circuit. Charge separation in disturbed weather regions leads
to current flow within the ionosphere, fair weather regions and
the planetary surface. The vertical conduction current density,
$J_{\rm c}$, in fair weather regions is  $\sim$2pA m$^{-2}$,
where the resistance of a unit area column of atmosphere, $R_{\rm c}$,
is about 100 to 300 P$\Omega$m$^2$ (\citealt{rycroft2000}). If
horizontal layers of cloud or particles are present, the electrical
conductivity is reduced because of the removal of the ions providing
the conductivity by the particles. Hence, for a passive particle
layer, this means that the layer also defines a region of reduced
conductivity. If a current passes vertically through the passive
particle layer (PPL), charging will result at the step change in
conductivity at the upper and lower layer boundaries. The charging can
be derived by assuming no horizontal divergence of the current (as is
observed, \citealt{gring1986}), and assuming Ohm's Law and Gauss' Law
in one dimension. For a conductivity $\sigma_{\rm t}(z)$ varying with
height $z$, the charge per unit volume $\rho_{\rm e}$ is given
by \begin{equation}
\rho_{\rm e} = \epsilon_0 J_{\rm c} \frac{d}{dz}\big(\frac{1}{\sigma_{\rm t}(z)}\big)
\label{eq:GH2}
 \end{equation} where $J_{\rm c}$ is the vertical current density and
$\epsilon_0$ is the permittivity of free space. Figure~\ref{fig:GH5}
shows calculations of the charging for a PPL of prescribed
concentration and size. This leads to a reduction in the concentration
of positive and negative ions in the same region. The gradients in
conductivity at the PPL boundaries allow the charge density to be
derived, either in terms of the mean charge calculated across the
particles, or as a particle charge distribution
(Fig.~\ref{fig:GH5}). The charging expected at the PPL edges is
clearly evident and similar charging effects have been observed at the
boundaries of layer clouds in the terrestrial atmosphere
(\citealt{nicoll2010}).
\paragraph{Conditions for global circuits:}
The existence of global circuits in planetary atmospheres has been
suggested through possible analogies with the earth system, in which
current flows between charge-separating and non-charge-separating (or
''fair weather'') regions, through the enhanced conductivity zones
provided by the planetary surface and the upper atmosphere
(\citealt{aplin2006, aplin2013}). Entirely different electrical
processes may be involved, such as in the global circuit suggested for
Mars (\citealt{filling86,farrell2001}) which is driven by dust, or be
associated with volcanic dust electrification
(\citealt{houghton2013}).  The basic electrical requirements for a
planetary global circuit have been discussed by \cite{aplinh2008},
which are
\begin{itemize}
\item upper and lower conductive regions
\item charge-separating processes
\item current flow
\end{itemize}
Implied necessary conditions are (1) a sufficiently strong
gravitational field to retain a gaseous atmosphere, and (2) proximity
to energetic sources of radiation (e.g. a host star or a binary
companion) which can form ionized layers in the atmosphere 
ultraviolet and X-ray regions of the spectrum to create an
ionosphere. Table~\ref{tab:GH} summarizes the possible approaches
which might be used to detect these necessary requirements.

\begin{table}
\caption{Possible detection methods for the key requirements of a global circuit in a planetary atmosphere.}
\begin{tabular}{l|c|c|c|c}
\hline
{\it  Requirements:} & \multicolumn{2}{c|}{\bf Charge generation} & {\bf Lower conductive} & {\bf Upper conductive} \\
&Electrical     & Precipitation & {\bf  surface or region} &  {\bf  region}\\
& discharges &                       &                & \\
\hline
{\bf Schumann} & $\checkmark$         &               &       $\checkmark$   & $\checkmark$  \\
{\bf resonances} & & & \\
\hline
{\bf Radar} &  & $\checkmark$ & $\checkmark$ & \\
\hline
{\bf Broadband} & $\checkmark$ & & \\
{\bf radio emission} & & & \\
\hline
{\bf Optical} & $\checkmark$ & & \\
\hline
\end{tabular}
\label{tab:GH}
\end{table}

Of these requirements, providing evidence in a planetary atmosphere of
current flow is a particularly key aspect. In the terrestrial
atmosphere, current flow was originally established using a surface
electrode with an appreciable collecting area
(\citealt{wilson1906}). Use of similar surface mounted electrodes is
unlikely to be practical in space missions, hence other approaches
suitable to the single burst of measurements made by descent probes
entering an atmosphere need consideration.  If horizontal layers of
cloud or particles are present in an atmosphere, which are passive
electrically, (i.e. not able to generate electrification internally),
Eq.~\ref{eq:GH2} indicates that seeking charging at the edges of
particle layers provides an opportunity for the existence of vertical
current flow. PPL edge charging can, in principle, be determined using
a descent probe able to measure charge and detect the presence of
particles, for example using the combination of electrical
(\citealt{nicoll2013}) and optical (\citealt{harrison2014}) detectors
used in the terrestrial atmosphere. Through deploying such sensing
technology on a suitable platform, vertical current flow in a
planetary atmosphere in the solar system may be inferred without the
need for surface measurements.

In summary, the bigger picture here concerns the relationship between
physical processes external to an atmosphere and active processes
within it. Future work in this area therefore needs to consider: 
\begin{itemize}
\item The range of charge separation processes which can occur in different
planetary environments and the controlling influences on current flow,
which may be internal or external in origin. Charge separation occurs
between the same material (e.g. the dust electrification on Mars),
different phases of the same substance (e.g. water-ice-hail
interactions on Earth), or between different substances and phases. 
\item 
In the last set of circumstances, account of the local atmospheric
chemistry and its influence on charging will be needed. Some
consideration should be given to the nature of the charge separation,
and whether simple electrical analogies in terms of constant current
or voltage sources are appropriate. 
\item In terms of the current flow,
there may be significant external influences, including the triggering
of lightning-like discharges by external variations (e.g. Owens et al,
2014). For some planetary body configurations, there may also be
direct tidal effects on the conductive regions in the atmosphere or
other coupled interactions such as those between Saturn's
magnetosphere and Titan.
\end{itemize}

\subsection{Electrical charging in volcanic plumes \& Volcanic Lightning Experiments}\label{ss:volc}



\begin{figure}[h]
\centering
{\ }\\*[-1.5cm]
\includegraphics[width=.8\textwidth]{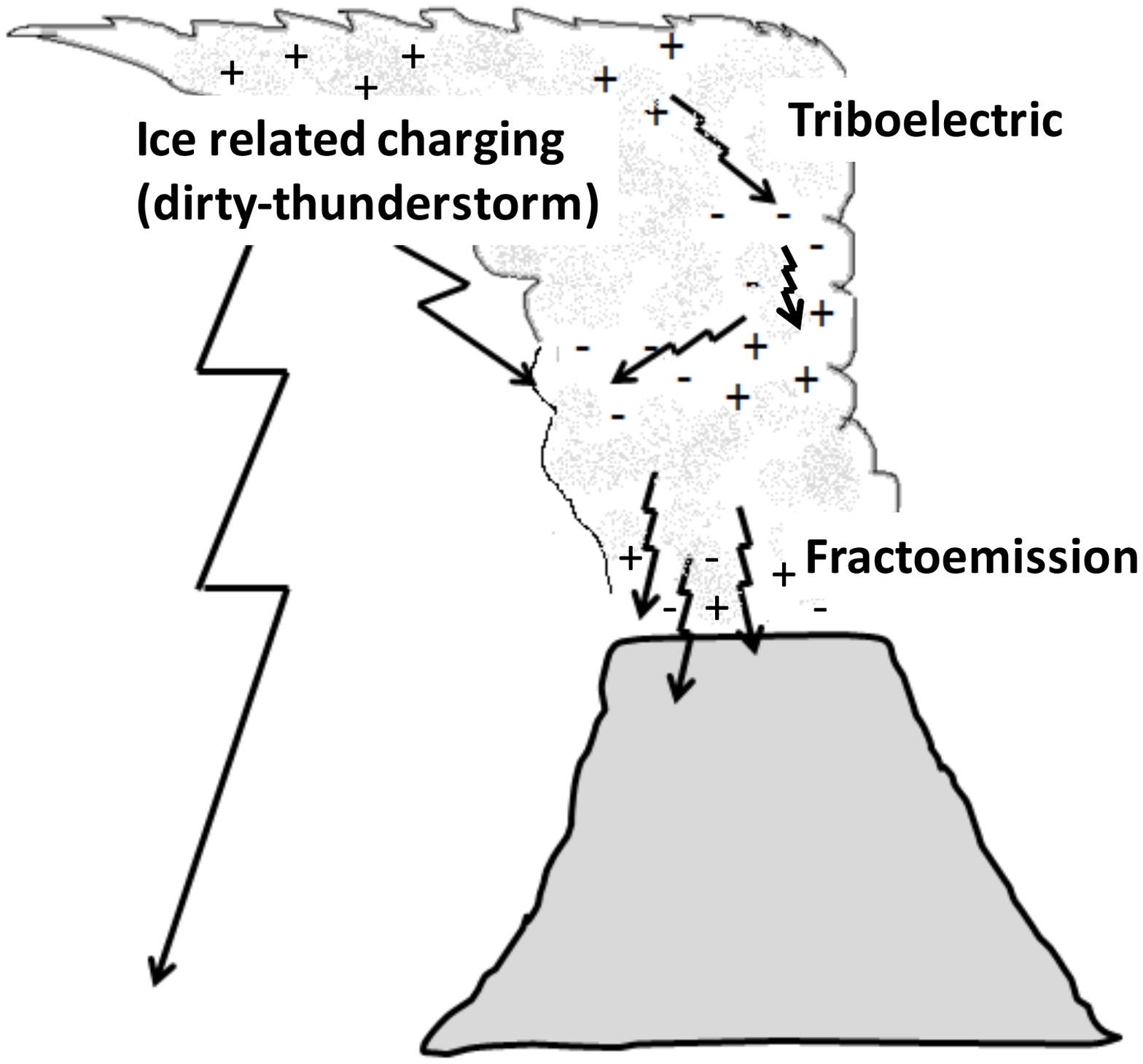}\\*[-6cm]
\caption{Sketch 
of volcanic charge generation mechanisms thought to be active in volcanoes on Earth.  Fractoemission, caused by the fragmentation of magma, is thought to occur close to the vent,  whereas triboelectric charging (frictional contact charging) can occur throughout the plume, wherever particles are present.  The dirty thunderstorm mechanism requires ice particles in the plume and  is only likely to be important for plumes which reach altitudes with temperature that allow freezing to occur.}
\label{kan2pic}
\end{figure}

\paragraph{Electrical charging in volcanic plumes:}
Volcanoes generate some of the most violent forces in nature, and are
not only present on Earth but on several of the planets and moons in
our solar system, e.g. on Venus and Io (\citealt{shal2015}), or more
generally, volcanism can occur on rocky planetary objects with a hot
core. The set of presently known extrasolar planets contains also
planets (e.g. 55 Cancri e, \citealt{dem2011}) that may be classified as volcanic due to
their proximity to their host star and their high bulk density that
indicates a rocky bulk composition. On Earth, volcanic lightning is
often present during eruptions (see \citealt{2006AGUFMAE53A0286H,
2010BVol...72.1153M} for reviews), providing strong evidence for the
electrical charging of volcanic ash as well as demonstrating that
charge separation sufficiently large to initiate breakdown within the
volcanic plume environment.  Numerous mechanisms have been suggested
by which volcanic ash in Earth-based volcanoes can become electrified
including fractoemission (\citealt{2000JGR...10516641J}), contact or
triboelectrification (\citealt{houghton2013}), and
thunderstorm-style ice-contact charging ('dirty thunderstorm'
mechanism;
\citealt{pontikis2}), each of which may occur at different altitudes
throughout the plume (Fig.~\ref{kan2pic}). Understanding the
relative importance of these mechanisms in generating volcanic
lightning during an eruption is required in order to explain
observations of volcanic lightning and why some eruptions produce
lightning and not others.  On Earth, volcanic lightning provides the
ability to detect explosive volcanic plumes remotely, as well as
estimates of the minimum plume height to be made in the absence of
other observational methods such as radar and lidar
(\citealt{2010ERL.....5d4013B}).  Electrostatic forces may also play
an important role in modulating the dry fallout of ash from volcanic
plumes, potentially important for modelling of ash transport downwind
of volcanic eruptions (\citealt{2010ERL.....5b4004H}), although much
future research is required in this area.

Away from Earth, active volcanism exists on several bodies in our
solar system.  Volcanic eruptions on Venus are typically associated
with fluid lava flows - there is no evidence of the explosive ash
eruptions that occur frequently on Earth which are often associated
with active volcanic lightning.  Conversely, Io, one of Jupiter's many
moons often exhibits signs of explosive eruptions. Io's eruptive
columns reach to hundreds of km altitude in contrast to Earth based
plumes which may reach up to up to 40km in rare circumstances
(\citealt{opp2003}).  The existence of volcanoes on
other bodies in the solar system (e.g. Venus, \citealt{airey2015}) suggests the possibility of charging
mechanisms associated with such volcanic activity, which may or may
not be similar to those on Earth.  This leads to the possibility that
studying volcanic lightning on Earth may provide insight into dust
charging processes in environments where mineral dust is common such
as in the atmospheres of brown dwarfs or extrasolar planets as
detailed in Sect.~\ref{s:exopl}.

\begin{figure}[h]
\centering
{\ }\\*[-2cm]
\includegraphics[width=1.0\textwidth]{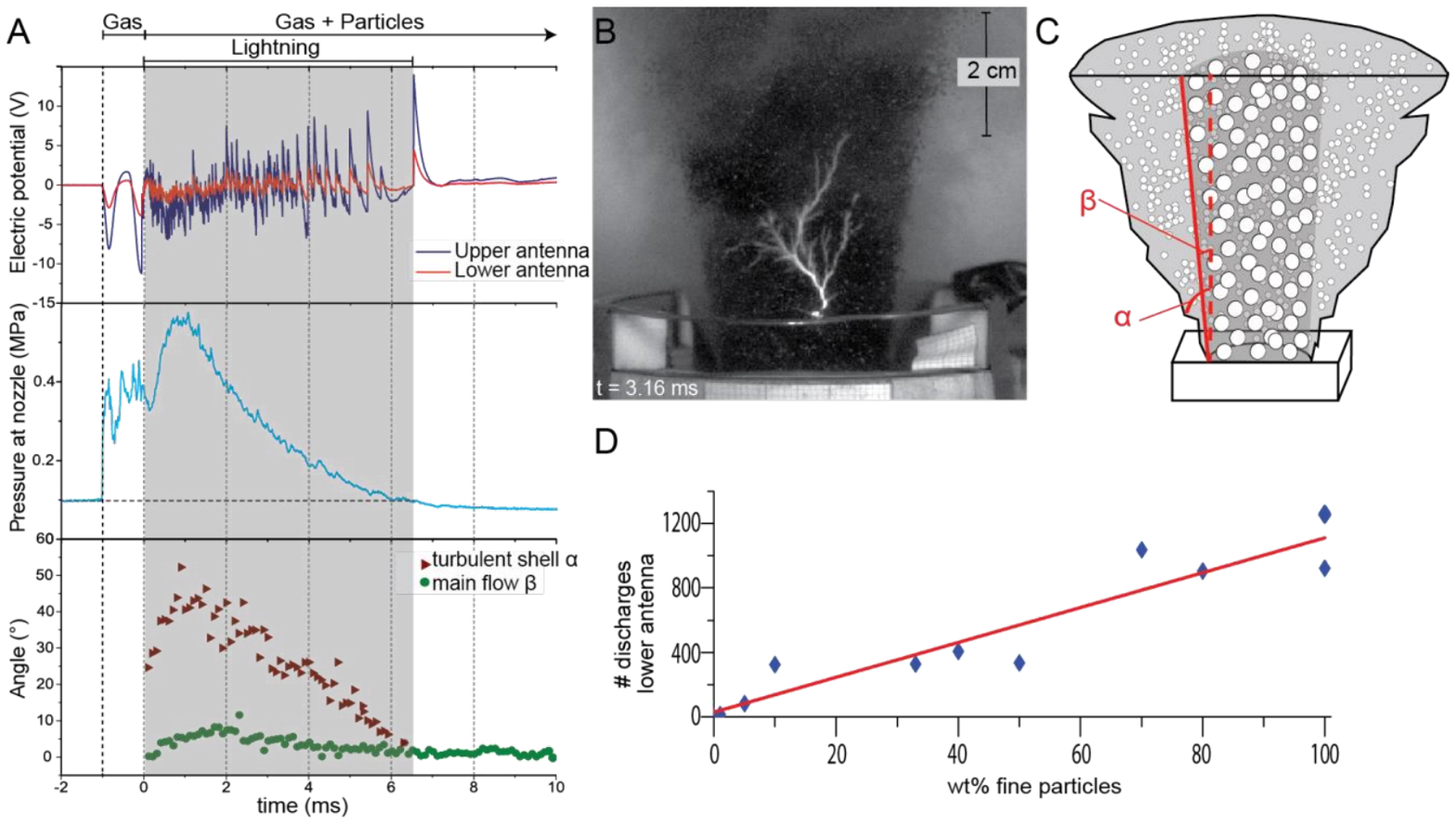}\\*[-12.5cm]
\caption{Results of a rapid decompression experiment with volcanic ash (250 $\mu$m). {\bf Panel A:} Electric potential recorded by the antennas, pressure at the nozzle and angle of the core of the flow ($\beta$) and the surrounding turbulent shell ($\alpha$) with respect to the vertical. Shaded area indicates the time window of lightning occurrence. {\bf Panel B:} Rest-frame of the high-speed videos showing the particle-laden jet is well-constrained and surrounded by the turbulent sheath of finer ash and lightning flashes are recorded. {\bf Panel C:} Schematic section of the jet showing the main flow core (coarser particles; dark grey shadow), the turbulent shell (finer particles; light grey shadow) and the respective opening angles ($\beta$ and $\alpha$) to the vertical. {\bf Panel D:} Number of discharges > 0.2 V recorded at the lower antenna in experiments with bimodal glass beads (500 and 50 m) as a function of the wt.\% of finer particles.}
\label{cimarellipic}
\end{figure}

\paragraph{Volcanic Lightning Experiments:}
Explosive volcanic eruptions are commonly associated with intense
electrical activity and lightning. A number of techniques have been
used to study the electrical activity of volcanic plumes including
close range VHF lightning mapping arrays
(e.g. \citealt{2007Sci...315.1097T,2013JVGR..259..214B}), long range
VLF lightning observations (e.g. \citealt{2010ERL.....5d4013B}) and
optical lightning detection using high speed cameras
(\citealt{2014EGUGA..16.9004C}). Direct measurement of the electric
field near the vent, where the electrical activity in the volcanic
plume is first observed is difficult, but a handful of 
studies exist including those by
\cite{1965Sci...148.1179A,1991Natur.349..598G,1998James,2002BVol...64...75M}. Lab based
experiments are also essential to studying volcanic charge generation
mechanisms in a controlled environment, and can allow different charge
mechanisms to be examined individually. Laboratory experiments by
\cite{2000JGR...105.2819B} and \cite{2000JGR...10516641J} have studied the
fractoemission mechanism, whereby James et al. generated silicate
particles by fracture during collisions between pumice samples. During
the experiments there was evidence of ion release during the fracture
process. Triboelectrification processes have also been studied in the
lab using both silica beads (\citealt{2009PhRvL.102b8001F}) and volcanic ash
(\citealt{houghton2013}), where it has been demonstrated that the
particle size distribution has important effect on the magnitude of
the charge generated.

\cite{cim2013} have achieved an analog of volcanic
lightning in the laboratory during rapid decompression (shock-tube)
experiments of gas-particle (both natural volcanic ash and glass
beads) mixtures under controlled conditions.
Experiments show that more discharges are generated for finer starting
material and that there is no correlation between the number of
discharges and the sample chemistry (\citealt{2011Geo....39..891T}).
The experiments highlight that clustering of particles trapped in the
turbulent eddies of the jet provides an efficient mechanism for both
charge generation (tribocharging) and lightning discharge as observed
in volcanic plumes. Clusters form and break-up by densification and
rarefaction of the particle-laden jet. A cluster's lifetime is
regulated by the turbulence time scale and its modification during the
evolution of the jet flow. Cluster generation and disruption provide
the necessary conditions for electrification of particles by
collision, local condensation of electrical charges and its consequent
separation, thus creating the electric potential gradient necessary to
generate lightning discharges. Clustering can be particularly
effective in the presence of prevalently fine ash-laden jets exiting
volcanic conduits\footnote{The volcano conduit is the pipe that
carries magma from the magma chamber, up through the crust and through
the volcano itself until it reaches the surface.} thus facilitating
ash aggregation in the plume (\citealt{2011Geo....39..891T}). Further
charging by the formation of hydrometeors (i.e. water droplets or ice
particles) in the upper regions of the plume
(\citealt{2012BVol...74.1963V}) could provide additional mechanisms of
plume electrification, although the presence of ice particles in the
plume (from low latitude volcanoes where surface temperatures are
high (\citealt{aiz2010})) can be ruled out in
many monitored eruptions that produced electrical discharges, thus
confirming the primary role of particle self-charging in the
generation of volcanic lightning.  The experiments show the direct
relation between the number of lightning discharges and the abundance
of fine particles in the plume as observed in the case of 2010
Eyjafjallaj\"{o}kull eruption in Iceland, as well as in many other
ash-rich eruptions or explosive episodes, independently from their
eruption magnitude and magmatic composition. Improved lightning
monitoring at active volcanoes may provide first-hand information not
only on the location of the eruption but more importantly on the
presence and amount of fine ash ejected during an eruption, which is a
fundamental input in ash-dispersion forecast models. Multiparametric
observations of volcanic plumes are therefore needed to fully
understand the favourable conditions for volcanic lightning generation
and to correctly interpret electrification and discharge phenomena to
understand plume properties. Newly designed shock-tube experiments
open new perspectives in the investigation of self-charging mechanism
of particles that are relevant for atmospheric phenomena on Earth
(such as dust storms and mesocyclones) and other planetary bodies, as
well as industrial processes involving granular materials.

\subsection{Kinetic gas-chemistry during discharges in solar-system planet atmospheres}

Atmospheric discharges have been detected in all gaseous giants of our
Solar System (\citealt{yair2012}) and are therefore likely to be
present in extrasolar planets as suggested
in \citep{helling2011a,aplin2013,hellingjsd2013,bailey2014}.
Transient Luminous Events (TLEs) occur in the Earth atmosphere (see
Sect.~\ref{s:eldis_solsys}\ref{ss:Ute}) where they influence the local
gas composition, and with that, potential observational features.

A number of models to study in detail the non-equilibrium kinetic chemistry of
TLEs have been developed (\citealt{gordillo-vazquez2008,gordillo-vazquez2009, 
grodillo-vazquez2010a, parra-rojas2013a,parra2015}). These studies have allowed the optical signatures
and spectra of TLE optical emissions (from the UV to the NIR) to be
quantified as should be seen from ground, balloons, planes and from
space (e.g. \citealt{
gordillo-vazquez2012}) illustrating good agreement with available
observed spectra.

Kinetic gas-chemistry models have been developed to calculate the TLE-induced
changes in the electrical conductivity
(\citealt{grodillo-vazquez2010a}) of the Earth upper atmosphere
showing good agreement with available measurements. The importance of
some key kinetic mechanisms (electron detachment from O$^-$) has been
shown to explain the inception of delayed sprites
(\citealt{luque2012}). The impact of lightning on the lower ionosphere
of Saturn and the possible generation of halos and sprites has been
modelled by \cite{dubrovin2014}. This allowed to study the coupling between
atmospheric layers in Saturn and Jupiter due to lightning-generated
electromagnetic pulses and to predict 
different possible optical emissions from elve-like events
triggered by lightning in the giant planets (\citealt{luque2014}).
The extension of such an approach to extrasolar atmospheres
requires a dedicated kinetic gas-chemistry network which is able to
handle a considerably wider range of chemical compositions and
temperatures than for the solar system planets
(see, e.g. the STAND2015 network from \citealt{rimmer2015}).

\subsection{Future Studies}
On Earth the quasi-static and the radiation components of the
lightning electric field have comparable effects on the secondary
TLE-discharges in the upper atmosphere. However, in planets with
larger typical distances, the radiation field can be stronger than the
quasi-static field (\citealt{luque2014}). The radiation field is
responsible for ring-shaped expanding emissions of light at the lower
edge of the ionosphere. It is therefore speculated that giant TLEs may
exist in giant planets. This new area of research has introduced many
open questions, such as:
\begin{itemize}
\item 
Can lightning-related TLEs occur on Saturn and Jupiter? What kind of TLE could be observable, what would be the required sensitivity and appropriate wavelength range? 
Could the optical flash emission on Saturn and Jupiter originate from other discharge processes than conventional lightning discharges? 

\item
Can lightning-related TLEs take place in the upper layer of the Venusian atmosphere? 
How would lightning influence  the chemical composition and electrical properties of the Venusian upper atmosphere?

\item
No direct optical lightning observation is available for the atmospheres of Neptune and Uranus, only indirect radio detection possibly associated to electric discharge events. What could be the lightning mechanisms in Neptune and Uranus?


\item What would be the possible atmospheric optical and chemical signatures in the case that lightning activity exists in extrasolar planets and brown dwarfs atmospheres? 
\end{itemize}



\section{Electrification on the Moon and on asteroids}\label{s:solsys}

Charged dust grains and dusty plasmas are known to constitute the
near-surface environment of airless bodies such as the Moon,
asteroids, comets, Saturn's rings and many planetary moons.  Our solar
system, being exposed to a variety of plasma conditions and solar
activity, provides a natural laboratory to study dust charging and
dynamics. Charging of neutral dust particles occurs when dust grains
are exposed to space plasma, for example, through interactions with
the solar wind. These plasma interactions are believed to be the
reason for many of the observations reported in the literature
(e.g. spokes in Saturn's B ring and dust streams ejected from
Jupiter \cite{honary2004}).
 
In dusty plasmas, dust particles have the ability to alter the
properties of various plasma waves and instabilities
(e.g. \citealt{dangelo1993, kopnin2009, rao1993, rao1995}). In some
cases the presence of dust can affect the instability
(e.g.  \citealt{sen2010}), whereas in other cases the presence of dust
can drive new unstable modes (\citealt{rao1990}).  Both high and low
frequency modes can be excited. High frequency modes are excited
because the dust can modify the relative drift between the plasma
species (electrons and ions) or simply reduce the electron
density. Low frequency modes (both electrostatic and electromagnetic)
occur when the dust dynamics are considered. One of the interesting
modes is dust-ion acoustic instability which is driven by the relative
drift between the dust and the plasma (\citealt{rao1993}). An example
of such a scenario exists in Saturn's E-ring where the plasma
co-rotates with the planet while the dust follows Keplerian
orbits. \cite{rosenberg1993} has shown that the relative speed between
the dust and the plasma to drive the instability is of the order of
the ion thermal speed. By introducing the magnetic field new modes
called dust-magneto-acoustic waves are excited according to the theory
(e.g. \citealt{rao1995}) which is the generalisation of the
electrostatic dust-ion acoustic wave, first reported
by \cite{shukla1992}.
 
Beyond the macroscopic behaviour of dusty plasmas, understanding dust
charging in the space environment is important for several
reasons. The variable exposure of the moon to solar wind, UV
radiation, terrestrial magnetospheric plasmas, and meteoroid impacts results in a
time-dependent, complex plasma environment. The charging, possible
subsequent mobilization, and transport of fine lunar dust have
remained a controversial issue since the Apollo era, and have been
suggested to lead to the formation of a dusty exosphere, extending
tens to hundreds of kilometres above the surface. Recent international
interest and potential return to the moon in the near future has been
declared by major space agencies around the world (NASA, ESA, JAXA,
Russia, China). The success of these missions depends largely on the
ability to understand and predict the effects of dust on the lunar
environment in order to prepare crews and equipment to withstand such
a harsh environment. Whilst NASA's Lunar Atmosphere and Dust
Environment Explorer (LADEE) (launched on 6th Sept. 2013) is the first
dedicated mission to make measurements of lunar dust composition,
other missions are planned. For example, there are dust detectors on
the Russian lander mission to the moon's South pole (Luna Glob, 2016)
and a joint Russian-Indian (Lunar-Resurs) mission in 2017/18.
 
Asteroids and comets are similarly complex environments, of interest
because they are formed from material originating from the time when
the Solar System was formed.
Precise isotope ratio measurements give insights into the formation of
our planetary system. Carbonaceous compounds from some primitive
asteroid, that have not been affected by weathering other than in
interplanetary space, could have contributed to the origins of life
through delivery of organic compounds to Earth.
There is therefore substantial scientific interest in measuring the surface
material of asteroids and comets.  Examples of successful missions
include NASA Deep Impact and Stardust (see review
by \citealt{verv2013}) and the European Rosetta mission's Philae
lander touched down on the comet 67P/Churyumov-Gerasimenko in November
2014 (e.g. \citealt{todd2007}). The mass spectrometry needed to
understand the rocky particles on the surface of asteroids (regolith:
dust, soil, broken rock, and other related materials and is present on
Earth, the Moon, Mars, some asteroids, and other terrestrial planets
and moons) is too sophisticated for comparatively simple
spacecraft-borne instrumentation, and this has motivated several
sample return missions aiming to return regolith to Earth for more
detailed analysis. The Japanese Hayabusa mission collected a sample
from asteroid Itokawa in 2005, NASA's Osiris-Rex mission 
visits asteroid Bennu in 2016, and a European mission, Marco Polo-R
was also recently studied in detail (\citealt{michel2014}).

\subsection{Charge effects on the Moon}

In its orbit, the moon is exposed to the incoming solar wind when it
is not in the Earth's magnetosheath where most of the plasma interacts
with the moon surface, forming a wake region behind the lunar
obstacle. The exposed (sunlit) surface is charged positively to about
+5V due to high photoelectron current but on the shadow side, the
inability of ions to fill in the plasma void results in regions with
energetic electrons, which will subsequently charge the surface
negatively to few hundred volts in normal conditions or up to few
thousand volts in extreme cases.  The charging from galactic cosmic
rays is negligible in comparison to the effects of the solar
wind. Like the Moon, asteroids have dusty plasma environments, with
similar charging mechanisms such as from space plasmas or the solar UV
flux (\citealt{lee1996}).
 
\begin{figure}
\begin{center}
\hspace*{-0.5cm}\scalebox{0.6}{\includegraphics[angle=0]{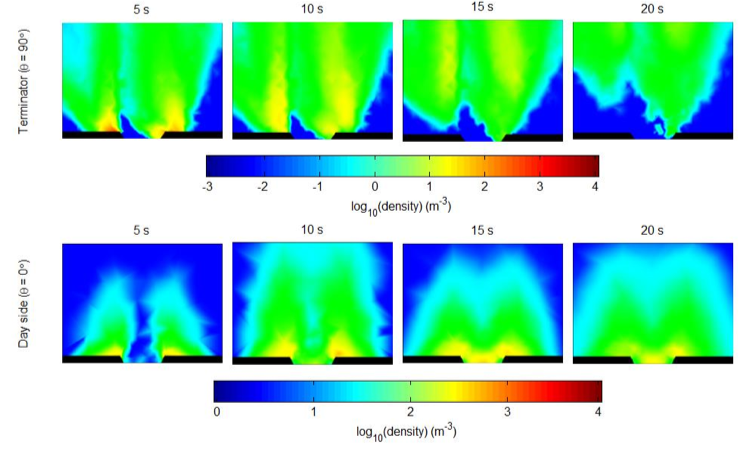}}
\caption{ Time-sequence simulations of the release of dust outside the rim of a lunar crater For two regions: terminator (top panels) and dayside (bottom panels). The crater of 5m diameter is modelled as a opening in the bottom of the panels which represents the surface of
the moon. The x and y axes represent the lunar surface and the height
of the simulation which are 45m and 60m respectively.}
\label{fig:moon}
\end{center}
\end{figure}

At midpoint between the sunlit and the night side of the moon, the
solar wind passes through almost parallel to the surface. At much
lower negative potential compared to the night side, this lunar
terminator region has been found to be the source of 'streamers' or
'horizon glow' as observed by astronauts during the Apollo mission.
It is found that the glow is produced by the scattering of sunlight by
dust particles originating from the surface, a result to be confirmed
by LADEE. It is thought that the dust on the lunar surface is charged
by the Sun's UV radiation and that other processes can contribute, such as
solar wind plasma, secondary electron emission and triboelectric
charging.  The repulsive electric field between the dust and the
surface causes the dust to levitate from the surface. Similar
mechanisms are expected to act on asteroids.  Although dust charging
and levitation have been extensively discussed
(e.g \citealt{whipple1981, goertz1989}), these processes are not yet
fully understood for the complex lunar surface where both the topology
and orbital configurations of the moon add to the complexity. Recent
development of 3D dusty plasma code based on Space Plasma Interaction
Software (SPIS) (\citealt{anuar2013, hess2014}) has provided a useful
tool to simulate many possible scenarios on lunar surface such as
lunar surface charging, shadowing phenomena and dust levitation.
Figure~\ref{fig:moon} presents time-sequence simulations of the
release of dust outside the rim of a lunar crater. At the terminator
(top panels) the presence of strong negative electric fields repels
dust particles, preventing them from reaching the basin of the crater.
On the dayside (bottom panels) dust is attracted towards the middle of
the crater basin due to the basin having a lower surface potential
than the rim surface.

 In the lunar environment a controversial and an open question is the
high altitude component of the lunar dust: what is the maximum height
that dust can be observed? The topology and the orbit of the moon
itself pose interesting questions such as:
\begin{itemize}
\item What is the charge density distribution on the surface as a function of local time and how does it change along the orbit as the Moon enters the earth's magnetosphere?
\item What is the plasma density distribution above the moon surface, and how does it change with height and time as a result of interaction with dust particles? 
\item What is the configuration of the local small-scale electric fields? How do the vertical and horizontal components of the surface electric fields evolve during the passage of the lit - dark boundary, and along the lunar orbit? 
\item How do magnetic anomalies alter the surface electric fields and plasma?
\end{itemize}

\subsection{Charge effects on asteroids}

In the case of an asteroid, there are possible electrostatic effects
on, firstly, surface material, and, secondly, through the
electrostatic effects of a spacecraft visiting an asteroid, which
could influence the outcomes of sample return missions. Asteroids
become charged by cosmic rays, the solar wind and photoelectron
emission. For asteroids in the solar system, the charging effects of
cosmic rays are negligible in comparison to the solar wind's effect,
on the nightside, and photoelectron charging on the dayside. The most
direct mechanism for charging effects on asteroids views electrostatic
processes as one type of ``space weathering''
which is a broad term for surface modification of these bodies.  Space
weathering is relevant for asteroid sample return missions since it
refers to processes that could physically and chemically modify the
sample from its ``pristine'' state, thought to be representative of the
early Solar System. As photoelectric levitation of dust particles on
the surfaces of asteroids is expected to occur, charging effects could
modulate the size distribution, by redistribution of  regolith,
for example, through ''ponding'' in craters. Modification of the size
distribution could also have more complex effects for the asteroid's
density and orbital evolution (\citealt{aplin2014}).

\begin{figure}
\begin{center}
\scalebox{0.5}{\includegraphics[angle=0]{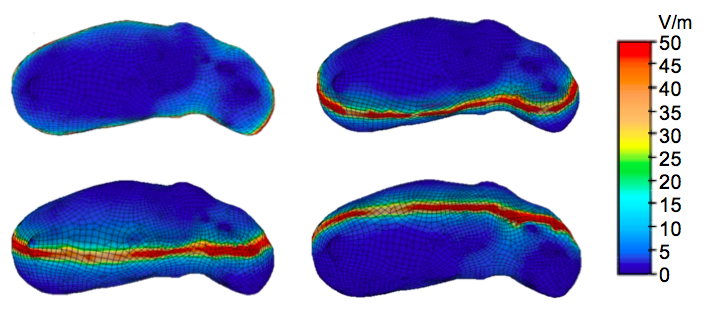}}
\caption{Four viewing angles of 3D electrostatic modelling of the surface electric field for the asteroid Itokawa (after \citealt{aplin2011}). The highest electric fields exist at the terminator region. Effects of surface topography on the electric field can also be seen.}
\label{f:Itokawa}
\end{center}
\end{figure}

Recent models (Aplin et al, 2014) considered the electrostatic
implications of a spacecraft visiting an asteroid, and found that
photoelectric shadowing from the spacecraft itself was
substantial. This shadowing will generate electric fields in the
sampling region, a hitherto neglected process that could modify the
sample to be collected. \cite{aplin2014} demonstrated that simple
isolated electrodes mounted on a spacecraft could measure the
screening from the spacecraft, and, with careful choice of position,
these electrodes could also measure the minimally disturbed electrical
environment.  Figure~\ref{f:Itokawa} shows modelled electric fields at
the surface of the asteroid Itokawa. The dayside is assumed to be at
+5V, and the nightside at -1000V (\citealt{aplin2011}) resulting in
high electric fields at the terminator. Further work is needed to
consider other electrostatic effects of a sample return mission, for
example the mechanical lofting of particles from spacecraft touchdown,
which would become charged. Triboelectric (frictional charging)
effects could also be significant. Although both Martian analogue and
lunar material tribo-charge efficiently (e.g. \citealt{aplin2012,
forward2009}), triboelectric effects have not been considered in the
asteroidal (like in Fig.~\ref{f:Itokawa}) or lunar
environment. Triboelectric charging could occur both from collisional
processes between lofted regolith, and potentially more significantly
for human exploration and sample return, from interactions between
spacecraft and the environment, such as sampling mechanisms or rovers.

There is clearly much work to be done in understanding the mechanisms
involved in dust charging and their effects. For the asteroid case,
dust needs to be included in the simulations as well as more realistic
representations of the spacecraft geometry, so that shadowing effects
can be studied more carefully.




\section{Charge processes in Extrasolar atmospheric environments}\label{s:exopl}
Charge processes and their effects occur in many astrophysical
environments. This section focuses on extrasolar objects where charge
and discharge processes introduce feedback cycles similar to those
discussed previously.  This section summarises a field
of research on cool extrasolar objects which starts to emerge as the
result of recent progress in X-ray and radio observations of brown
dwarfs.  Brown dwarfs are objects with mass intermediate between stars
and planets (Fig.~\ref{fig:SMDP}; for a review
see \citealt{hell2014}).  Since they are not sufficiently massive for
hydrogen burning in their core, they cool during their entire life time. Brown
dwarf atmospheres therefore evolve from the state of a warm stellar
atmosphere into an atmosphere as cool as the atmospheres of solar
system planets. The oldest brown dwarfs are amongst the oldest objects
in our Universe. Very-low mass stars and brown dwarfs are collectively
known as ultracool dwarfs.

Charge processes are important also in star and during planet
formation in protoplanetary disks. Ionisation processes are suggested
to help the first steps of planet formation as demonstrated by
microgravity coagulation experiments at the International Space
Station (\citealt{konop2005}; see also
Sect.~\ref{s:eldis_solsys}\ref{ss:volc}), and to allow the star to
continue to accrete mass through the propolanetary disk.

Planets and stars do have magnetic fields. The magnetic field strength
and geometry differ for different stars (\citealt{donati2009}) which
has implications for e.g. the size of a planetary magnetopause or the
high-energy radiation impact into a planetary atmospheres
(\citealt{vid2014, see2014}). Stellar magnetism changes with mass and
rotation (as indicator for age for some stars) of the objects
(\citealt{donati2008,morin2008,morin2010,vid2014b}) introducing an
additional complexity in the astrophysical context of atmospheric
electrification. Brown dwarfs can have magnetic field strengths of
~1000G (= 0.1 T).

This section first summarises recent multi-wavelength observation of
brown dwarfs as the best detectable ultra-cool and planet-like objects
(Sect.~\ref{s:exopl}\ref{ss:littlef}). Section~\ref{s:exopl}\ref{ss:BDion}
addresses ionisation mechanisms in ultra-cool atmospheres, and
Sect.~\ref{s:exopl}\ref{ss:inuts} summarised recent ideas for ionising
protoplanetary disks through which stars grow and planets
form.

\subsection{Multi-wavelength observations of activity on ultracool dwarfs }\label{ss:littlef}

\begin{figure}[htb]
\begin{center}
\includegraphics[width=0.75\columnwidth]{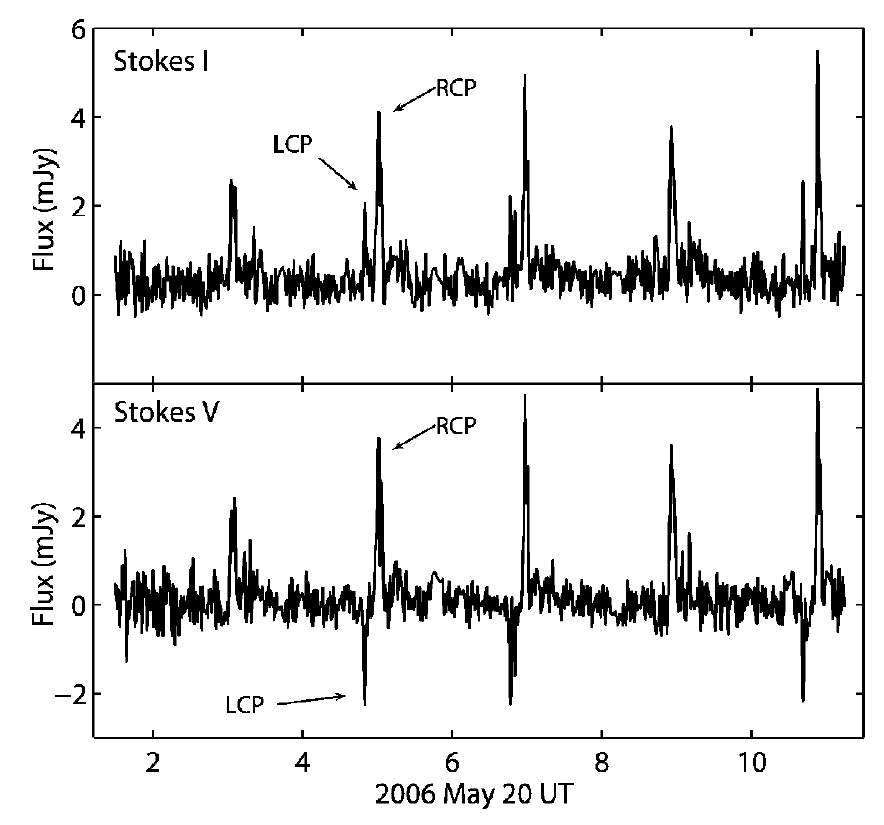}
\caption{Light curves of the total intensity (Stokes I) and the
  circularly polarized (Stokes V) radio emission detected at 8.44 GHz
  from TVLM 513-46546, an M9.5 dwarf, taken from
  \protect\cite{hallinan2007}. Right circular polarization is
  represented by positive values, and left circular polarization is
  represented by negative values in the Stokes V light curve. Bursts
  of both 100\% right circularly polarized emission (an example is
  high-lighted as 'RCP') and 100\% left circularly polarized emission
  (an example is highlighted as LCP) are detected with a periodicity
  of 1.96 hr.}
\label{f:hallinan2007}
\end{center}
\end{figure}

Below the mid-M spectral type stars (T$_{\rm eff}<$3200K; for
definition see Appendix~\ref{s:AGl}), a strongly declining H$\alpha$
emission indicates a weakening of chromospheric
activity\footnote{Chromospheric activity in form of H$\alpha$, X-ray
  or Ca II K\&K line emission results from the interaction of the
  stellar radiation field with a hot plasma above the atmosphere of a
  stellar object. The hot plasma that forms the chromosphere is the
  result of magnetic wave dissipation into thin gases and/or the
  deposition of excess radiation energy.}  (\citealt{gizis2000,
  kirkpatrick2000, liebert2003, reiners2008, williams2014}).  X-ray
observations support this finding: Whilst X-ray detections are common
for late-M spectral types (T$_{\rm eff}>$3400K), 
the X-ray luminosity declines steeply for L-type brown dwarfs (T$_{\rm eff}<$2000K; \citealt{williams2014}).  By contrast, brown dwarfs are very bright radio emitters, but
no correlation between X-ray and radio emission exist as it is
established for stars and solar events (G\"udel-Benz relation,
\citealt{benz1994}).  Since the discovery of the first radio emitting
brown dwarf by \cite{berger2001}, numerous surveys have detected radio
emission at GHz frequencies in nearly 200 objects
(\citealt{berger2002,berger2006,berger2010,hallinan2006,hallinan2007,hallinan2008,mclean2012}),
including emission in the coolest brown dwarfs with spectral types as
late as T6.5 (\citealt{route2012}). In 12 of these objects, the radio
emission is highly polarised, coherent and pulses on the rotation
period of the dwarf. These properties suggest that the source of the
radio emission is the electron cyclotron maser instability (CMI;
\citealt{wu1979}). The electron cyclotron maser mechanism has been
shown to be responsible for the auroral kilometric radiation on Earth
(see, \citealt{trak2008,2008PPCF...50g4011S,vor2011}). Figure~\ref{f:hallinan2007}
shows a light curve of a M-dwarf (TVLM 513-46546) of spectral typ M9.5
which shows clear and periodically repeating emission peaks at
8.44GHz.

In the solar system, planets have been extensively shown to be closely
associated with auroral emission, caused when electrons moving along
the magnetic field lines impact the atmosphere. \cite{nichols2012}
show that a model designed to explain Jupiter's aurora
(\citealt{cowley2001}) is able to explain the observed radio fluxes in
the ultracool dwarf 'pulsars'\footnote{Classically, a pulsar (a
  pulsating radio star) is a neutron star that is highly magnetized
  and rapidly rotating. The emitted radiation can only be observed
  when the beam is pointing towards Earth. The term 'ultracool dwarf
  pulsar' borrows this idea of beamed, lighthouse like radiation.} of
order MW Hz$^{-1}$. It could therefore be possible that the radio
emissions of some ultracool dwarfs are powered by auroral currents.
If this is true, there are profound implications for the importance of
ionisation processes on ultracool dwarfs. The current system described
by \cite{nichols2012} requires, however, both a seed ionisation in the
atmosphere, and a plasma in the magnetosphere to operate.  In turn,
the impact of auroral electrons on the atmospheres is likely to be
dramatic; whilst Jupiter's aurora increase the atmospheric
conductivity by a factor of 1000
(e.g. \citealt{strobel1983,millward2002}), the radio power of
ultracool pulsars is 10,000 times that of Jupiter. Jupiter's seed
plasma is largely driven by the solar wind and the volcanically active
Jupiter moon Io.  Brown dwarfs will not have such external plasma
sources, unless they are part of a binary system with mass transfer.

It is likely that these aurora are linked to optical
variability seen in the ultracool dwarf pulsars. This association is
suggested by the fact that five of the six ultracool dwarf pulsars
that have been observed at optical wavelengths show periodic
variability on the same period as the radio emission. For comparison
$\sim5$\% of randomly chosen ultracool dwarfs show periodic
variability. Whilst the exact mechanism producing this optical
variability is not yet clear, multi-colour observations of one
ultracool dwarf pulsar has ruled out starspots as the cause
(\citealt{littlefair2008}).


\paragraph{Near-IR Signature of Chromospheric Activity in Brown Dwarfs:}\label{ss:sora}

Recent near-IR observations with the AKARI
satellite can only be explained under the assumption that a
chromosphere comparable to the solar chromosphere is present.
Theoretical studies of brown dwarf atmospheres predict that such
low temperature atmospheres are dominated by molecules and dust, and
that they can be well modelled by simple radiative equilibrium
assuming thermodynamic equilibrium. However, AKARI observations in the
near-infrared wavelength range suggest that also chromospheric
activity plays an important role for the atmospheric structure in
particular for early-type brown dwarfs (\citealt{saro2014}).
\begin{figure}[htb]
\begin{center}
{\ }\\*[0cm]
\includegraphics[width=\columnwidth]{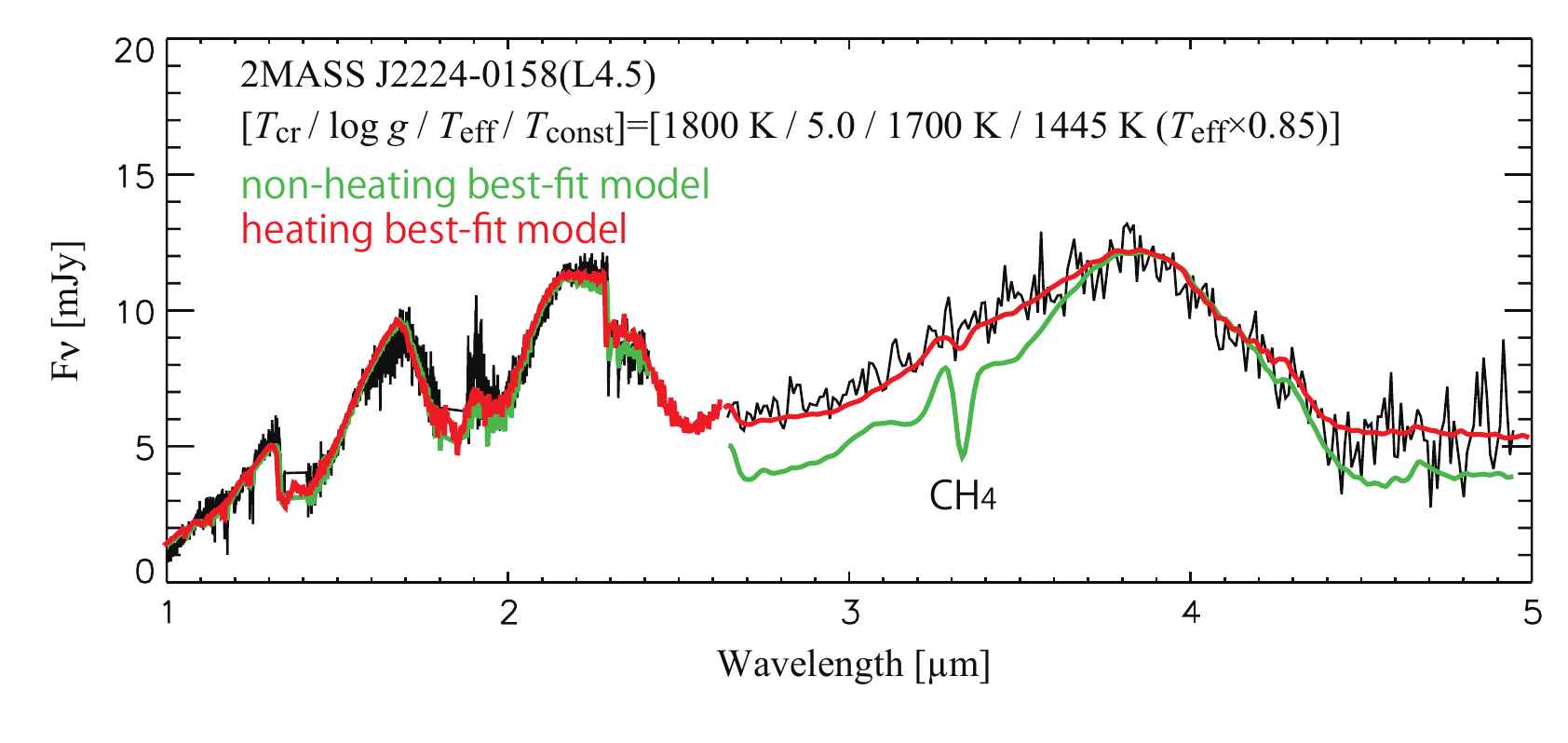}\\*[-0.5cm]
\caption{Comparison of the model spectrum (red and green smooth lines) with the observed spectrum for the L4.5 brown dwarf 2MASS J22240158 (thin black spiky line), which is well explained by the heating model
atmosphere (red line) taking into account the heating in the upper
atmospheres.}
\label{f:saro2014}
\end{center}
\end{figure}
Deviations between theoretical model spectra and observed spectra
around 3.0 and 4.5 $\mu$m, which is sensitive to the upper atmospheric
structure of brown dwarfs, suggest an additional heating source in the
upper atmosphere. The comparison of the model spectrum with the
observed spectrum for a L4.5 type brown dwarf with moderate H$\alpha$
emission, 2MASS J2224-0158, is shown in Fig.~\ref{f:saro2014} as an
example. \cite{saro2014} construct a simple
model that includes heating due to chromospheric activity which
results in a dramatic change of the chemical structure of the
atmosphere. The resulting model spectra of early-type brown dwarfs
with chromospheric heating considerably improves the match with the
observed spectra. This result suggests that chromospheric activity is
essential to understand the near-infrared spectra of brown dwarfs, and
that MHD processes can heat the upper atmosphere. A similar conclusion
was reached by \cite{schmidt2015} who photometrically examine a sample
of 11820 M7-L8 dwarfs. \cite{roba2015} have used a grid of model
atmopshere simualtions to demonstrated that it is reasonable to expect
the formation of an ionosphere and, therefore, also a chromosphere in
ultra-cool atmospheres such as on brown dwarfs.

\paragraph{Irradiated brown dwarf atmospheres: }\label{p:casewell}

Only a handful of systems are known where a brown dwarf is heated by a
hot companion.  These brown dwarfs have close orbits of a few days,
and they transit their host star, giving a measure of the brown
dwarf's radius, which is inflated by the energy input from its star.
\begin{figure}[htb]
\begin{center}
\scalebox{0.4}{\includegraphics[angle=0]{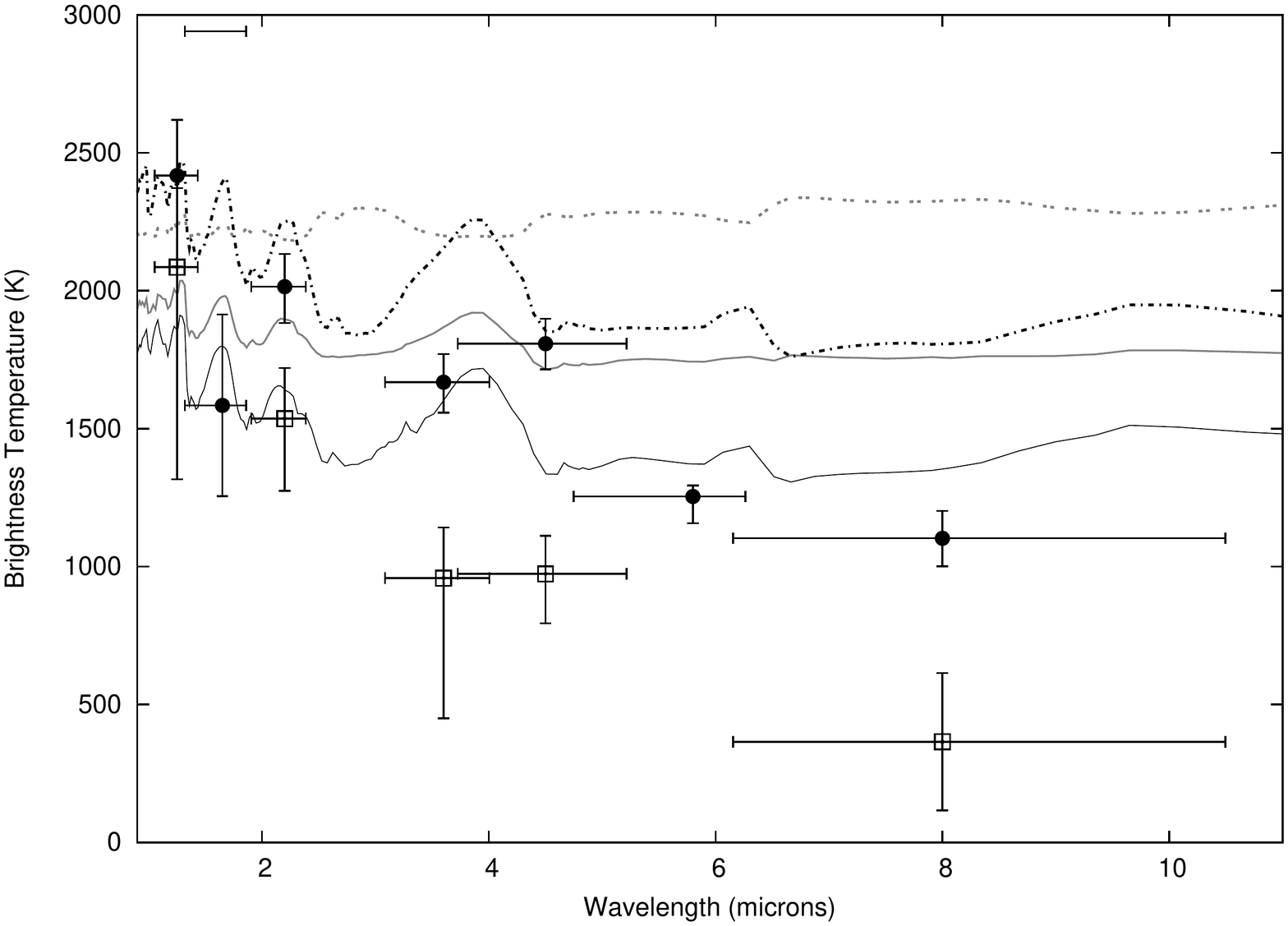}}\\*[-0.8cm]
\caption{Brightness
  temperatures for WD0137-349 on the irradiated (circles) and
  unirradiated (open boxes) sides of the brown dwarf. The errorbars on
  the X scale represent the widths of the filters used. Solid lines
  are models that use full circulation and energy transport from the
  heated to non-heated side of the brown dwarf. Dotted lines show the
  zero circulation models. The grey lines are models containing TiO
  and black lines for the non-TiO model. The models represent the flux
  on the the dayside only. The H and [5.8]$\mu$m temperatures for the
  unirradiated side are upper limits only (diamonds), derived from
  the white dwarf's flux (\citealt{casewell2014}).}
\end{center}
\end{figure}
White dwarf - brown dwarf binaries provide a case where the brown
dwarfs are not outshone by their companions, and therefore an
opportunity to study irradiated brown dwarfs.
In five of these systems, WD0137-349B (\citealt{maxted2006}), NLTT5306
(Steele et al. 2013), SDSS141126+200911 (Beuermann et al. 2013),
WD0837+185 (Casewell et al. 2012) and GD1400B (Farihi et al. 2004) the
brown dwarf is known to have survived a phase of common envelope
evolution.\footnote{The phase of common envelope evolution in the
  binary star evolution involves the brown dwarf being engulfed by,
  and immersed in, the expanding atmosphere of the white dwarf
  progenitor as it evolves away from the main sequence.  These systems
  are very close, and tidally locked resulting in one side of the
  brown dwarf continually being heated by its companion.} 
WD0137-349 is the best studied system. It is photometrically variable
in all wavelengths from the $V$ band though to 8.0 $\mu$m. These
variations are on the orbital period of the system and peak at 4.5
$\mu$m. Converting the dayside and nightside flux to brightness
temperature (peak blackbody temperature with that flux at that
wavelength) shows a temperature difference between the two hemispheres
of $\sim$500 K, and a possible temperature inversion in the
atmosphere. \cite{casewell2014} compare the observed photometry fluxes (i.e. radiative fluxes measured in a certain wavelength interval)  to models of irradiated brown dwarfs
and show that the data are best fit by models that incorporate
full energy circulation around the brown dwarf, but do {\it not} contain a
temperature inversion. However, at 2$\mu$m ($K$ band) and at 4.5
$\mu$m, the flux of the brown dwarf is still much brighter than the
model. 
\cite{casewell2014}  suggest that UV irradiation can cause
photochemical reactions in the upper brown dwarf's atmosphere that
produce large hydrocarbon molecules causing the brown dwarf to be
brighter at 2$\mu$m and 4.5 $\mu$m as were demonstrated for CR
impact by \cite{rim2014}.

\subsection{Ionisation processes in ultra-cool atmospheres}\label{ss:BDion}

Brown dwarf and planets atmospheres are
spectroscopically characterised by a rich ensemble of molecules
(e.g. SiO, TiO, VO, CO, H$_2$O, FeH) which lead to the conclusion that
such atmospheres are too cool for thermal ionisation to significantly
influence the local chemistry or energy content. But brown dwarfs and
extrasolar planets exist in a larger diversity and, hence, are
exposed to very different environments: The cosmic ray flux will be
different in an interstellar environment than in the solar system
(compare Sect.~\ref{s:eldis_solsys}). The chemical composition of the gas in
atmospheres outside the solar system causes the formation of mineral
clouds where the cloud particles are composed of a mix of silicates,
iron and metal-oxides (\citealt{helling2003, helling2009,
  helling2009b}), very similar to volcano ash. Extrasolar mineral clouds are much larger than terrestrial clouds due to the larger extension of the atmospheres of brown dwarfs and giant gas planets.  Also such clouds are
susceptible to charge and discharge processes through cosmic rays
(\citealt{rimmer2013}) and turbulence-enhanced dust-dust collisions
(\citealt{helling2011a}). The study of the break-down condition in
non-solar system atmospheres suggest that different intra-cloud
discharge processes dominate at different heights inside mineral
clouds: local coronal (point discharges) and small-scale sparks at the
bottom region of the cloud where the gas density is high, and flow
discharges and large-scale sparks near, and maybe above, the cloud top
(\citealt{hellingjsd2013}). \cite{bailey2014} apply scaling laws to
demonstrate that discharge will propagate farther in brown dwarf
atmospheres than in atmospheres of giant gas planets.

Brown dwarfs can be irradiated by a binary-companion
(\citealt{maxted2006, casewell2013},
Sect.~\ref{s:exopl}\ref{ss:littlef}) resulting in similar global
circulation patterns as demonstrated for irradiated giant gas planets
(e.g. \citealt{knut2007, dobbs2013}). If such a wind of sufficient
high speed hits a sufficiently pre-ionised gas, Alfv{\'e}n ionisation
can produce bubbles of gas with a degree of ionisation of $\sim$ 1
(\citealt{stark2013}). The surrounding atmosphere remains in
its low degree of ionisation leading to an atmosphere with a time-dependent and spatially
intermittent degree of ionisation. A sufficient degree of ionisation is
the precondition to understand the magnetic coupled atmospheric gas
responsible for radio emission in brown dwarfs (Sect.~\ref{s:exopl}\ref{ss:littlef}),  and from 
magnetically driven mass loss of extrasolar planetary atmospheres
(\citealt{tanaka2014}). Other mechanisms for planets to lose mass are
related to their host star's radiation field (\citealt{murray2009}).

An important input for understanding ionisation processes is the
global atmospheric structure and results from radiative-hydrodynamic
simulations are therefore discussed in more detail below.

\paragraph{Large-scale modelling of globally circulating extrasolar atmospheres: }

A prominent sub-class of
extrasolar planets are the short-period gaseous planets. 
They are (and due to observational constraints will remain) the best characterized of all
extrasolar planets. The short periods ($\sim$3 days),
gaseous nature, and largely circular orbits suggest that the rotation
rate of these planets is tidally locked to their orbital period. The
result is a stationary day-night heating pattern across the
surface. The proximity to their host stars means that the hot daysides
will be highly irradiated, reaching temperatures of several thousand
degrees.

The large longitudinal temperature gradient between the day and night
hemispheres drives atmospheric dynamics that  transports heat
from day to night sides.  The efficiency of this advective transport is
a subject of extensive multi-dimensional radiative hydrodynamical
studies. Phase-curve observations, consisting of infrared measurements
of the planetary flux throughout the entire orbit, do suggest that the
night-side of the planet is somewhat cooler then the day
(e.g. \citealt{knut2007}). However, the night-side temperatures are still
on the order of a thousand degrees, much larger then one would expect
simply from the internal cooling of the planet.
This suggests that the
atmosphere is actually fairly good at transporting energy across the
entire planetary surface.

The winds driven by the extreme temperatures on short period planets
are unlike any seen in any solar systems. The coupling between the
(slow) planetary rotation and the temperature differential results in
the development of a broad, supersonic, super-rotating equatorial jets
(\citealt{tsai2014}). Gas velocities at pressure levels of 0.1 bars in the
well-known HD189733b can reach 5km/s (e.g. \citealt{dobbs2013}). 

 Thus, longitudinal transport and mixing, in contrast to vertical
convection/turbulence as is important in Jupiter, plays a much larger
role.  The complexity of the atmospheric dynamics requires coupling
together a dynamical model (solving the fluid equations) to a
radiative model (involving the detailed opacities). Currently, models
utilize molecular opacities, primarily due to species such as CO,
H$_2$O, and CH$_4$. However, transit observations
(\citealt{pont2013, sing2014}) taken for different wavelength bands
suggest a
cloud coverage throughout the atmosphere. While in hindsight based on
observations of brown dwarfs or our gas giant planets this is not
surprising, it makes the coupled problem significantly more
complex. As the efficiency of energy transport by the gas flow depends
on both the fluid velocity and the cooling timescale the growth of
cloud layers and the associated change in opacity will modify
radiative timescale and may have important consequence for energy
re-distribution. Conversely, the changes in the dynamics will alter
the growth efficiency of clouds. Unfortunately, as with clouds on
Earth, it is not at all clear if this will result in a net cooling or
heating of various regions. 
The precision with which the local atmospheric properties, like gas
temperature, cloud properties, gas composition, is modeled are crucial
and efforts are ongoing in the community by, for example, improving
the treatment of the equation of state that cover a temperature range
of 250K$\,\ldots\,$6000K and a gas pressure range of 10$\mu$bar to
10Mbar. The equation of state provides the abundances of opacity
species (ions, atoms, molecules, cloud particles) that influence the
local temperature through radiative transfer effects.  Theoretical
calculations now suggest that clouds should be very prevalent
throughout these atmospheres (\citealt{Lee2015}).

MHD simulations allow first insights into magnetic coupling effects
despite containing much less information about atmospheric processes
than radiative-hydrodynamic circulation models. Such MHD simulations
have also been inspired by studies of protoplanetary disks (site of
planet formation) and solar physics
(e.g. \citealt{rogers2014}). \cite{tanaka2014}, for example, have used
an open magnetic flux-tube model in their ideal MHD simulation to
demonstrate that under this conditions the planet could form a wind
driven by Alfv{\'e}n waves. \cite{murray2009} present an extensive
study of UV and X-ray driven mass loss from irradiated extrasolar
giant gas planets.

\subsection{Discharges  in protoplanetary disks}\label{ss:disks}

\paragraph{Discharge in magneto-hydrodynamically turbulent gases:}\label{ss:inuts} 

Magneto-hydrodynamical turbulence is suggested as a mechanism to
sustain the ionisation in a cool atmospheric gas.  The energy
dissipated from MHD turbulence is to ionise the gas. If this energy is
large enough, a positive feedback loop develops where this local
ionisation serves as driving mechanisms for magneto-hydrodynamical
turbulence.  Magneto-hydrodynamical turbulent motion in weakly ionized
media continuously creates local electric field even in comoving
frame of the media.  If the electric field is larger than the critical
electric field for electron avalanche, discharge occurs in such a
weakly ionized media.
This idea has been proposed for the turbulence driven by
magneto-rotational instability in protoplanetary disks
by \cite{inut2005} who found that energetic electrons in a
Druyvesteyn distribution may produce impact ionization in particular
conditions of the dusty gaseous disks:

The atmospheres of gaseous planets and brown dwarfs are non-uniform in
chemical composition for various reasons (formation of dust grains and
their sedimentation, temperature stratification, occasional accretion of planetesimals etc., see Sect.~\ref{s:exopl}\ref{ss:BDion}).
The convection of those objects can be ``double diffusive convection''
where the structure is destabilized by the diffusion of an elemental abundance gradient.
 
Recent numerical simulations show that the double diffusive convection
evolves into a multi-layer structure where double diffusive convection
is confined into thin layers and usual convection occupies most of the
volumes, which results in very small energy flux in the radial
direction (e.g. \citealt{rosenb2011a,rosenb2011b}).  
The
effect of magnetic field can be important in the thin layers, since
magnetic tension force is inversely proportional to the length scale
of the eddy.  If the ionization degree is kept high enough in the thin
layers, the magnetic field possibly lowers the convective energy flux
even further, and hence, slows down the gravitational contraction of
those objects.  Therefore, any processes that may increase the
ionization degree is important in the theory of long-term
evolution of very cool brown dwarfs and gaseous planets.  A simple
energetics argument based on the order of magnitude calculations shows
that the energy required for keeping the ionization degree
sufficiently high (magnetic Reynolds number >1) is substantially small
($<10^{-4}$) compared to the available energy as a turbulent
convective motion.  The critical electric field for impact ionization
in the astrophysical dusty plasma is calculated in detail as a
function of gas density in \cite{okuz2014} who also
show that the resultant Ohm's law is highly non-linear and requires a new
method to handle the magneto-hydrodynamics in particular regime.  To
determine the viability of the proposed process theoretically,
magneto-hydrodynamics numerical simulations incorporating the
micro-physics of electron impact ionization are required (e.g \citealt{mura2012a, mura2013}).

\paragraph{Observation of Lightning in Protoplanetary Disks by Ion Lines:}

\begin{figure}
\scalebox{0.25}{\includegraphics[angle=0]{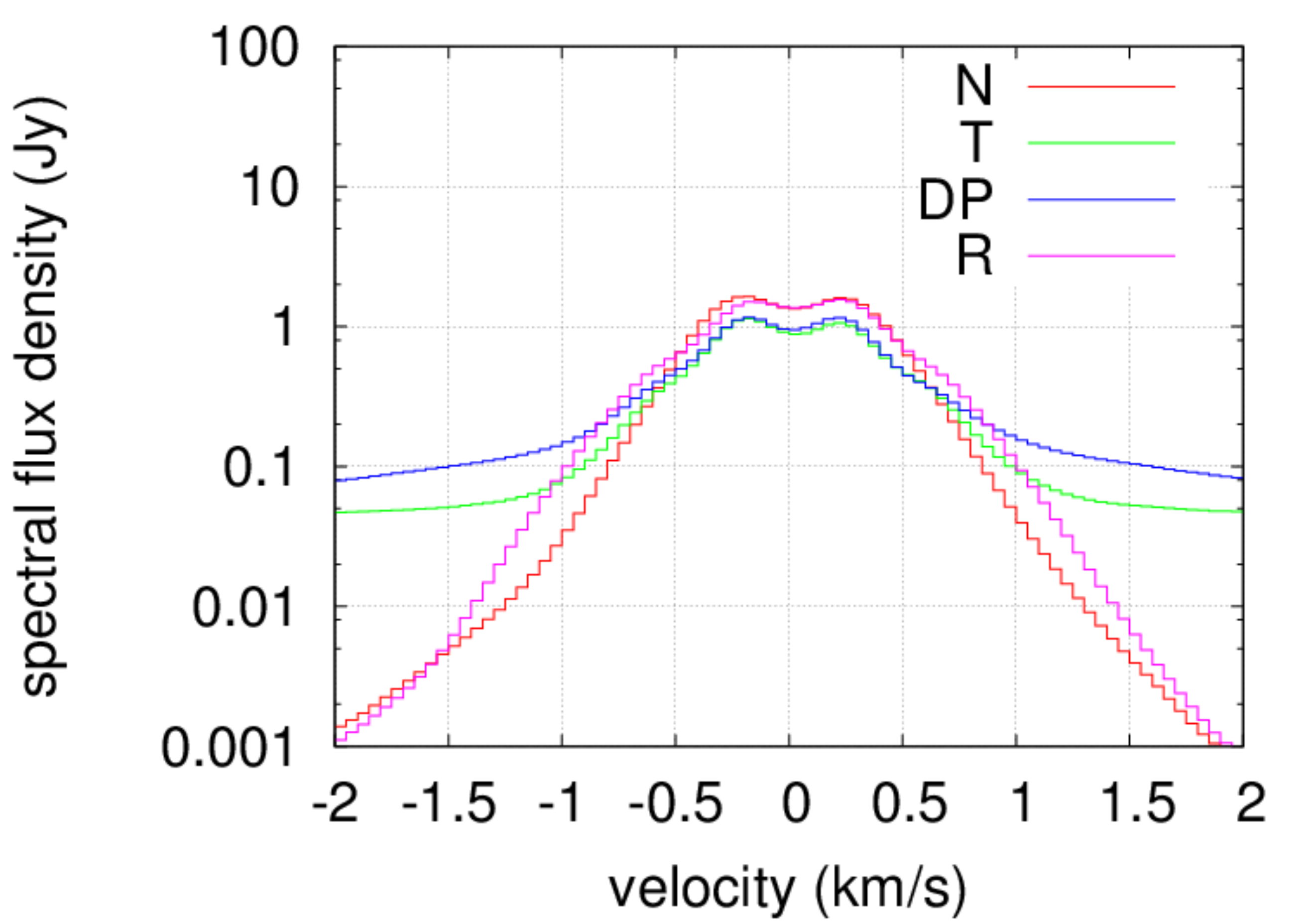}}
\scalebox{0.25}{\includegraphics[angle=0]{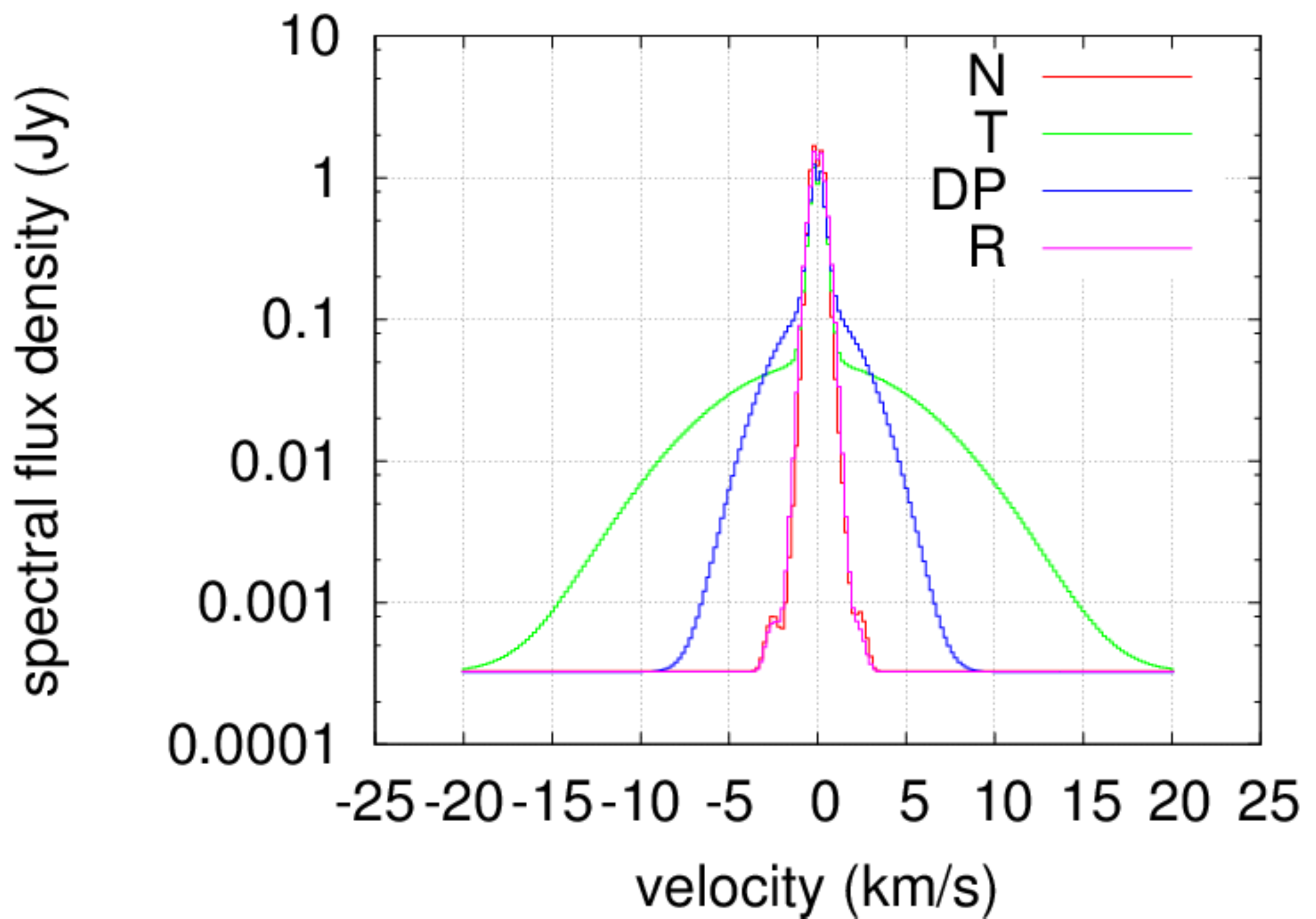}}
\caption{HCO$^+$ (3-2) line profiles [Jy]  at 267.56GHz (1.1205 mm)  without (left) and with (right) lighting.  Lighting is here understood in analogy to Earth lighting as a large-scale discharge process. Different line styles (N, T, DP, R) indicate different break-down models. The disk inclination of 7deg at a distance of 56 pc is similar to TW Hya. A minimum-mass solar nebula was applied and a lighting region 25AU $\ldots$ 50 AU was considered. The line flux considerably increases if lighting occurs. The ALMA sensitivity limit is 0.01Jy for this line.}
\label{Fig:muran1}
\end{figure}

Lightning, a large-scale discharge process in analogy to
Earth-lightning, in protoplanetary disks has been studied as a
candidate mechanism for chondrule formation, it provides a unique
window to probe the electromagnetic state of the protoplanetary
disks. Evidence for strong (500-1000G
(0.05-0.1T; \citealt{wasil2000}), transient magnetic fields is found
in meteorites. As a consequence, multiple lightning models have been proposed for protoplanetary disks
(\citealt{gibbard1997possibility,weidenschilling1997production,desch2000generation,muranushi2010dust}). \cite{muranushi2010dust}
calculate the charge distribution of dust in a protoplanetary disk
where a magneto-resonance instability produces an electromagnetic
field. If the electric field potential is large enough for an ensemble
of insulated but charged dust particle, a field break-down will occur
similar to the field-breakdown in dust clouds of brown dwarfs and
extrasolar planets (\citealt{hellingjsd2013}). Different break-down
models ({\bf T}ownsend, {\bf D}ruyversteyn-{\bf P}enning, {\bf
R}unnaway) can be tested which lead to different values for the
break-down field influencing the shape of the line profile
(Fig.~\ref{Fig:muran1}, N - no field).

\begin{figure}
\hspace*{-1.5cm}
   \begin{tabular}{ccccccc} \includegraphics[angle=0,width=5.8cm]{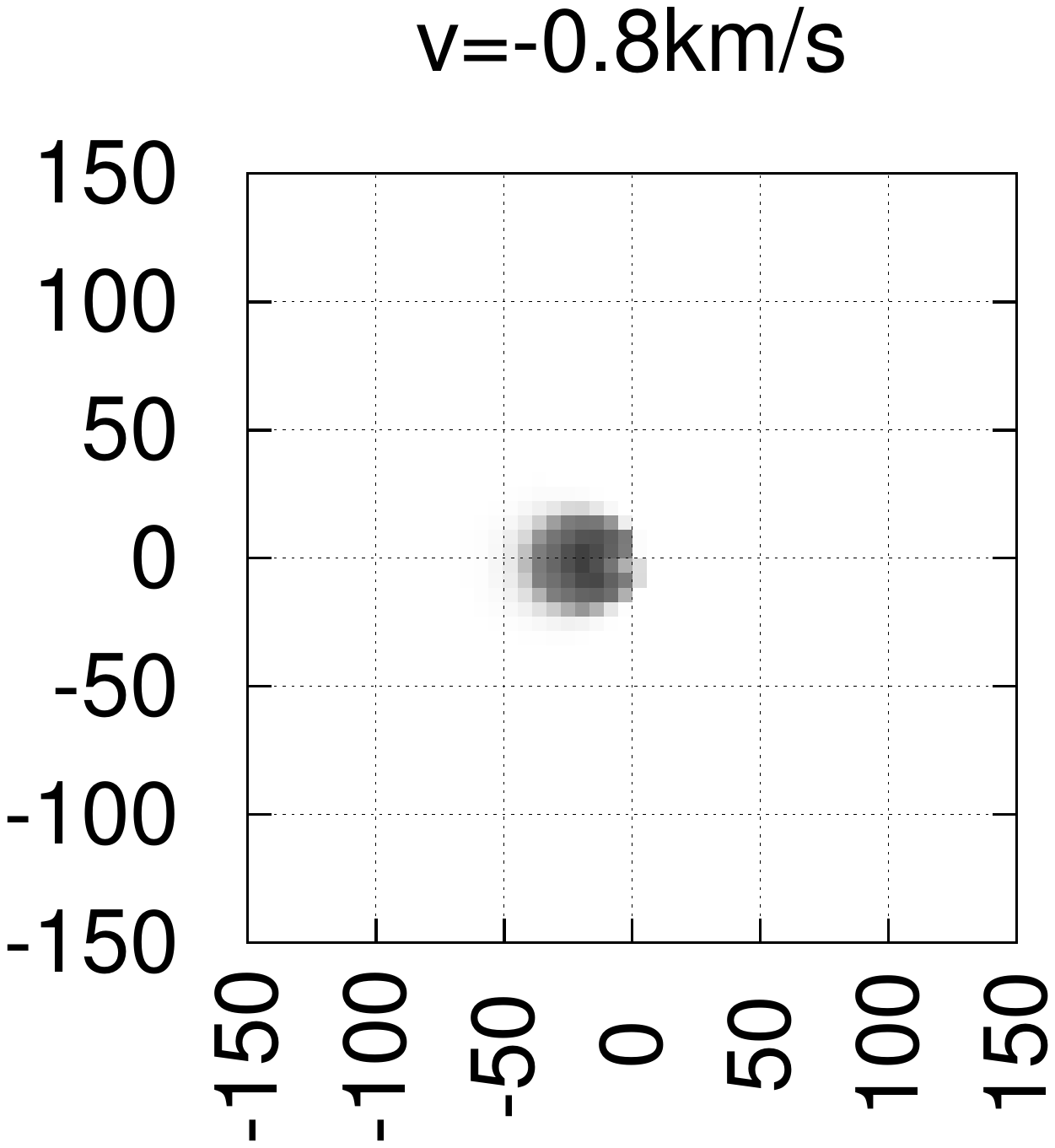} \hspace{
   -3.8cm}& \hspace{ -3.8cm}
   \includegraphics[angle=0,width=5.8cm]{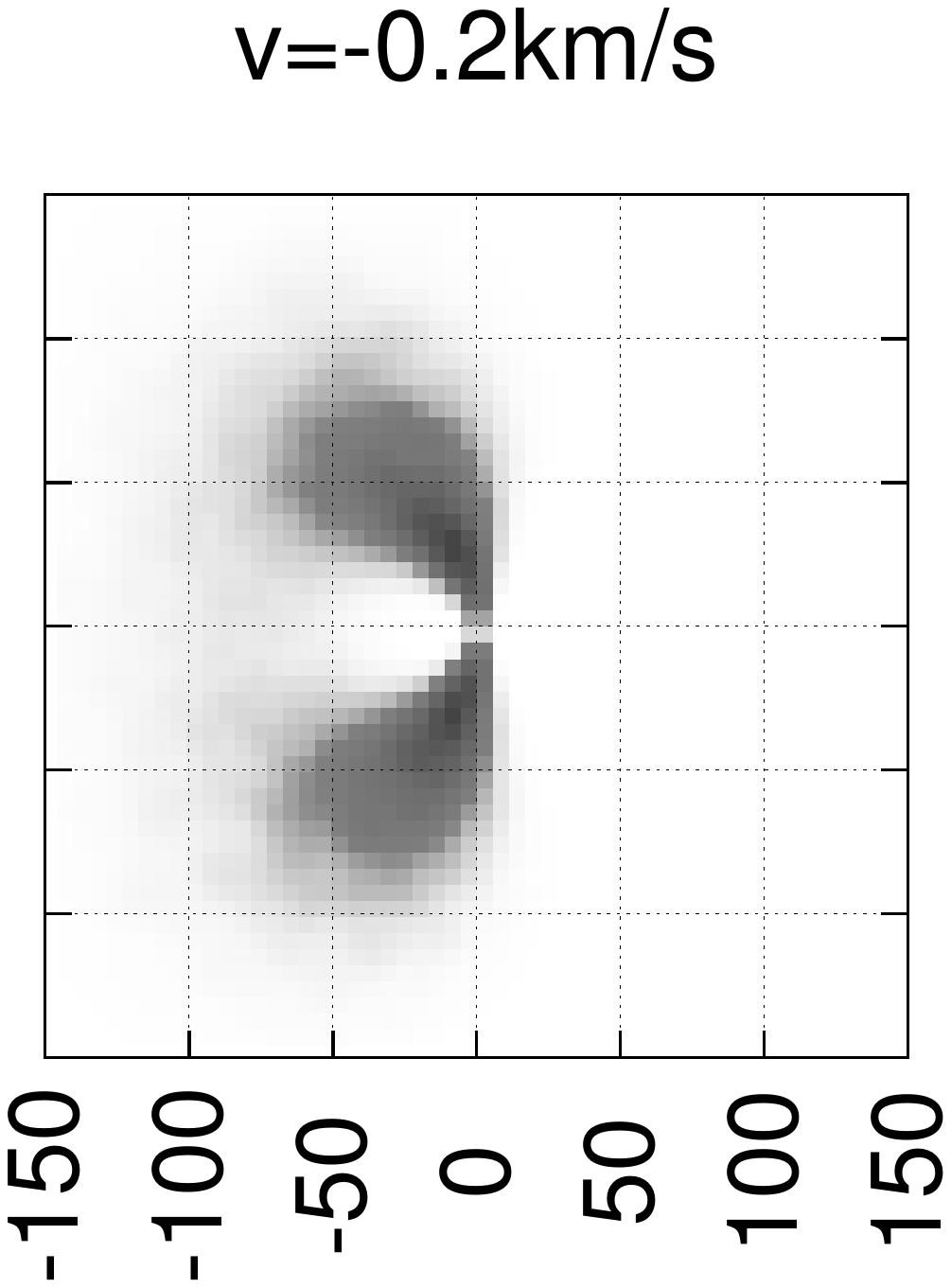}  \hspace{ -3.8cm}& \hspace{ -3.8cm}
   \includegraphics[angle=0,width=5.8cm]{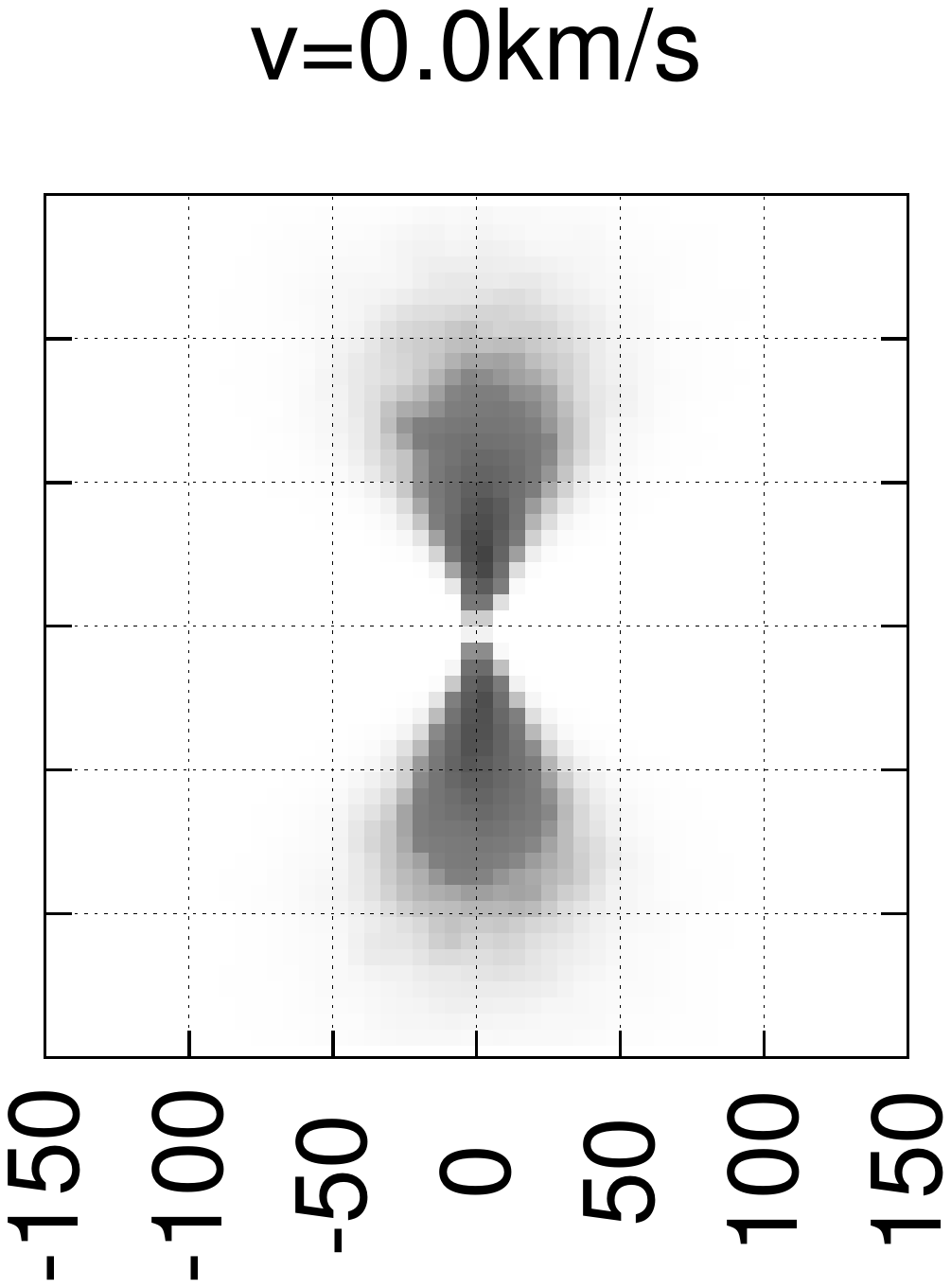}  \hspace{ -3.8cm}& \hspace{ -3.8cm}
   \includegraphics[angle=0,width=5.8cm]{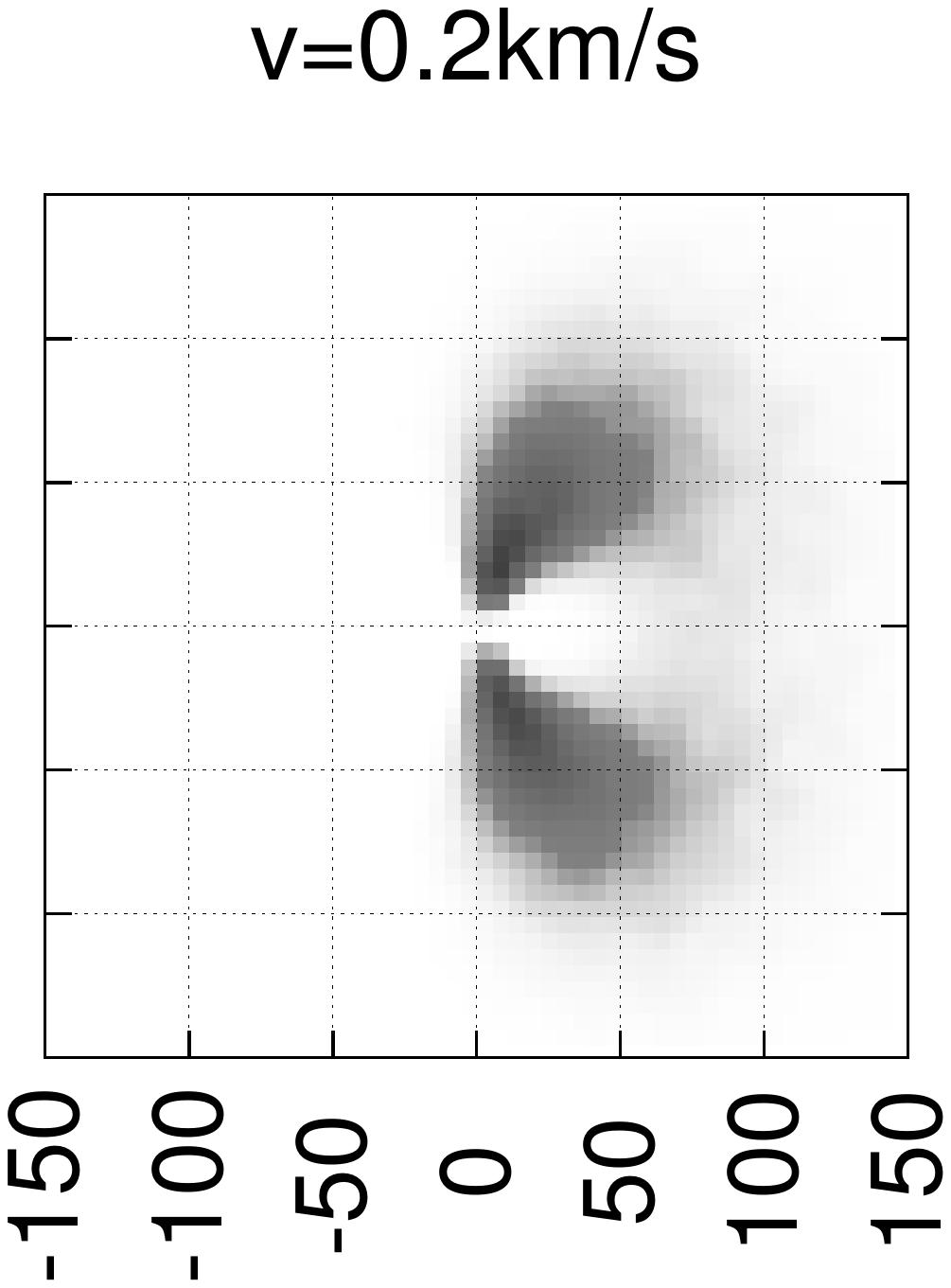}  \hspace{ -3.8cm}& \hspace{ -3.8cm}
   \includegraphics[angle=0,width=5.8cm]{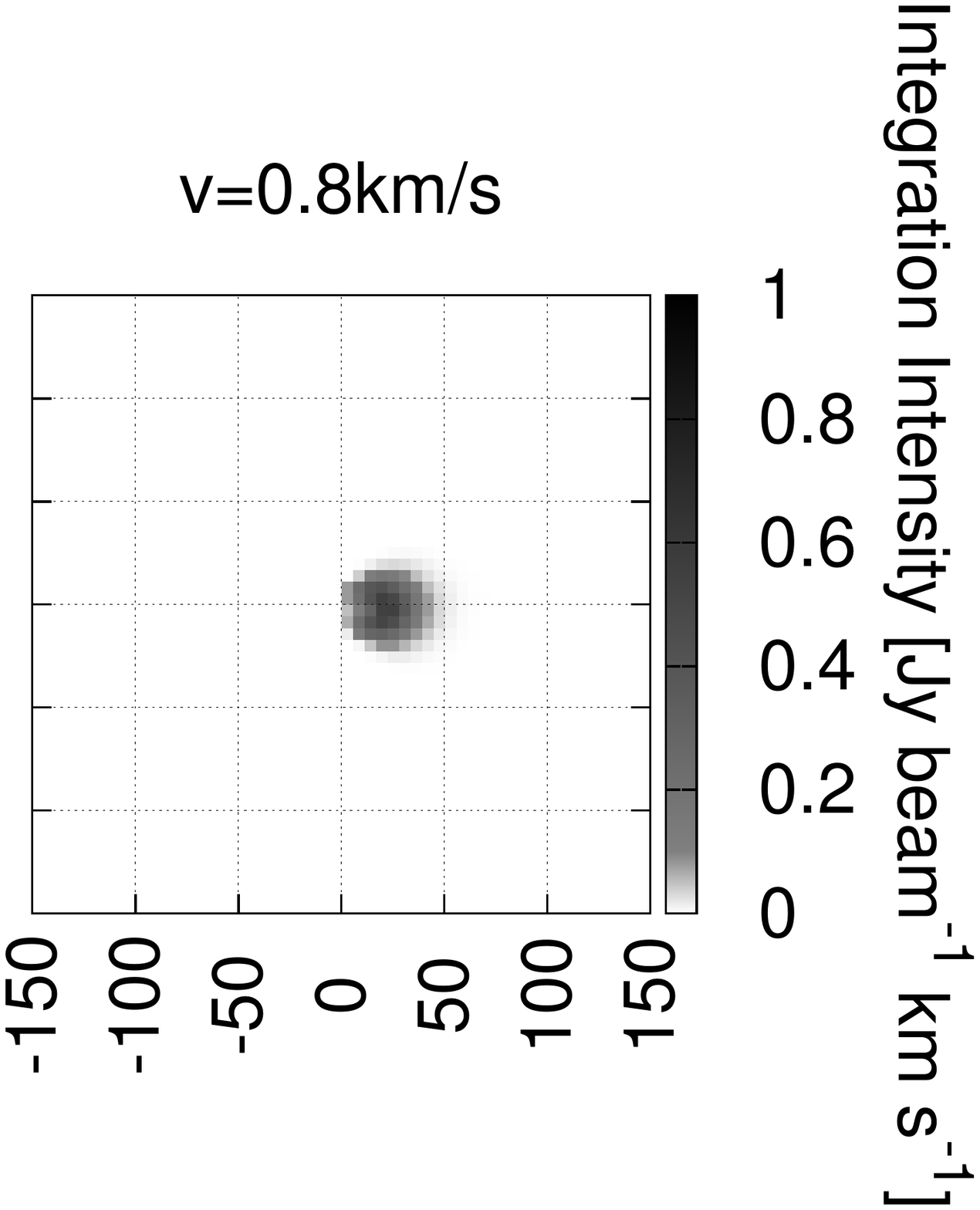}  \\*[-0.7cm]
   \includegraphics[angle=0,width=5.8cm]{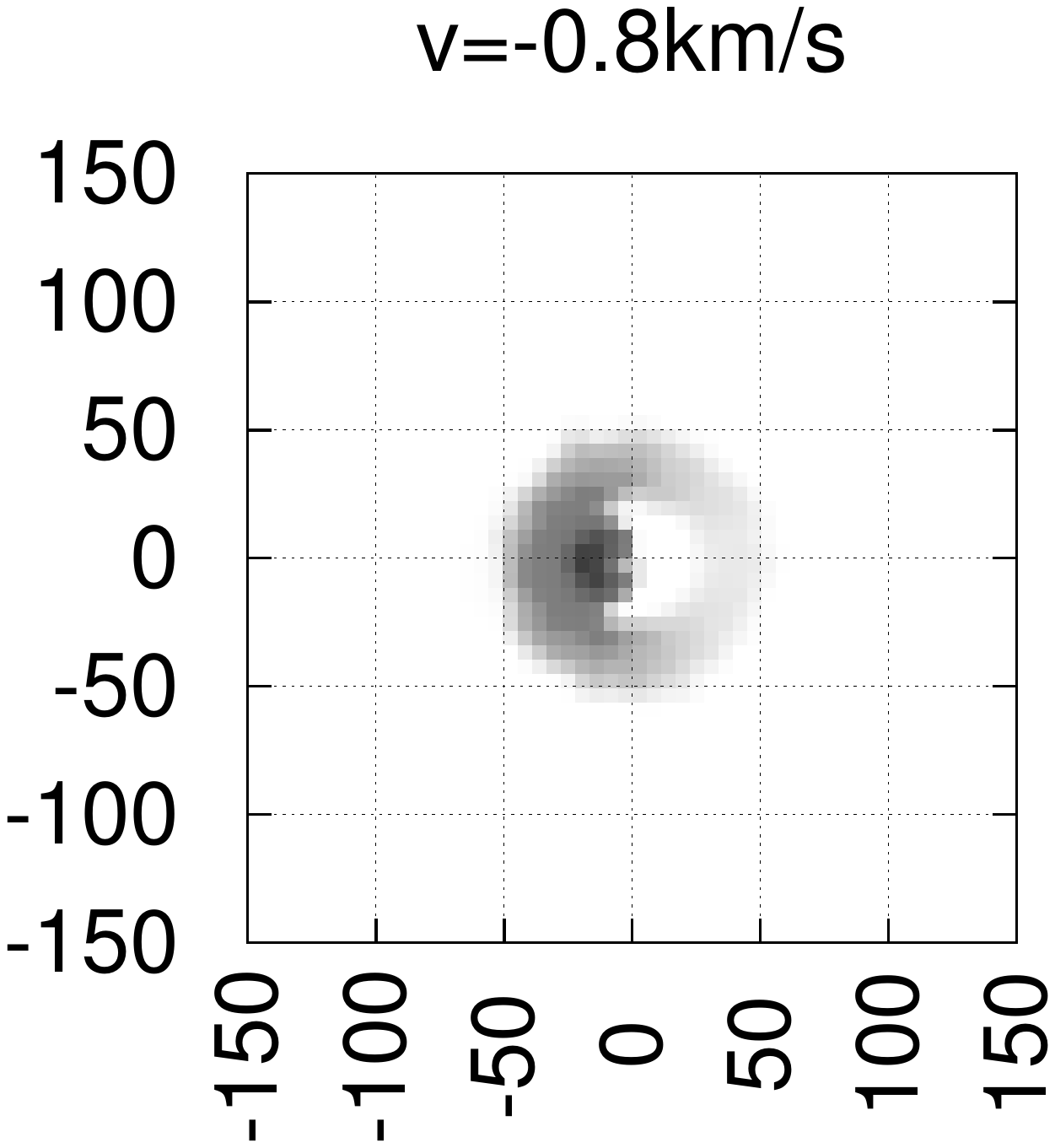}  \hspace{ -3.8cm}& \hspace{ -3.8cm}
   \includegraphics[angle=0,width=5.8cm]{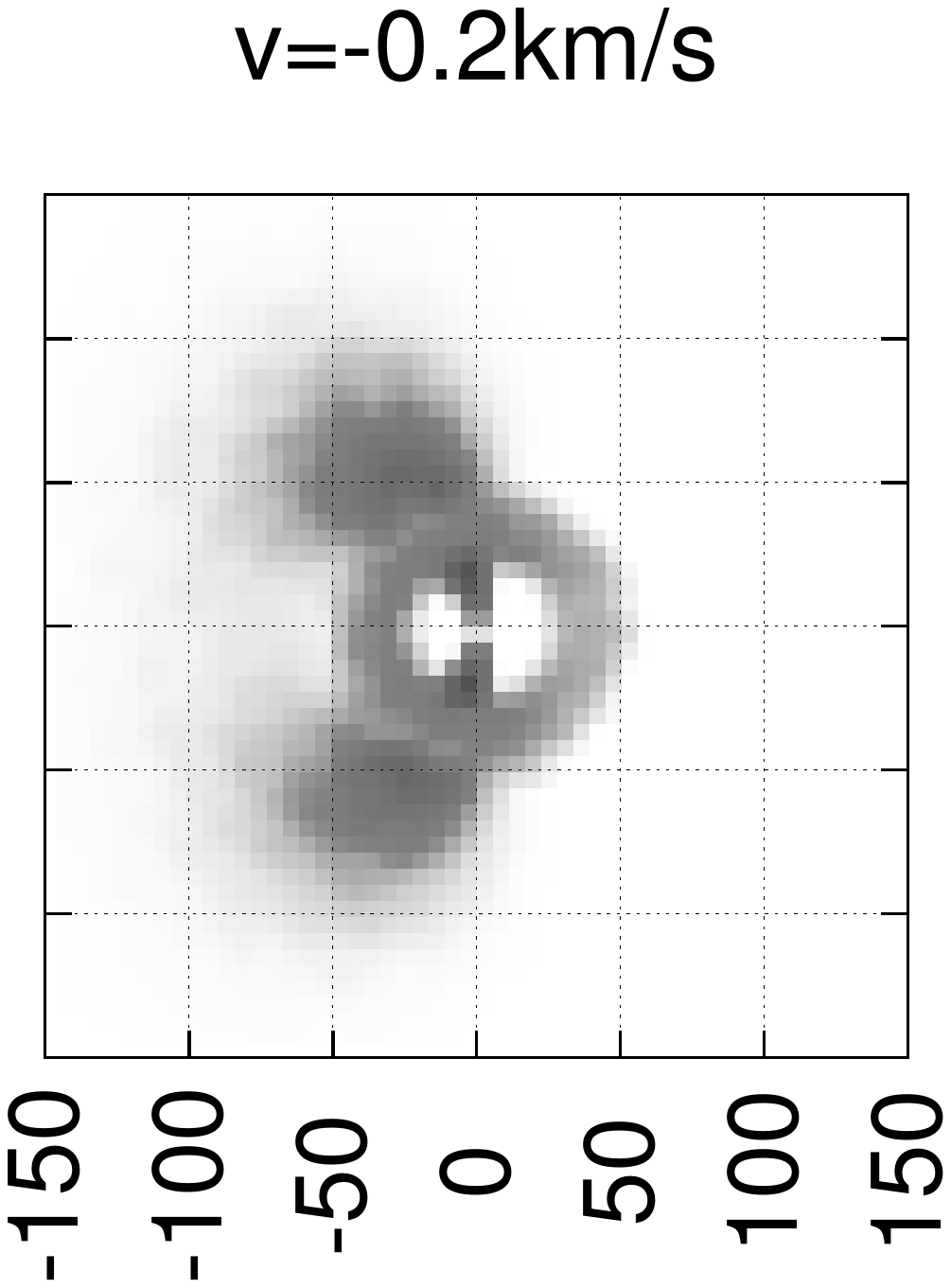}  \hspace{ -3.8cm}& \hspace{ -3.8cm}
   \includegraphics[angle=0,width=5.8cm]{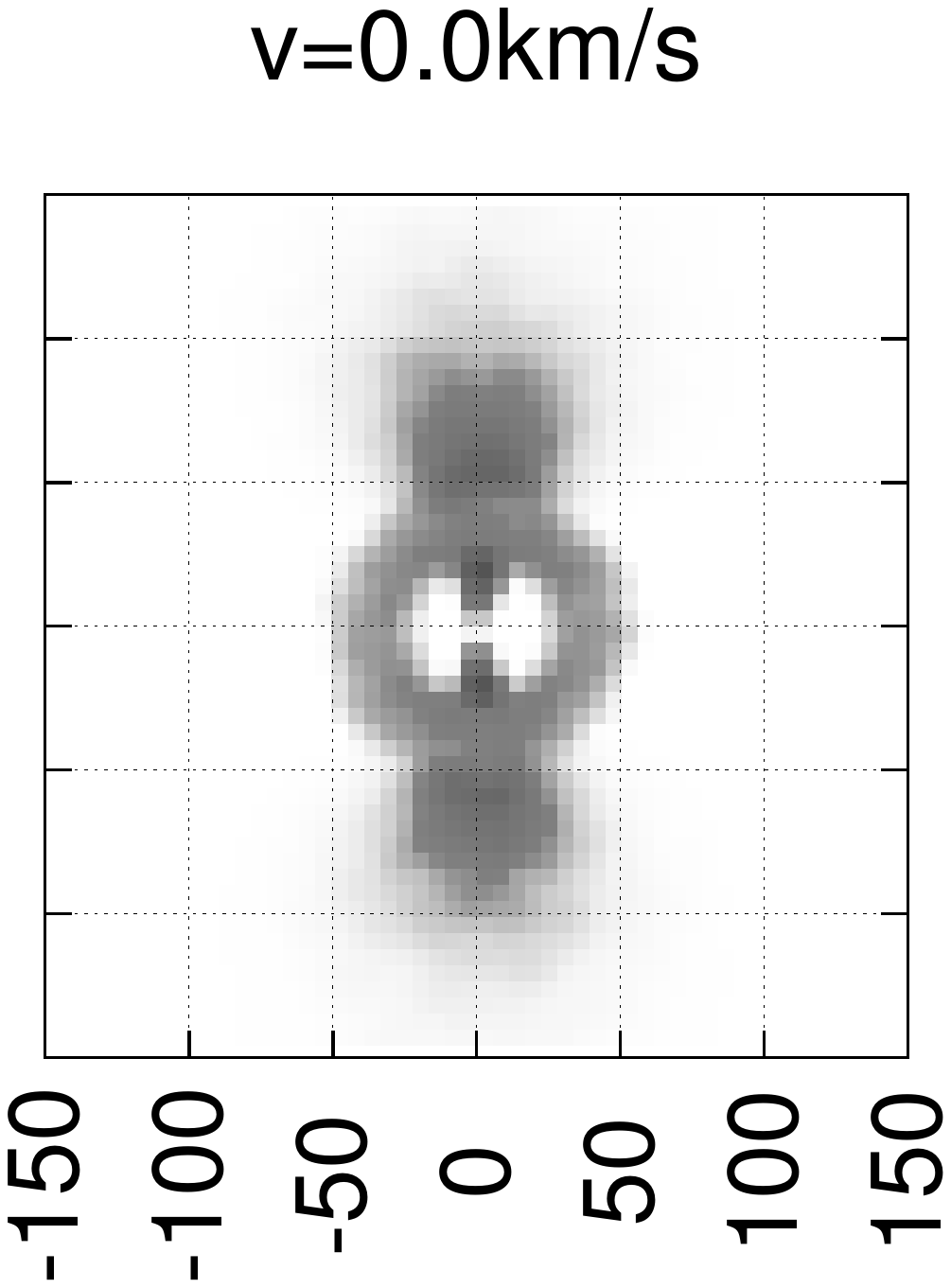}  \hspace{ -3.8cm}& \hspace{ -3.8cm}
   \includegraphics[angle=0,width=5.8cm]{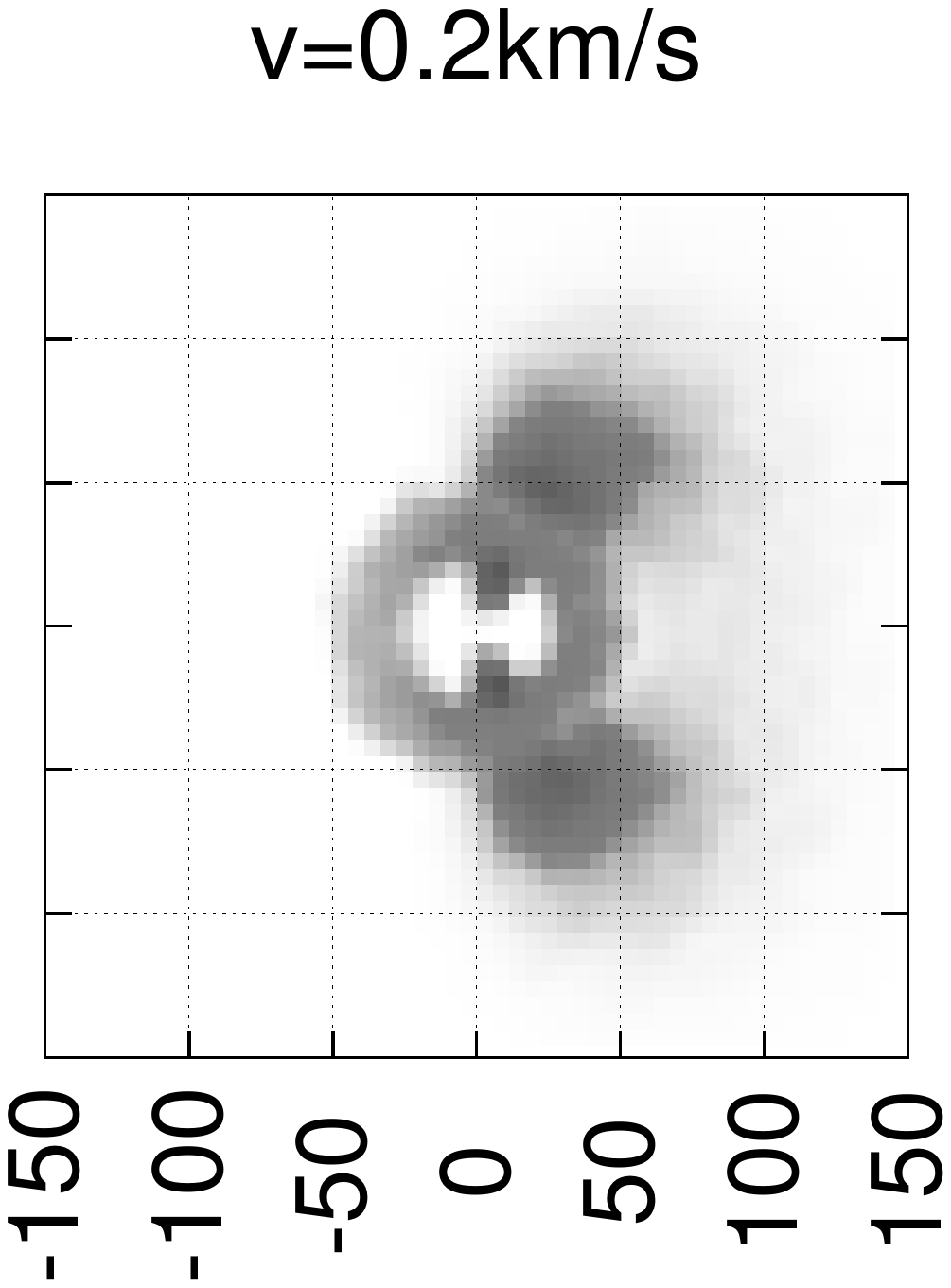}  \hspace{ -3.8cm}& \hspace{ -3.8cm}
   \includegraphics[angle=0,width=5.8cm]{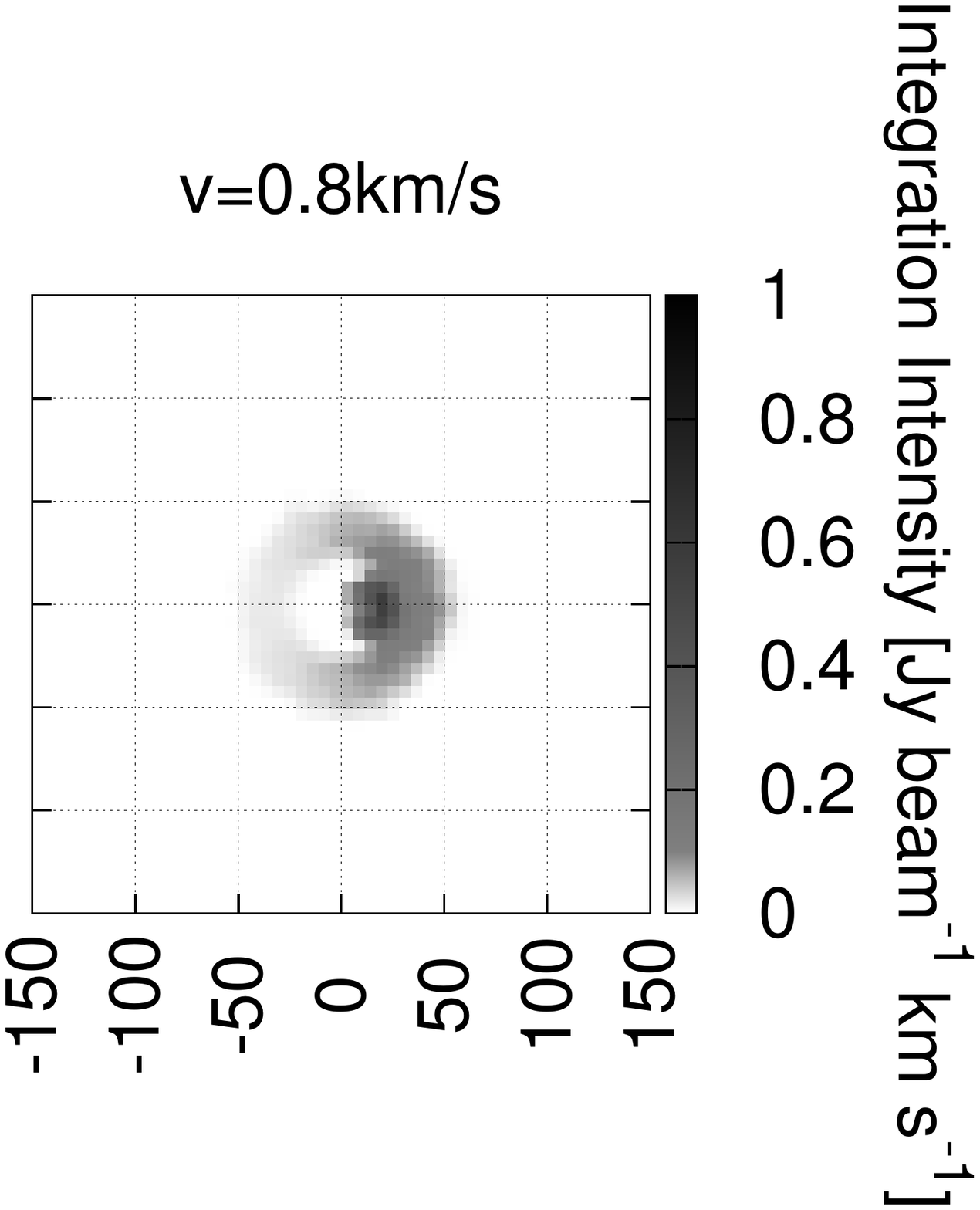}  \\*[-0.7cm]
   \includegraphics[angle=0,width=5.8cm]{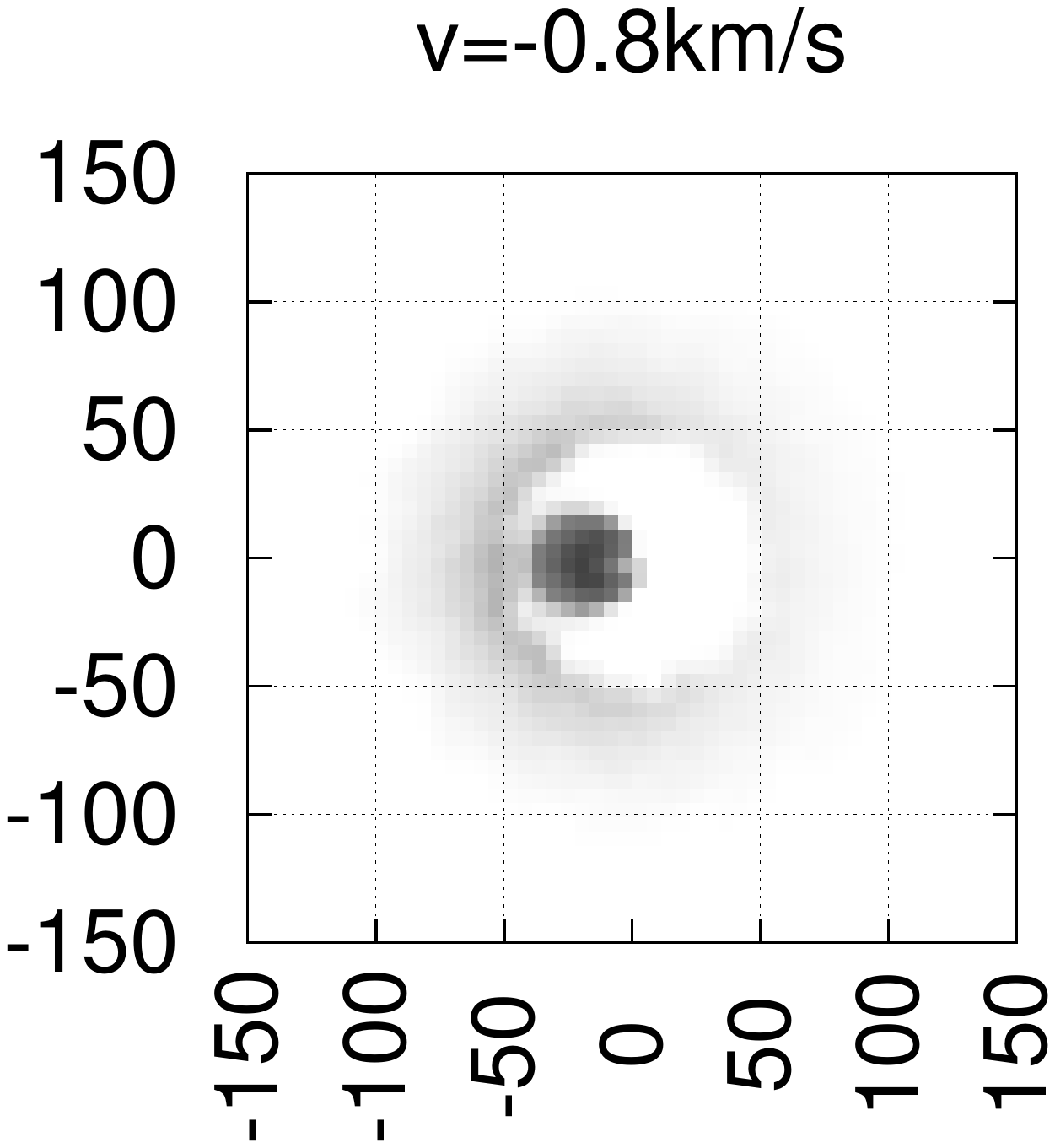}  \hspace{ -3.8cm}& \hspace{ -3.8cm}
   \includegraphics[angle=0,width=5.8cm]{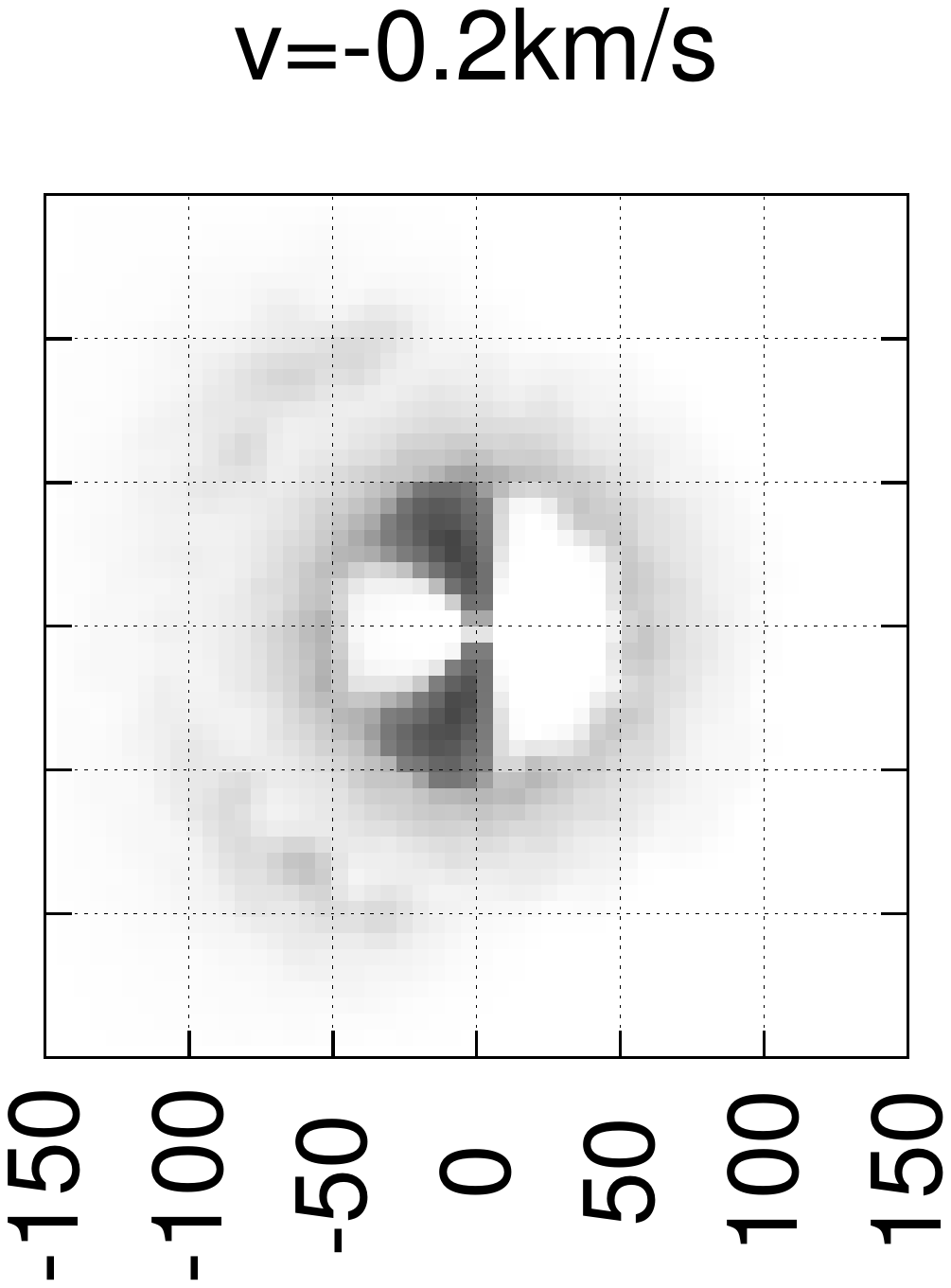}  \hspace{ -3.8cm}& \hspace{ -3.8cm}
   \includegraphics[angle=0,width=5.8cm]{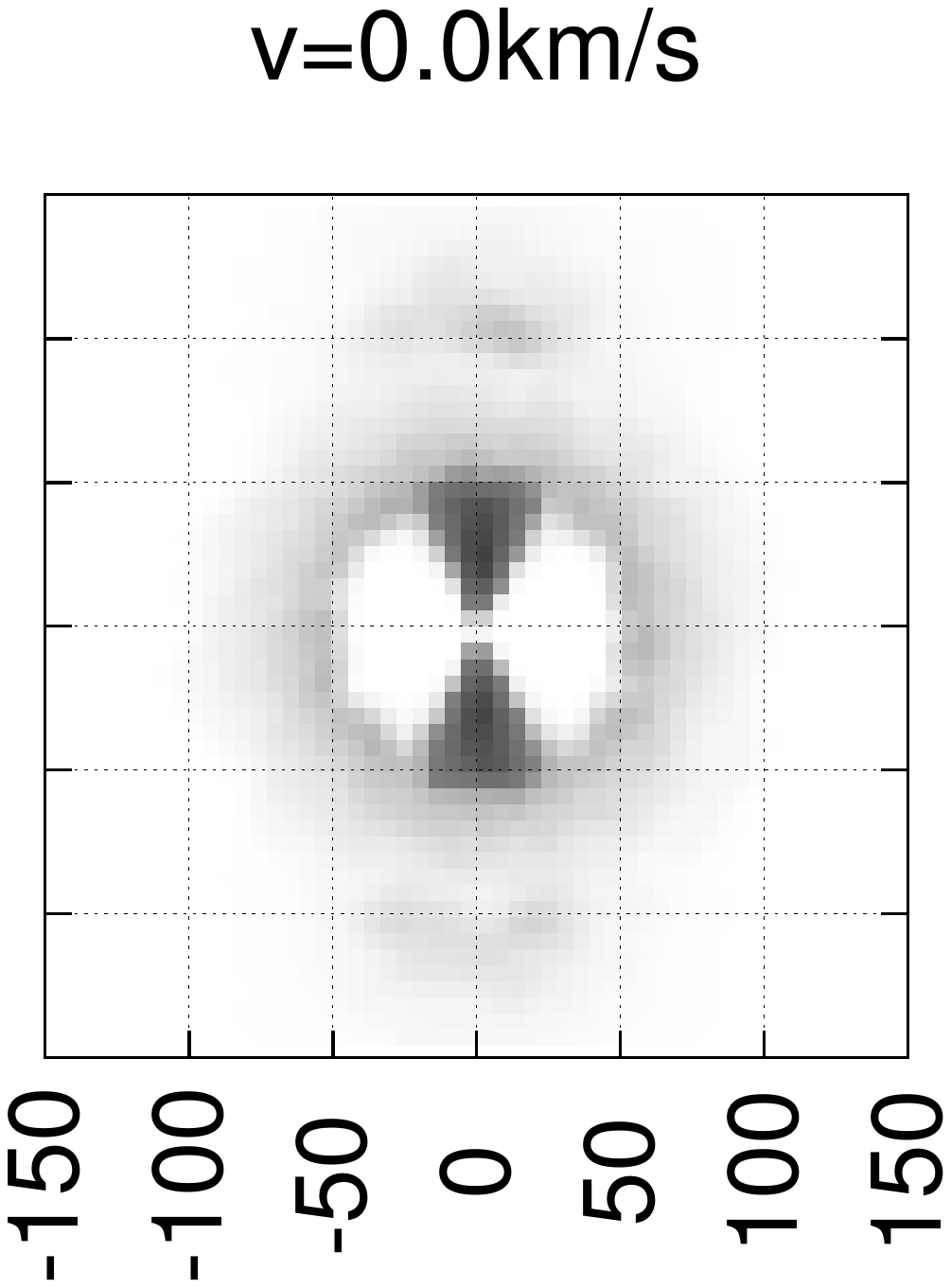}  \hspace{ -3.8cm}& \hspace{ -3.8cm}
   \includegraphics[angle=0,width=5.8cm]{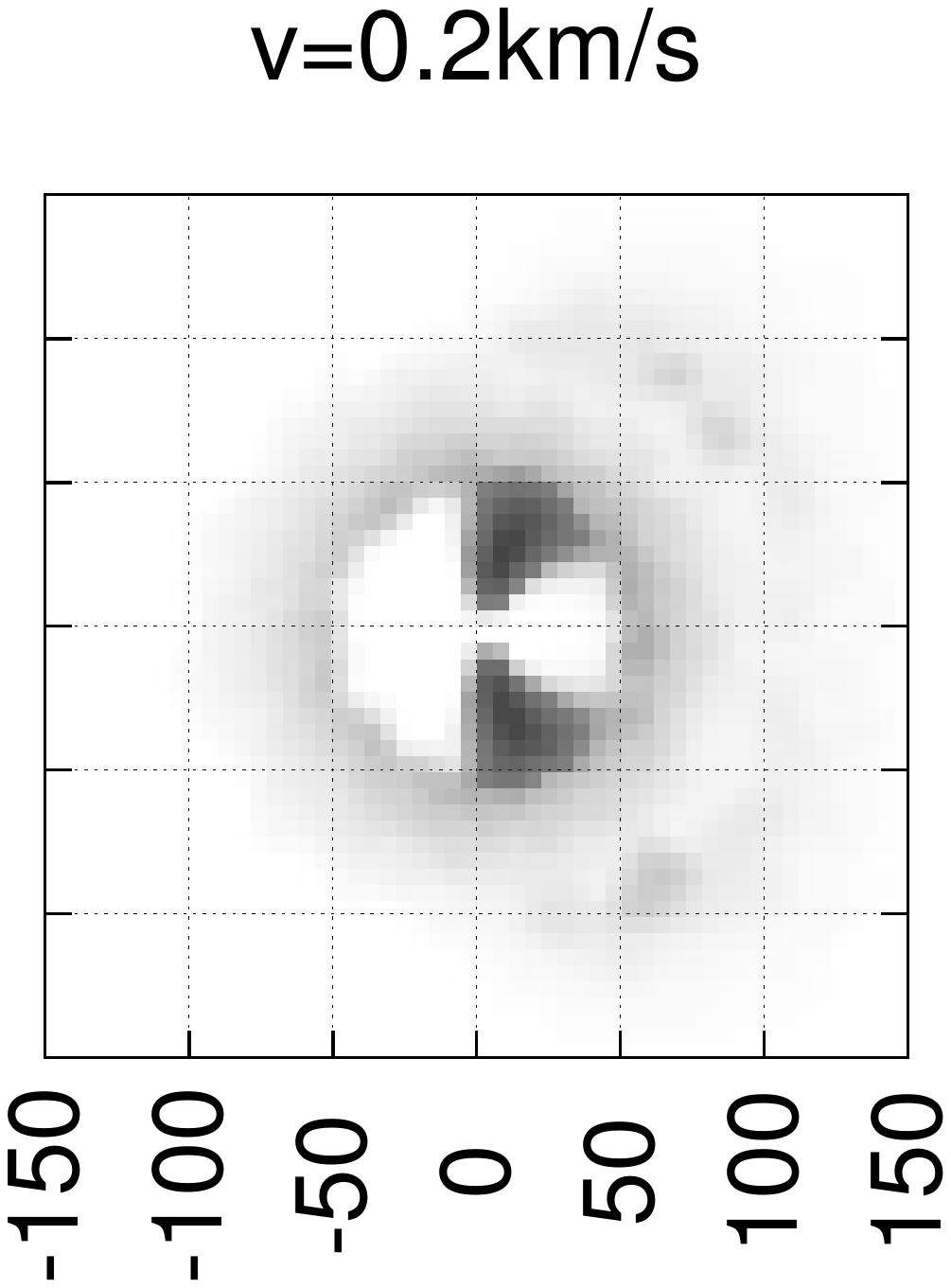}  \hspace{ -3.8cm}& \hspace{ -3.8cm}
   \includegraphics[angle=0,width=5.8cm]{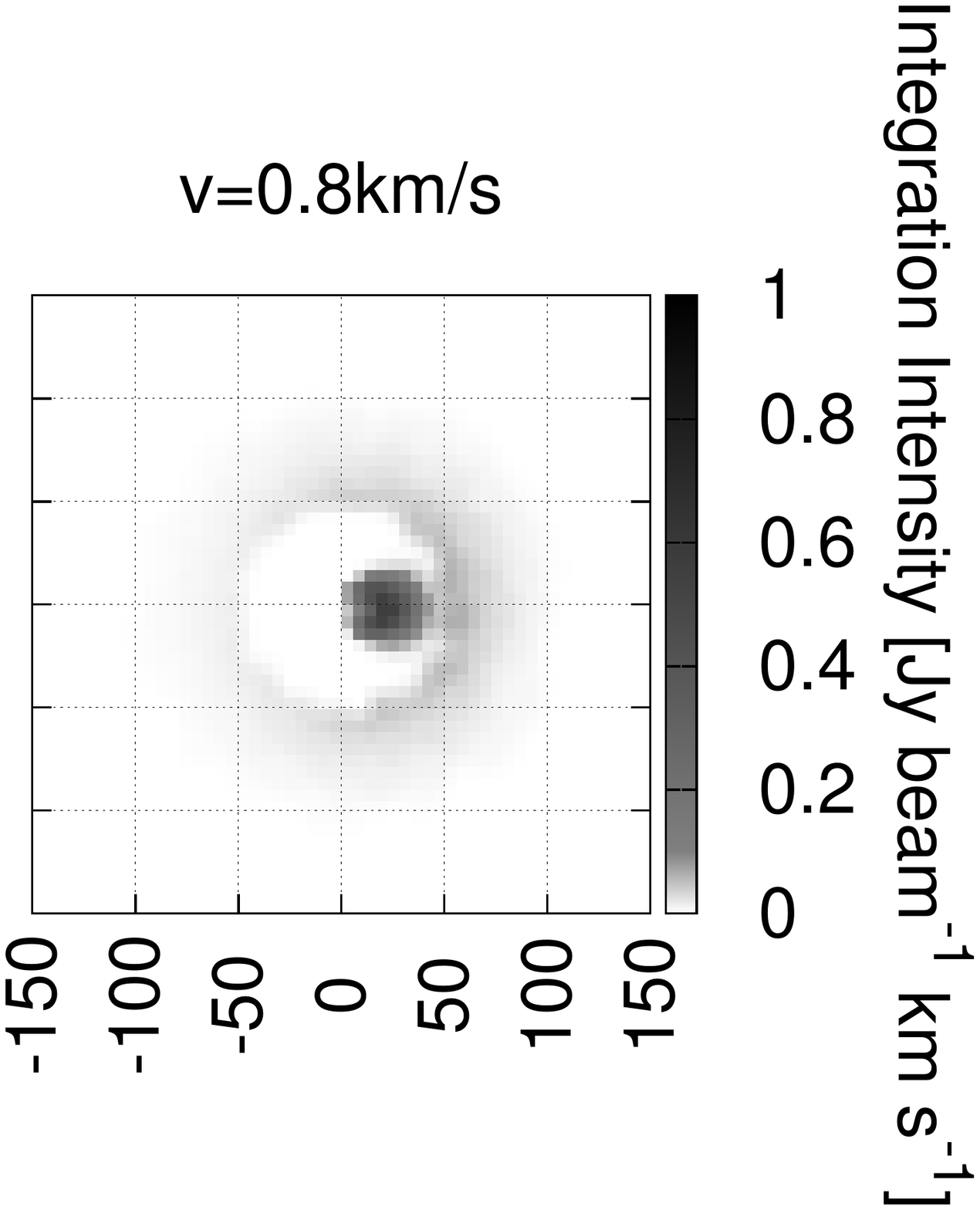}
   \end{tabular}
\caption{Simulated integrated emissions maps [Jy beam$^{-1}$ km s$^{-1}$] for the HCO$^+$ line at 267.56GHz (1.1205 mm) without (top) and with (middle and bottom row, classic Townsend breakdown model) lighting. The lighting is considered to occur in a region of 25AU$\ldots$50AU for the middle row,
and of 50AU$\ldots$100AU for the bottom row. These ALMA channel maps
were simulated for 10mJy per beam of 0''.65x0''.44. HCO$^+$ appears in a larger
fraction of a protoplanetary disk through the effect of lighting.}
\label{Fig:muran2}
\end{figure}

 The electric field accelerates the free electrons that ionize the
surrounding gas but also the positively-charged ion species to the
energy comparable to the electrons. Because the ionization energy is a
universal constant for each individual species, each ion will move
with a characteristic, constant velocity in the lighting zone that is
larger than the thermal gas velocity.
This will be unique observational feature to detect and distinguish breakdown models in
protoplanetary disks. 

In this model, it is assumed that the fractional abundances of HCO$^+$
relative to H$_2$ is $9\cdot 10^{-9}$. The value is taken from
r=100au, z=3h from the XR+UV disk chemistry model of \cite{walsh2012}. The
underlying assumption is that HCO$^+$ has been gradually produced by
XR+UV, and although the HCO$^+$ molecules may experience a sudden
accelerated by the lightning electric fields, this does not contribute
to the change of the number density of HCO$^+$ in the present model.

Velocity distribution of the ion species (e.g. HCO$^+$, DCO$^+$ and
N2H$^+$) can now be derived and the line profiles simulated
(Fig.~\ref{Fig:muran1} for HCO$^+$).  The two-dimensional
position-velocity images with lightning assumed to occure in a certain
disk region (middle, bottom) is shown as simulated ALMA\footnote{Atacama Large Millimeter Array, www.almaobservatory.org} channel maps
in Fig.~\ref{Fig:muran2}. The change in the line profiles depending on
the presence of a large-scale discharge is demnstrated in
Fig.~\ref{Fig:muran1}. We found lightning features of 10-100mJy appear
in line profile. Using ALMA, full-disk lightning will produce
100$\sigma$ signals at 56pc (TW Hya, Fig.~\ref{Fig:muran2}) and
20$\sigma$ signals at 140pc (Orion nebula; see \citealt{mur2015}).

\bigskip

\subsection{Future studies}

Combining the expertise available on solar system and terrestrial
atmosphere research and electrical phenomena therein will help
answering some of the following question, but might also be inspired by these questions:

\begin{itemize}
\item In how far can terrestrial and solar system lightning statistics guide our expectation for extrasolar environments like on extrasolar planets, brown dwarfs and in proroplanetary disks? 
\item How are brown dwarf (and extrasolar planet) atmospheres affected by the irradiation of their companion (or host star)? 
\item Going beyond the Solar System, clouds are present in brown dwarf and exoplanetary atmospheres. What kind or which combination of atmospheric electricity phenomena could explain the required levels of ionized atmosphere to provide an explanation for the continuous radio emissions, the 656 nm and X-ray emissions?
\item What would be the possible atmospheric optical and chemical signatures in the case that lightning activity exists in exoplanetary and brown dwarfs atmospheres? 
\end{itemize}


\section{Conclusion}\label{s:concl}

Electrification processes and electrical phenomena are ubiquitous:
dust charging and discharging is linked to electric gas breakdown in
planetary atmospheres inside and beyond the solar system where it is
involved in global circuits and the occurrence of plasma
processes. Charge processes play a major role in modifying the ambient
chemical composition and the transport properties of neutral gas also
in protoplanetary disks where planets form. Charged relativistic
grains are suggested as potential primary particles for ultrahigh
energy Cosmic Rays (\citealt{2014arXiv1412.0578H}). The following set
of challenges has emerged as common for the themes of this paper
 which have been guided by the
workshop 'Electrification in dusty atmospheres inside and outside the
solar system' held in September
2014\footnote{http://leap1.sciencesconf.org/}:

\begin{itemize}
\item[(a)] An increased population of ion, free electron and radicals
  lowers the chemical potential for gas-species reactions, leading to
  potentially observable spectroscopic fingerprints, and
\item[(b)] also increases the thermal and electrical conductivity of
  the gas to a certain threshold, enabling more energetic phenomena such as lightning to take
  place or accretion to proceed during star- and planet formation.n
\item[(c)] The presence of free charge may be transient (as in
  lightning) but the electrostatic influence can endure:
  Charging of dust and aerosols can influence electrostatic character
  of the ambient atmosphere on longer length and time scales to
  produce small non-thermal populations of energetic particles.
\item[(d)] Finite enhanced electrical conductivity can allow magnetic
  relaxation, and access to stored magnetic energy as a general source
  of excitation which in unavailable to neutral gases.
\item[(e)] Non-thermal electrons may facilitate
  chemical reactions in ways that are classically (i.e.
  gas-thermodynamically) unlikely: For example, dissociative
  electron attachment can produce oxygen radicals at little energetic
  cost, leading to oxidative reactions proceeding at a rate
  inconsistent with ambient temperatures, or the formation for complex carbohydrates. 
\item[(f)] Charged dust may evolve differently compared to neutral dust:
  Long-range organization produced by 
  electrostatic effects could produce coherent dust dynamics that
  would not be possible if only fluid mechanics dominates.
  surface charging, leading to elongated growth (non-zero
  eccentricity: polarization of light is observable), or destruction
  of part of the grain population by Coulomb explosion.
\end{itemize}

These themes lead to the need of 
\begin{itemize}
\item Further research in dust charging mechanisms in the context of volcano lightning, atmospheres of Brown Dwarfs and exoplanets, and protoplanetary disks.
\item New instruments in future space missions to test new findings about electrical activity in Solar System Planets. Result would provide models that  could mimic  electrical activity in Brown Dwarfs and exoplanets.
\item Further research about the role of dust in electrical discharges in the upper atmosphere of the Earth.
\item New key insights in charge mechanisms of the Moon and asteroid
  fine dust grains by interacting with the solar wind and UV
  flux. This contributes to the fundamental knowledge of the Moon
  electric environment and will be very useful for further man
  missions to the Moon and unmanned missions to the asteroids.
\item 3D simulations of extrasolar atmospheres including chemical and
  electrical feedback of clouds in a magnetised gas.
\end{itemize}

\bigskip
{\bf Acknowledgment} ChH highlight financial support of the European
Community under the FP7 by an ERC starting grant 25743. DAD gratefully
acknowledges support from EPSRC via grant numbers EP/K006142/1 and
EP/K006088/1.  FJGV thanks the Spanish Ministry of Economy and
Competitiveness (MINECO) under projects FIS2014-61774-EXP and
ESP2013-48032-C5-5-R and the EU through the FEDER program.  We thank
all the participants to the workshop {\it Electrification in dusty
  atmospheres inside and outside the solar system} held in September
2014 in the Scottish Highlands for their input and inspiration. We
thank the Royal Astronomical Society, the ERC and the IoP
Electrostatics Group for financial support. We thanks Sarah Casewell
and Alejando Luque for their help in preparing Table 1. Gabriella
Hodos{\'a}n is thanked for helping with the literature
collection. Keri Nicoll is thanked for her inspiring feedback on
Sect.~\ref{s:eldis_solsys}\ref{ss:volc}.  Most of the literature
search has been performed using ADS.

\appendix

\section{Glossary}\label{s:AGl}
\textbf{AC}: alternating current\\
\textbf{asteroid}: small rocky bodies of inner solar system, ranging in size from 10m to 900m in diameter \\
\textbf{aurora}:  large diffuse light-emitting structures in the lower ionosphere (> 90 km) generated when energetic particles displaced from the ionosphere collide with ground state neutral species and excite them. The excited species (oxygen atoms and nitrogen molecules) emit light when returning to ground state\\
\textbf{carbonaceous compound}: material rich in Carbon; in an astrophysical context, such compounds usually are associated with primitive solar system remnants \\
\textbf{conduit, volcano conduit}:  the pipe that carries magma from the magma chamber, up through the crust and through the volcano itself until it reaches the surface.\\
\textbf{chondrule}: molten or partially molten droplets that appear as spherical, solid  inclusions of different chemical composition than  the matrix of their parent asteroid. They represent one of the oldest solid materials within the solar system.\\
\textbf{cosmic rays}: Ionized nuclei and electrons that are distinguished by their high energies. The ionized nuclei have energies ranging from 10$^6$ eV to greater than 10$^{20}$ eV and comprise 99\% of cosmic rays. They originating either from the Sun or outside the Solar System likely from Super Novae or Gamma Ray Bursts.\\
\textbf{cyclotron maser instability}: the mechanism whereby a population of relativistic electrons drift along an ambient magnetic fields, producing coherent radiation that reflects the magnetic field strength. \\
\textbf{DC}: direct current\\
\textbf{Debye length}: the scale-length associated with the violation of charge neutrality in a plasma, due to thermal fluctuations causing charge separation \\
\textbf{double diffusive convection}: a form of convection (i.e. hydrodynamic bulk motion) that is driven by two distinct gradients in fluid composition arising from two different species abundances. \\
\textbf{Druyvesteyn distribution}: a driven-equilibrium distribution function that takes into account the presence of large-scale electric fields in a plasma, as well as interactions with neutrals \\
\textbf{effective temperature, T$_{\rm eff}$ [K]} is a measure for the total radiation flux emitted at all wavelength $\lambda$ [\AA] (T$_{\rm eff}=F_{\rm tot}/\sigma$ with $F_{\rm tot}=\int F_\lambda d\lambda$; $F_\lambda$ [erg/s/cm$^2$/\AA] -- radiative flux; $\sigma$ [erg cm$^{-2}$s$^{-1}$K$^{-4}$] - Stefan-Bolzmann constant)\\
\textbf{electrical conductivity}: a material property that characterises the ease with which electricity can be passed through it \\
\textbf{extrasolar}: outside or beyond the Solar System \\
\textbf{fair weather current}: atmospheric current of ions present during undisturbed weather condition\\ 
\textbf{floating potential}: the electric potential (or voltage) that spontaneously arises on a surface immersed in a plasma, due to the difference in mobility between electrons and heavier ions  \\
\textbf{fractoemission}: the emission of particles (charged, neutral and photons) during and after fracturing of surfaces \\
\textbf{Geiger counter}: an instrument for measuring ionizing radiation,  detects alpha particles, beta particles and gamma rays using the ionization produced in a Geiger-M\"uller tube\\
\textbf{Hydrometeors}:  water droplets or ice particles\\
\textbf{ion acoustic wave}: a sound wave carried by the motion of plasma ions, as opposed to the electrons \\
\textbf{isotope ratio}: means of quantifying the relative abundance of isotopes (which are elements which have nuclei that differ in the number of neutrons, but which are otherwise chemically identical).  \\
\textbf{jet}: directed and confined stream of fluid or gas \\
\textbf{Jy}: Janskys (symbol: Jy) are the unit for the observed spectral flux density: 1 Jy = $10^{−26}$ W m$^{−2}$ Hz$^{−1}$. The unit is named after Karl G. Jansky, an US radio astronomer. His discovery of  the radio waves emitted by the Milkyway initiated radio astronomy as new research field.\\
\textbf{M-dwarfs}: the lowest mass (0.075-0.5M$_{\rm Sun}$), main sequence  stars; most common type of stars in the milky way\\
\textbf{magma}: fluid mixture of molten and semi-molten rock and volatiles produced by volcanism\\
\textbf{magnetic Reynolds number}: a dimensionless number equal to the ratio of advective to diffusive effects, where the latter are characteristic of the magnetic field. Hence a large magnetic Reynolds number ($\gg$1) means that the magnetic field plays a dominant role in the fluid evolution as diffusion is unimportant and the magnetic field is advected with the fluid flow.\\
\textbf{M (spectral type)}: Stars are grouped into spectral classes which link to their effective temperature, luminosity, evolutionary state. The spectral class M indicates the coolest stars on the main sequence where hydrogen burning assures the most stable phase in a star's life. Brown dwarfs are cooler than M-dwarfs and were classified as L, T and Y with Y being the coolest and most planet like.\\
\textbf{mesocyclone}: a rapidly rotating column of air, typically a few miles in diameter, readily identified by its characteristic radar signal and consistent with storm conditions\\
\textbf{mobility}: the drift  speed of a charged particle produced when subjected to a steady electric field \\
\textbf{near-IR}: electromagnetic radiation in the wavelength range 800nm to 5 microns \\
\textbf{Ohm's law}: relates the electrical current flowing between two points to the potential difference between those same points\\
\textbf{plasma void}: a finite region in a dusty plasma which is dust-free \\
protoplanetary disk: a region of dust and rocks orbiting a young star from which planets could be formed \\
\textbf{regolith}: layer of unconsolidated dust and fragmented rock that covers a terrestrial planet  \\
\textbf{shock tube}: a device designed to create shocks (i.e. sharp density and pressure discontinuities) in gases, usually in order to produce ionization fronts \\
\textbf{sounding}: a method to measure local temperature, humidity, wind etc. in the Earth atmosphere by means of radio sonds, laser beams (optical) or sound waves\\
\textbf{sprite}: a large (50 km high and 10-20 km wide) electrical discharge that occurs above thunderclouds at altitudes around 50-85 km  with a diffuse region (above 70-75 km) and a filamentary (streamer-like) region (below 70-75 km) [values are given for Earth]
\textbf{stratosphere}: major layer of the Earth's atmosphere, lying above the troposphere (0-11 km) and below the mesosphere (50-90 km) \\
\textbf{thunderstorm}: storm characterised by the presence of lightning \\
\textbf{triboelectrification}: the process whereby two surfaces can acquire or lose charges by mutual collision \\
\textbf{troposphere}: lowest layer of the earth's atmosphere, lying between 0 and 11km, in which most of the weather phenomena occur  \\
\textbf{turbulence}: chaotic flow in which the pressure and gas velocity change rapidly in space and time  \\
\textbf{volcanic ash}: fragments of rock created during a volcanic eruption, usually 2mm or less in diameter\\
\textbf{volcanic conduit}: passage or tube created by the flow of magma in a volcano \\
\textbf{volcanic plume}: the gas and ash cloud ejected into the atmosphere by a volcanic eruption 
\newpage


\newpage

\bibliographystyle{mn2e}
\bibliography{bib_master}

\end{document}